\documentclass{aa}

\usepackage{morefloats}
\usepackage[utf8]{inputenc}
\usepackage[T1]{fontenc}
\usepackage{csquotes}
\usepackage{amsmath}
\usepackage{amssymb}
\usepackage{pifont}
\usepackage{tabularx}

\usepackage{units}
\usepackage{lscape}
\usepackage{longtable}
\usepackage{lmodern}
\usepackage{subfig}
\usepackage{rotating}
\usepackage{pdflscape}
\usepackage{url}
\usepackage{footnote}
\usepackage{float}
\usepackage{tabularx}
\usepackage{breakurl}
\usepackage{tikz}
\usetikzlibrary{shapes,arrows}
\usetikzlibrary{svg.path}
\usetikzlibrary{chains}
\usetikzlibrary{calc}
\usepackage{graphicx}
\usepackage{txfonts}
\usepackage{hyperref}
\begin{document}

\authorrunning{Müller et al.}

   \title{Sharpening up Galactic all-sky maps with complementary data}

   \subtitle{A machine learning approach}

   \author{
          Ancla Müller\inst{1},
          Moritz Hackstein\inst{1},
          Maksim Greiner\inst{2},
          Philipp Frank\inst{3,4},
          Dominik J.\,Bomans\inst{1},
          Ralf-Jürgen Dettmar\inst{1}\and
          Torsten Enßlin\inst{3}
          }

   \institute{Ruhr-Universit\"at Bochum,
  Universit\"atsstra{\ss}e 150, 44801 Bochum, Germany, Fakult\"at f\"ur Physik und Astronomie, Astronomisches Institut\\
              \email{amueller@astro.rub.de}
      \and IInsight Perspective Technologies GmbH, Lichtenbergstraße 8, 85748 Garching, Germany
	\and
             Max Planck Institute for Astrophysics, Karl-Schwarzschild-Str. 1, 85741 Garching, Germany
     \and Ludwig-Maximilians-Universität München, Geschwister-Scholl-Platz 1, 80539 München, Germany
             }

   \date{Received June XX, 2018; accepted XXX XX, 2018}


  \abstract
   {Galactic all-sky maps at very disparate frequencies, such as in the radio and
$\gamma$-ray regime, show similar morphological structures. This mutual information
reflects the imprint of the various physical components of the
interstellar medium.}
  {We want to use multifrequency all-sky observations to test resolution improvement and restoration of unobserved areas for maps in certain frequency ranges. For this we aim to reconstruct or predict from sets of other maps all-sky maps that, in their original form, lack a high resolution compared to other available all-sky surveys or are incomplete in their spatial coverage. Additionally, we want to investigate the commonalities and differences that the ISM components exhibit over the electromagnetic spectrum.}
   {We built an $n$-dimensional
representation of the joint pixel-brightness distribution of $n$ maps
using a Gaussian mixture model and investigate how predictive it is. We study the extend to which one map of the training set can be reproduced based on subsets of other maps?}
   {Tests with mock data show that reconstructing the map of a certain frequency from other frequency regimes works astonishingly well, predicting reliably small-scale details well below the spatial resolution of the initially learned map.
Applied to the observed multifrequency data sets of the Milky Way this technique is able to improve the resolution of, for example, the low-resolution Fermi LAT maps as well as to recover the sky from artifact-contaminated data such as the ROSAT 0.855\,keV map.
The predicted maps generally show less imaging artifacts compared to the original ones.
A comparison of predicted and original maps highlights surprising structures, imaging artifacts (fortunately not reproduced in the prediction), and features genuine to the respective frequency range that are not present at other frequency bands.
We discuss limitations of this machine learning approach and ideas how to overcome them.
In particular, with increasing sophistication of the method, such as introducing more internal degrees of freedom, it starts to internalize imaging artifacts.
}
   {The approach is useful to identify particularities in astronomical maps and to provide detailed educated guesses of the sky morphology at not yet observed resolutions and
locations.}


   \maketitle
%

\section{Introduction}
Historically, research on astrophysical phenomena has been separated based on the different means of observation into the multiple areas of the electromagnetic spectrum.
There exist radio, microwave, IR, optical, X-ray, and $\gamma$-ray astronomy as individual disciplines.
Here, an integrated approach is followed in which we combine information on the Galactic sky covering the full width of the measurable electromagnetic spectrum.
This data set is jointly analyzed by investigating commonalities between different frequency bands using machine learning.
In particular, we inspect to which degree the Galactic diffuse X-ray and $\gamma$-ray skies are encoded in the other frequency bands, for example, in the radio and microwave regimes.\par
From stellar observations it is known that the Milky Way is composed of a thin and a thick disk whose structure is dominated by spiral arms and a bulge, as well as a bar in the central region (e.g., \citealt{MW2002}).
The disk is surrounded by a halo, which has at least the radius of the disk.
The gas and dust distribution partly coexists with the stellar structures, such that this coarse classification also holds for the diffuse emission.
Additionally, the halo is dominated by different diffuse small and large-scale structures (typically outflows from the Galactic disk) that are observable in different frequency ranges, but overall the physical origin is subject of current research.
For example, in the radio, X-ray, and $\gamma$-ray regime, large-scale structures such as the North Polar Spur or the Fermi bubbles expand above and below the plane \citep{su2010}.
The origin of the Galactic halo structures is under current research, as the mechanisms driving gas and energetic particles out of the disk into the halo are not identified entirely.\par
Within previous works only small parts of the whole electromagnetic spectrum of all-sky observations are used to, for example, investigate the correlation between the Galactic H\,I emission and the cosmic microwave background, which turned out to be nonsignificant \citep{2007PhRvD..76h7301L}.
\citet{su2010} found a correlation at the edges of the Fermi bubbles between X-ray and $\gamma$-ray observations.
The correlation between dust and H\,I has been investigated by \cite{dust2014} on a contiguous area centered on the southern Galactic pole with regard to the dust evolution within the diffuse interstellar medium.
Subsequently, \cite{dust2016}  focused on analysing Galactic outflows and halo material by also investigating the all-sky far-infrared and H\,I correlation.
\cite{1999ApJ...524..867F} presented predicted all-sky maps for the submillimeter and microwave emission determined via extrapolation over the diffuse interstellar dust emission using theoretical models.
\cite{2008MNRAS.388..247D} compiled a data set using the radio frequency maps available at that time and computed a model based on three components predicting the diffuse radio sky by principal component analysis.
In \cite{CO} the authors derived different all-sky foreground maps from the microwave regime and radio continuum observations via Bayesian analysis.
\par
In contrast to previous works, we analyze the multifrequency all-sky data sets covering the electromagnetic spectrum from radio to $\gamma$-ray frequencies without the use of theoretical models.
The importance of this endeavor has been outlined by, for example, \cite{doi:10.1146/annurev-astro-082214-122457} who stated that although the Galactic emission in separate frequency regimes is investigated in more and more detail, the physical and phenomenological connections across all frequencies are still studied insufficiently due to their complexity.
Especially, the authors state that while cosmic rays are major actors in the feedback between high-energy events in galaxies and the interstellar medium, only very little is known about how this feedback operates from the microscales of plasma physics and shocks to the large scales of galactic fountains and winds.\par
In \cite{Fermi1} the nonparametrically estimated $\gamma$-ray all-sky maps have been published as well as a physical two-component decomposition of their diffuse emission over the whole sky.
The decomposition has been achieved by pixelwise spectral fits of two template spectra, which were taken from spectrally extreme regions: the southern Fermi bubble and the disk cloud complex.
The authors claim that one component is predominantly composed of leptonic and the other of hadronic emission.
A comparison of these components to the rest of the electromagnetic spectrum can test this claim.
We investigate to what extent individual sky components, here the proposed leptonic and hadronic $\gamma$-ray emission, are encoded in the complementary all-sky multifrequency data.
We find that re-predicting a learned low-resolution $\gamma$-ray sky map from the complementary data sets yields a higher resolved map.
It contains physically plausible small-scale structures.
This prediction of unobserved features is possible as these are caused by processes in the interstellar medium imprinting onto observations over the whole electromagnetic spectrum.
The physical ingredients required to generate the $\gamma$-ray components, such as gas, dust, photon fields, and relativistic particles, reveal their presence also in other frequency ranges.
Furthermore, a comparison of predicted to observed $\gamma$-ray or X-ray maps identifies structures in particular frequency bands, which cannot be attributed to other frequencies.
Examples for this are the southern Fermi bubble in the $\gamma$-ray regime or the brightness of the Vela region and the North Polar Spur in the X-ray regime.


\section{Gaussian mixture models}
To investigate how the comprehensive diffuse all-sky radiation in one part of the electromagnetic spectrum relates to other parts, we use Gaussian mixture models (GMMs, \citealt{Bishop}). This is a machine learning method to abstract the measured information in a nearly model-free fashion.
GMMs are able to describe an empirical distribution function (usually being present in form of samples) by a mixture of continuous Gaussians.
They answer questions such as, given the measurement $x$ and $y$, what is a probable value for $z$ if the GMM was trained with samples of corresponding $x$, $y$, and $z$ values.
We demonstrate that a GMM is able to plausibly predict features of the X-ray and $\gamma$-ray sky.
For this, a set of $n$ available all-sky maps is ingested into a GMM by forming $n$-dimensional data vectors out of the $n$ diffuse brightnesses at all pixel positions of a sky pixelization.
We use $n \leq N=39$ maps dependent on the problem we analyze and train the GMM with the log-brightnesses (magnitudes\footnote{We here define magnitudes as the positive natural logarithm of the flux values of a map. We do not have the need to be specific about the flux value corresponding to magnitude zero; this can be left arbitrary and changing from map to map. We note that this convention differs from the usual astronomical one, where magnitudes refer to the negative flux logarithm for the base 10.}).
The GMM then yields a probability distribution for the magnitude vectors $d(x)$ at different sky positions $x$.
Thus, each pixel carries one such vector. 
The set of all these vectors will be one training data set for the GMM.
This GMM should learn the probability distribution of these magnitude combinations occurring on the sky.\par
A GMM represents a multidimensional probability distribution function (PDF) of some data vector  $d=(d_1 \; d_2\; ... \;d_n)^\textrm{T}$ by a sum of multidimensional weighted Gaussians
\begin{equation}
P(d|\pi, \mu, \Sigma)= \sum _{k=1}^K \pi _k \mathcal{N}(d|\mu _k,\Sigma _k)
\label{GMM_likelihood}
\end{equation}
with $\pi _k \, \in [0,1]$, $\sum _{k=1}^K \pi _k =1$ being the normalized weights of the $K$ Gaussian components, $\mu _k$ their means, and $\Sigma _k>0$ their strictly positive definite covariances.\par
In our case, we use a GMM where prior information is included into the training such that the likelihood in Equation \ref{GMM_likelihood} is multiplied by the prior information to a posterior given by
\begin{equation}
P(\pi, \mu, \Sigma |d) \propto P(d|\pi, \mu, \Sigma)\,P(\pi, \mu, \Sigma).
\label{posterior}
\end{equation}
Here, the inverse Wishard distribution is the conjugate prior for the covariance matrices given by
\begin{equation}
P(\Sigma_k) \propto |\Sigma_k |^	{-\alpha -1}\, \exp \{-\mathrm{tr} \Psi \Sigma_k^{-1} \},
\end{equation}
where $\Psi = \beta I_n$ with $I_n$ as the identity matrix. The free parameters $\alpha$ and $\beta$ are canonically chosen to always be 1 during the GMM training. \par
Here, the training of the GMM is based on the Expectation-Maximization (E-M) algorithm \citep{Bishop} which maximizes the natural logarithm of the posterior (Eq. \ref{posterior}) with respect to the weights $\pi _k$, the means $\mu _k$, and the covariance matrices $\Sigma_ k$ for each Gaussian component $K$.
These parameters will be updated to a predefined accuracy via iterative maximization steps.
In our case, we accept the found solution of the E-M algorithm when the difference between the current iteration solution of the posterior (Eq. \ref{posterior}) and the previous one is smaller than $10^{-6}$ in the dimensionless magnitude units we choose later on.\par
A trained GMM, one from which its parameters, the weights, means, and covariances of the Gaussian components are appropriately specified,
permits to inspect relations between different dimensions $x$ and $y$ of a data vector $d = (x,y)$. Here, $d$ splits into two parts $x=(d_1,...,d_m)$ and $y=(d_{m+1},...,d_n)$.
For example the conditional probability distribution (CPD) of $x$ given $y$ can be reconstructed by
\begin{equation}
P(x|y)= \frac{P(x,y)}{P(y)},
\label{cond}
\end{equation}
where
\begin{equation}
P(y)=\int \textrm{d}x\, P(x,y)
\end{equation}
as marginalization over $x$ can be calculated analytically due to the integrability of multidimensional Gaussians.
These CPDs summarize the knowledge on $x$ in case that $y$ is known and are therefore a mean to, for example, predict missing data in case of incomplete knowledge.
We are also able to compute the standard deviation pixel-wise for each prediction. However, the GMM is overall underestimating the uncertainties in the brightest regions, as discussed in Appendix \ref{App:std}.
%
%

\section{Verification} \label{Chap:Ver}
In the following, we present certain tests of the functionality of the GMM on simulated Galactic all-sky maps. These mock data sets are computed in Python \citep{Python} utilizing the versatile signal inference library NIFTy \citep{Nifty}.
For these tests we constructed simplified simulated Galactic emission structures.
We check whether the GMM is able to recover one map of the training set from a set of other maps.
We find that the method is reliable at least under the test conditions and that it is able to improve image resolution as well as to restore or recreate unobserved or defective areas.
\subsection{Simulation}
\label{sec:simulation}
\begin{figure}[h!]
\centering
 \subfloat[\label{MockI}]{\includegraphics[width=\columnwidth]{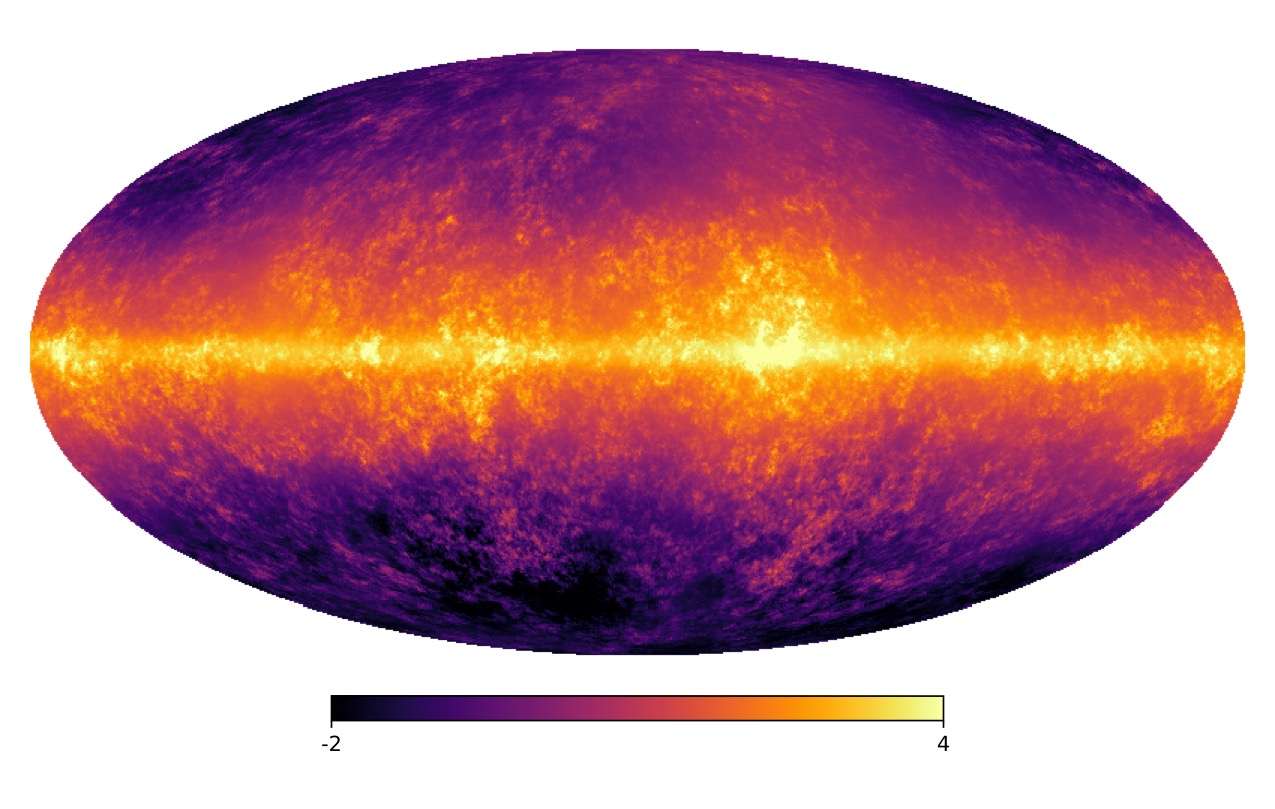}}
 \\
   \subfloat[\label{MockII}]{\includegraphics[trim=0 60 0 0, clip, width=0.5\columnwidth]{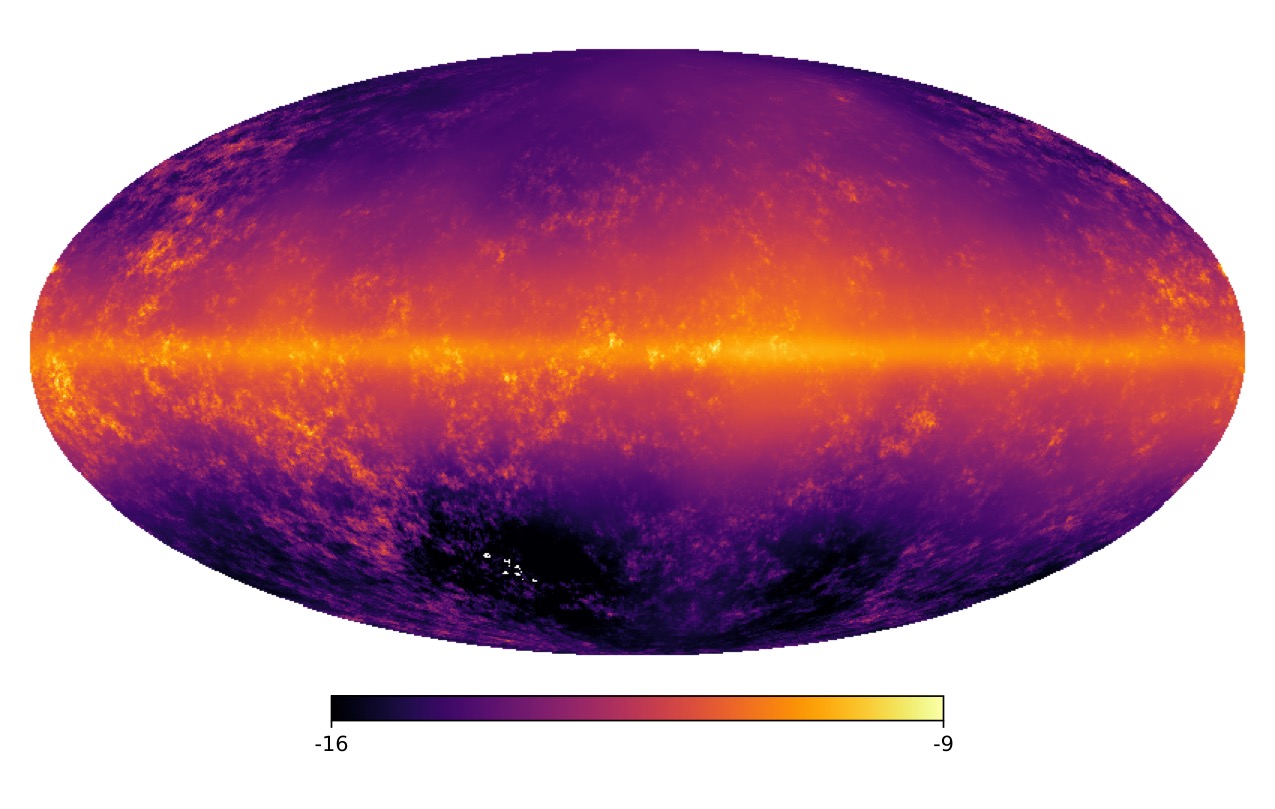}}
   \subfloat[\label{MockIII}]{\includegraphics[trim=0 60 0 0, clip, width=0.5\columnwidth]{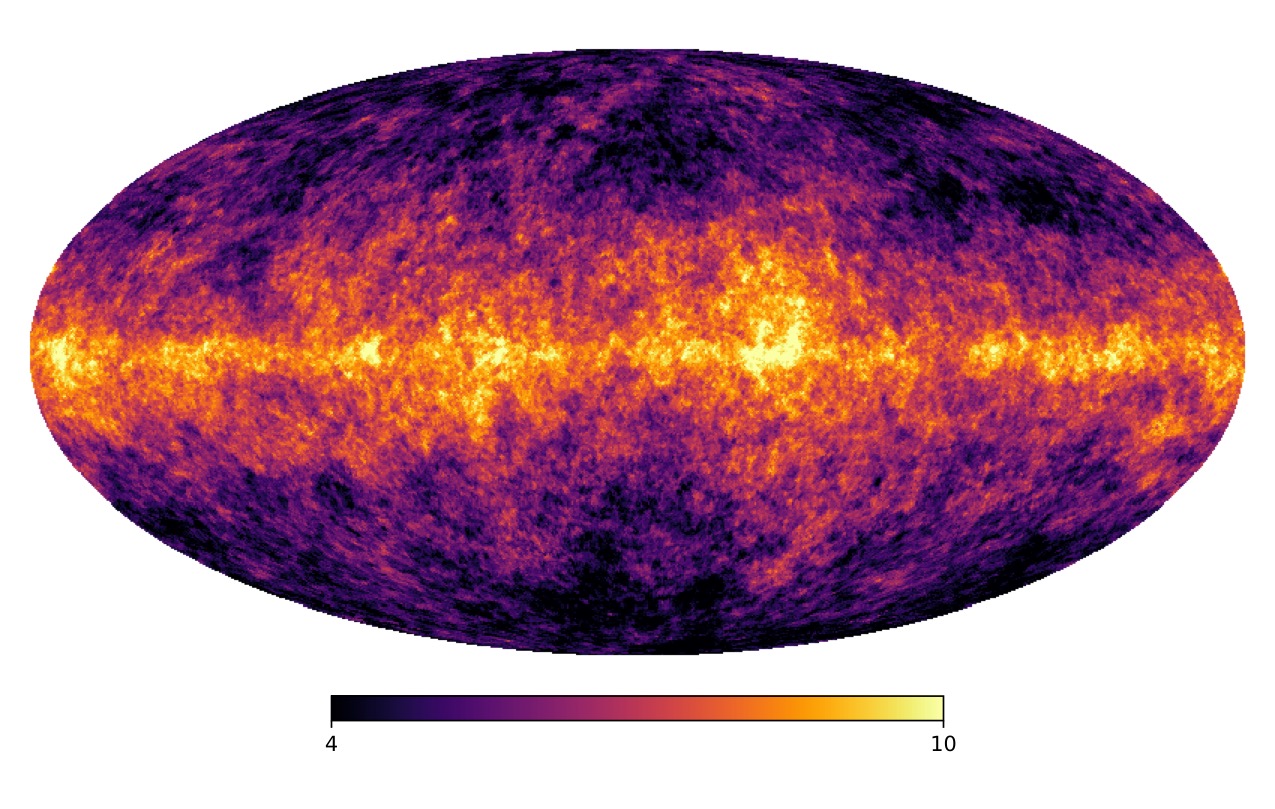}}
  \caption[]{Three of the six simulated Galactic all-sky maps; logarithmic scaling. Panel (a) shows the simulated Galactic all-sky map Mock\,I, panel (b) Mock\,II, and panel (c) Mock\,III.}
  \label{Abb:mock_org}
\end{figure}
We generated simulated Galactic all-sky maps such that the Galactic morphology of the diffuse interstellar medium (ISM) is roughly replicated. For this purpose, we consider physical causes mimicked by $I$ different Gaussian random fields on the celestial sphere that have angular power spectra given by
\begin{equation}
C_i(\ell) = \frac{a_{i}}{\left(1+(\frac{\ell}{\ell_{i}})^2\right)^{\frac{\alpha _i}{2}}}
\end{equation}
in which the multiplier $a_i$, the width $\ell_i$, and the exponent $\alpha_i$ can be adjusted with $i\in \{1, ..., I\}$.
This kind of decreasing power spectra is chosen to describe smooth flux distributions (diffuse emission structures) as, for example, discussed in \cite{power}.
From such power spectra with arbitrarily chosen parameters \mbox{$a_i=\{1, 0.5, 3\}$}, \mbox{$\alpha_i=\{3, 3, 4\}$}, and \mbox{$\ell_i=\{1, 2, 1\}$} respectively, three physical cause fields $g_i(x)$ are randomly drawn.
Further, we emulate several physical phenomena by analytical functions.
The Galactic disk profile is reproduced using a Gaussian along the Galactic latitude $l$
\begin{equation}
h(x) = e^{-\frac{l(x)^2}{L_1}}+e^{-\frac{l(x)^2}{L_2}} \label{disk}
\end{equation}
with $L_1=\pi / 1000$\,rad for the thin disk and $L_2=\pi / 20$\,rad for the thick disk component.
We combine the causes and the profile into physical components to
\begin{equation}
\rho_i(x) = e^{h(x) + g_i(x)}. \label{disk}
\end{equation}
Then, these components are combined into flux magnitudes, our observables, according to
\begin{equation}
d_k(x) = \ln \left[\sum _i B_{ki} \, \rho_i(x) \right].
\label{eq:mock}
\end{equation}
Here, $B$ is a mixture matrix that we set up to give the mock data a realistic-looking level of complexity.
We experimented with several such matrices until we found one that provides a sufficiently difficult problem to the GMM, while producing sky maps that a layman could confuse with Galactic all-sky observations.
For the mixture matrix $B$ as well as for the schematics of the mock setup see Appendix \ref{App:Mat}.\par
The resulting data set consists of six different simulated all-sky maps representing arbitrary frequency bands of which we show three in Figure~\ref{Abb:mock_org}.
First, Figure \ref{MockI} presents the Galactic all-sky map Mock\,I, whose reconstruction is to be achieved in the following analyses.
Figures~\ref{MockII} and \ref{MockIII} display two of the remaining five simulated maps.
Here, Figure~\ref{MockII} contains mainly the mixture of two of the causes, one providing the smooth Galactic latitude profile and the other creating few but pronounced filament structures.
The same holds also for Figure~\ref{MockIII}, where the third cause yields strong and detailed filaments that are also visible in Figure~\ref{Abb:mock_org}.\par
The mixture of the causes yielded very distinct and diverse mock data sets.
Although the mock sky fluxes are simply linear superpositions of Gaussian random fields (multiplied by a galactic profile function), the logarithm applied to these mixed fluxes rescales them nonlinearily. The GMM will be exposed to those log-fluxes, or magnitudes, without having a concept of this nonlinear operation built in.
Therefore, this mock data is challenging the abilities of the GMM with structures and internal relations that partly resemble the complexity of the Galactic reality.
\subsection{Reconstruction}
\begin{figure*}[h!]
\centering
   \subfloat[\label{fig:predfull}]{\includegraphics[trim=0 60 0 0, clip, width=\columnwidth]{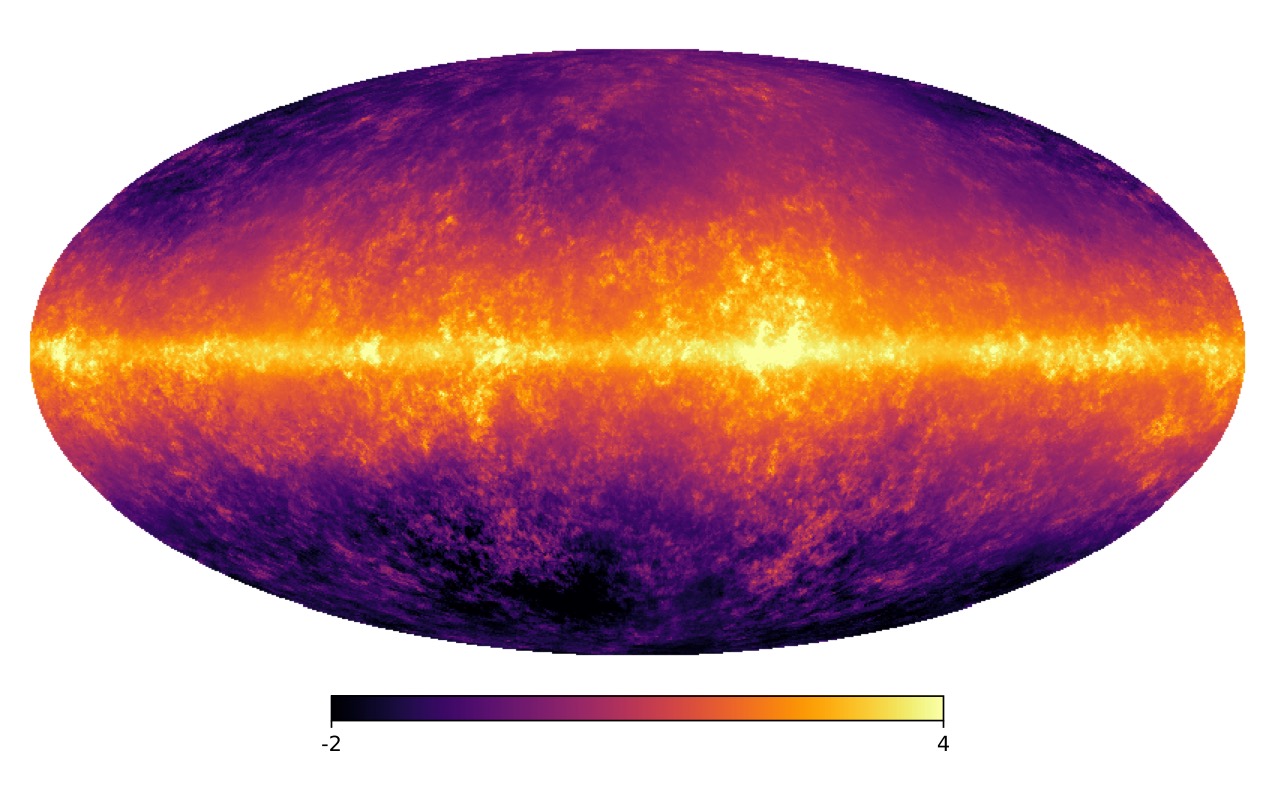}}
   \hfill
   \subfloat[\label{fig:org-predfull}]{\includegraphics
  [trim=0 60 0 0, clip, width=\columnwidth]{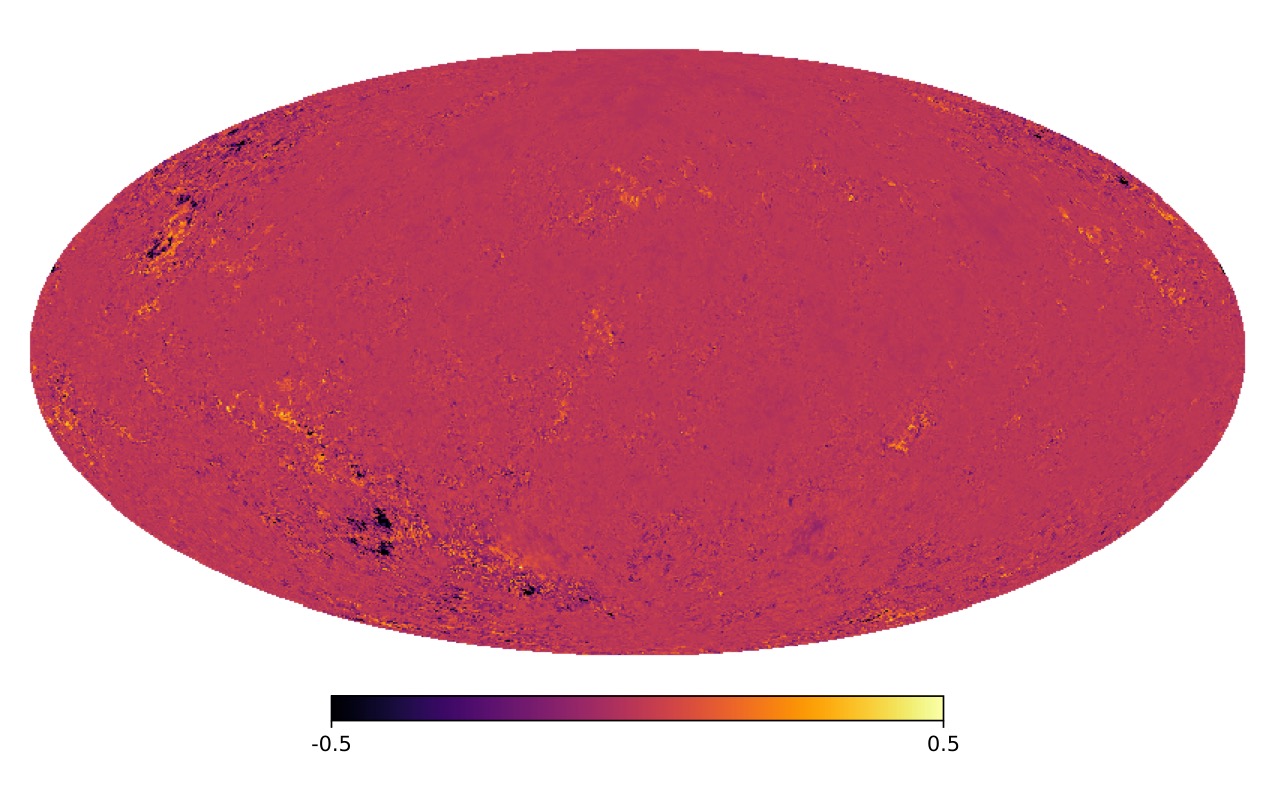}}
  \\
   \subfloat[\label{fig:smooth}]{\includegraphics[trim=0 60 0 0, clip, width=\columnwidth]{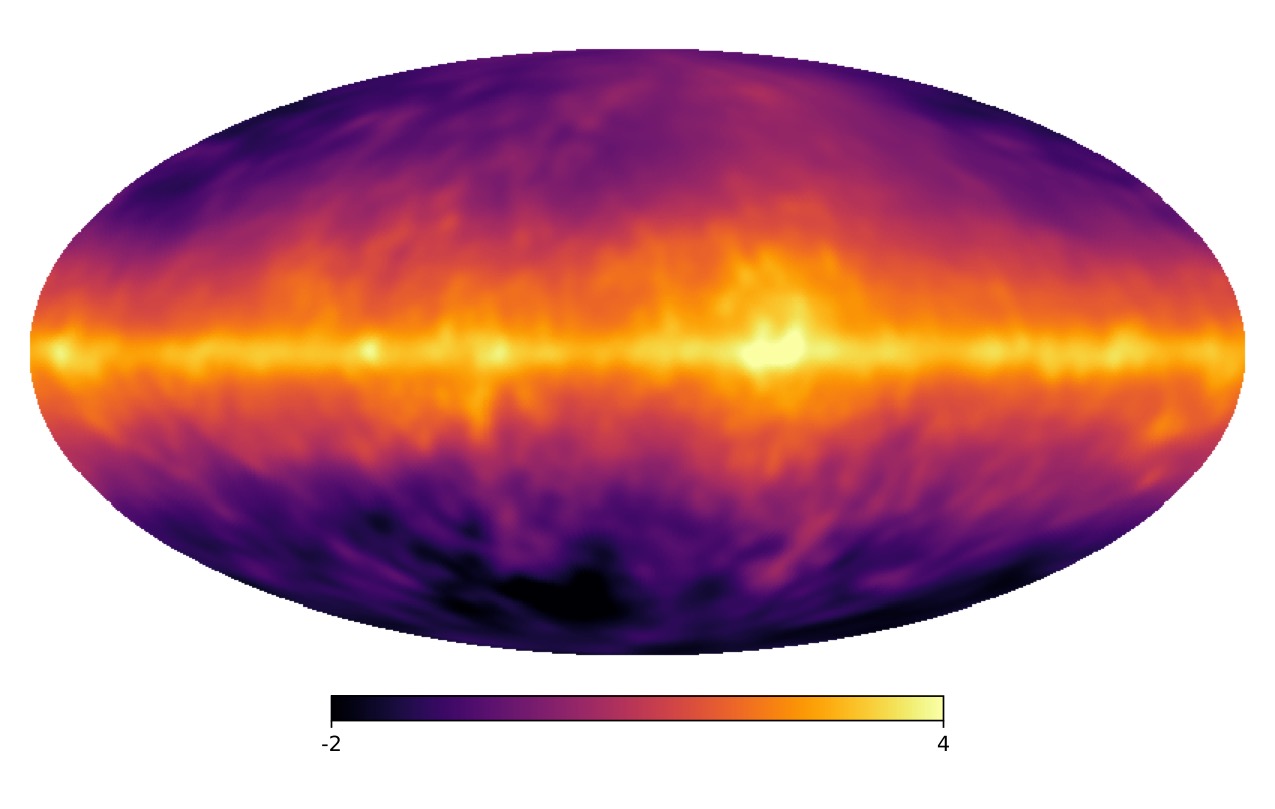}}
   \hfill
   \subfloat[\label{fig:org-smooth}]{\includegraphics[trim=0 60 0 0, clip, width=\columnwidth]{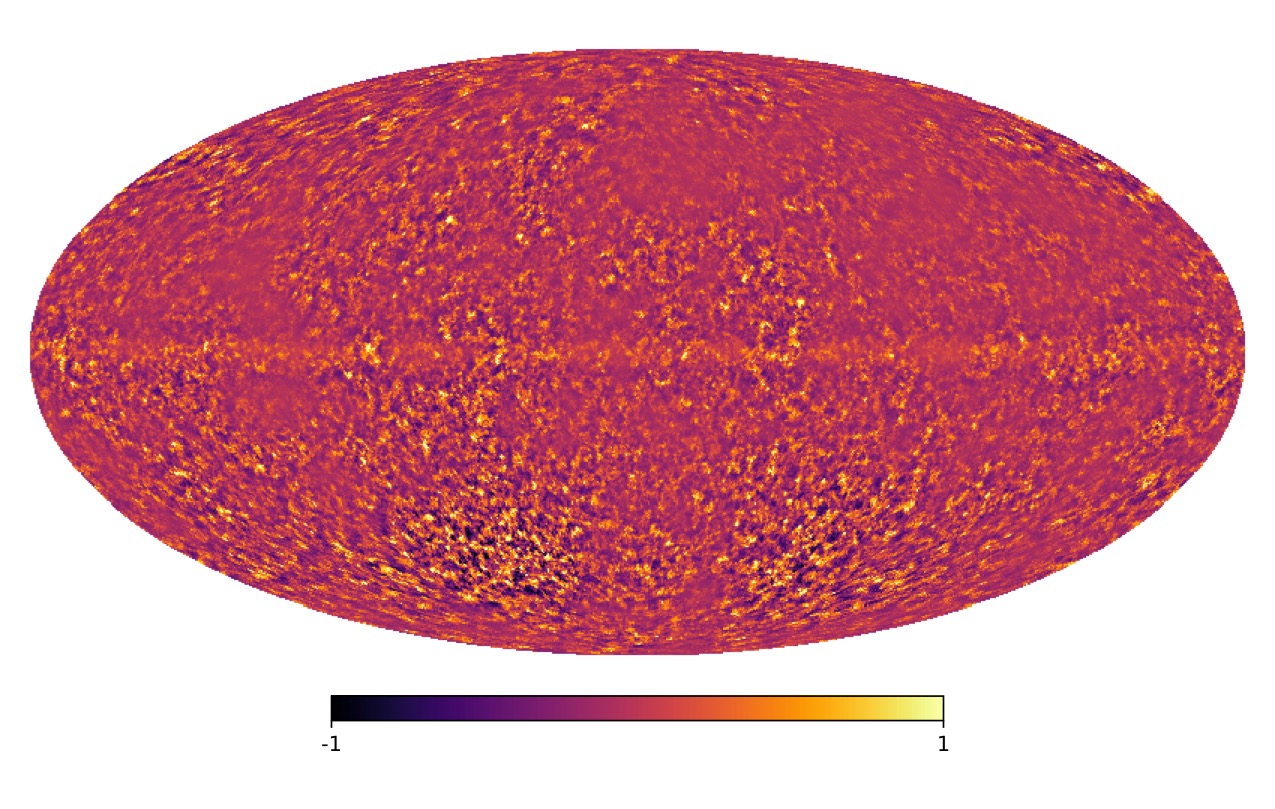}}
\\
   \subfloat[\label{fig:predsmooth}]{\includegraphics[width=\columnwidth]{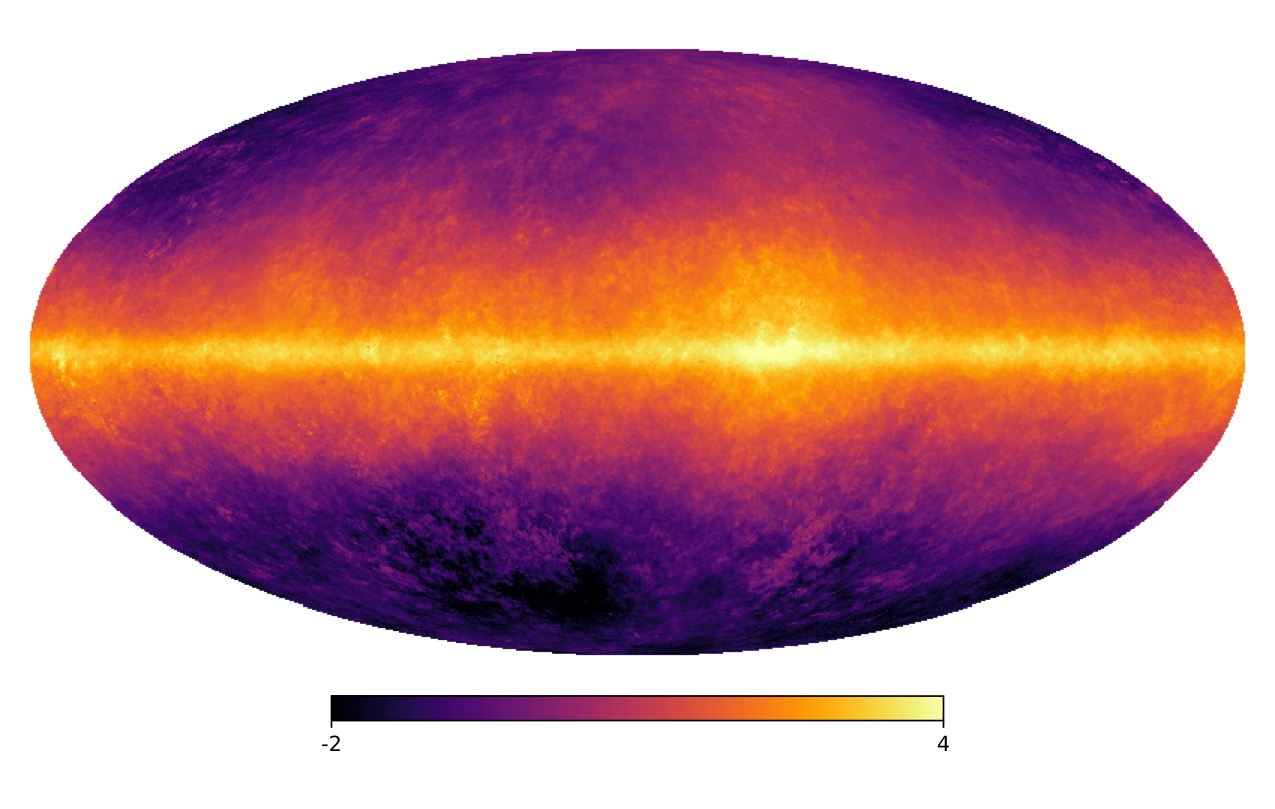}}
   \hfill
   \subfloat[\label{fig:org-predsmooth}]{\includegraphics[width=\columnwidth]{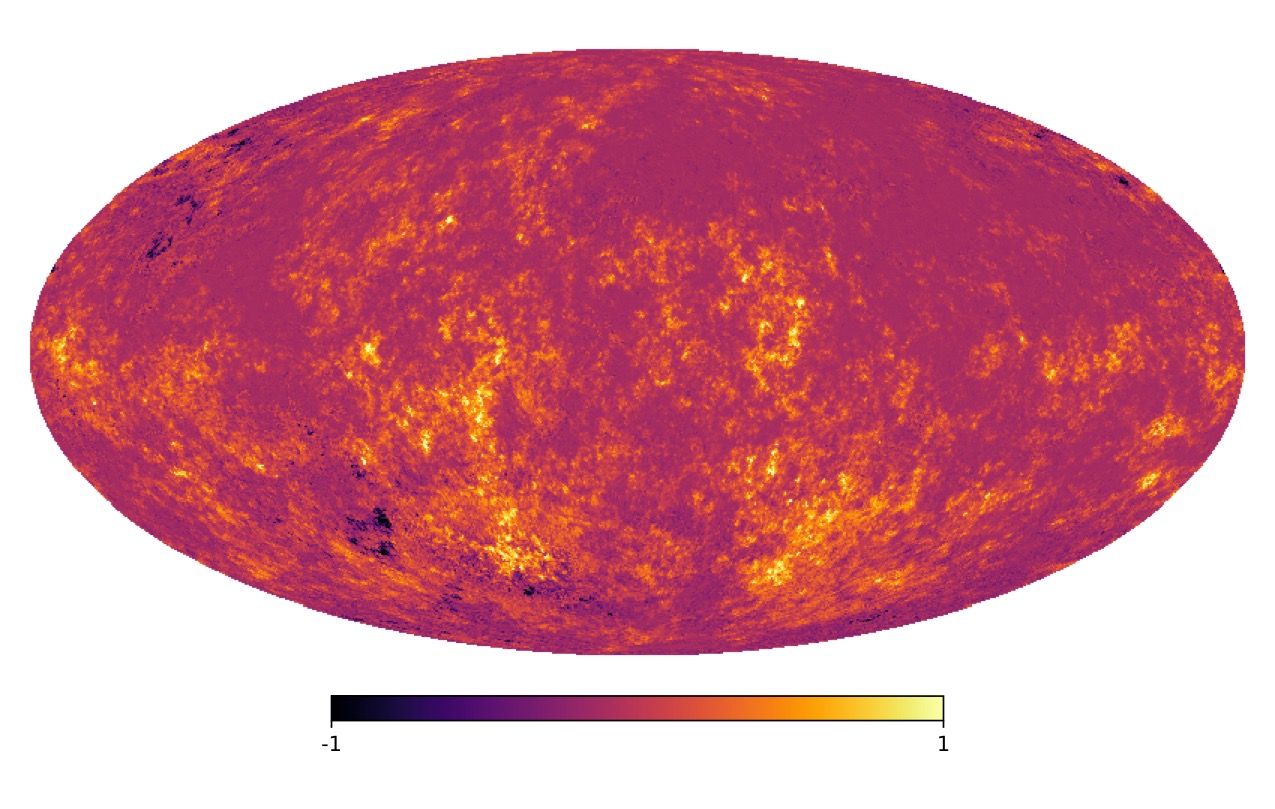}}
  \caption[]{Reconstruction and resolution improvement of Mock\,I. Note the different color bars for the maps on the left and right, respectively. The GMM is trained with $K=9$ Gaussians, and the predictions are determined from the CPD marginalized over the respective Mock\,I data set. Panel (a): Prediction of Mock\,I. Panel (b): Difference between the original Mock\,I map and the prediction, $\textrm{RMS}=0.05$. Panel (c): Smoothed map of Mock\,I. Panel (d): Difference between the original Mock\,I map and the smoothed one, $\textrm{RMS}=0.23$. Panel (e): Prediction of the smoothed Mock\,I map. Panel (f): Difference between the original Mock\,I map and the prediction based on the smoothed one, $\textrm{RMS}=0.17$.} \label{fig:predsmoothes}
\end{figure*}
First, we show that the computed probability distribution $P(d_1, ..., d_n)$ of the GMM with $K=9$ Gaussians trained with the six simulated all-sky maps at $\textrm{nside} = 128$ allows to reconstruct the original data.
For spherical all-sky maps in HEALPix format the parameter nside defines the number of pixels npix by $\mathrm{npix} = 12 \cdot \textrm{nside}^2$ \citep{HEALPix}.
We determine the expected magnitude value for each pixel from the deduced CPD (Eq.~\ref{cond}) of the GMM output.
The prediction (Fig.~\ref{fig:predfull}) based on the five remaining maps shows that the reconstructed map agrees well with the original map.
This is supported by the difference map shown in Fig.~\ref{fig:org-predfull}, which represents the pixel-wise subtraction of the predicted map from the original data.
While there are differences visible especially in the eastern side of the map, these differences are small compared to the original pixel values.\par
To quantitatively assess the difference between the original and the predicted map the root mean square (RMS) of the values in the difference map is computed, with an RMS of zero meaning complete agreement between the two maps.
By that, the RMS measures the accuracy of the prediction and assesses the quality of the different GMM input selections (see also Sec.~\ref{resolution} and \ref{restoring}).
We compare the RMS at the end of each subsection. \par
In case of this reconstruction we determined an RMS value of 0.05 for the difference map shown in Figure~\ref{fig:org-predfull}.
Comparison with the following RMS values shows the high accuracy of the reconstruction.
%
%
\subsection{Resolution improvement}\label{resolution}
In the following, the usage of the GMM for resolution improvement is discussed.
The selected mock sky map (Fig.~\ref{Abb:mock_org}) is smoothed using a Gaussian kernel with $\sigma = 0.03$\,rad simulating the situation of a lower resolved map.
This lower resolved map shall be improved by predicting small-scale structures using a GMM trained with the low-resolution map and the remaining high-resolution maps.
Since the original data of Mock\,I is known but not used for training, we can evaluate how effective the GMM is in sharpening up resolution.\par
The smoothed map of Mock\,I in Figure~\ref{fig:smooth} displays a significant loss of detail, which is reflected by the small-scale residuals in the difference map (Fig.~\ref{fig:org-smooth}).
We train the GMM with the smoothed map and the remaining five other maps and then use it to reconstruct Mock\,I.
While certain individual structures, which are not distinctively contained in the other data sets used in the GMM, cannot be reconstructed successfully (compare Figure~\ref{fig:org-predsmooth}), the overall resolution can be significantly improved with respect to the map used in the training, as it can be seen in Figure~\ref{fig:predsmooth}.\par
The original magnitudes are more accurately reproduced for the bright regions in the simulated Galactic disk than for the dimmer pixels at higher latitudes.
However, the magnitudes of the predicted small-scale structures of brighter pixels are underestimated and therefore not reproduced completely.
This underestimation of the magnitudes can be minimized by using more Gaussian components during the training phase.\par
We compute an RMS of 0.23 for the difference between the original Mock\,I and the smoothed sky map (Fig~\ref{fig:org-smooth}), and an RMS of 0.17 for the difference between the original and the predicted smoothed (but resolution improved) data set of Mock\,I (Fig~\ref{fig:org-predsmooth}).
\begin{figure}[h!]
\centering
\includegraphics[width=\columnwidth]{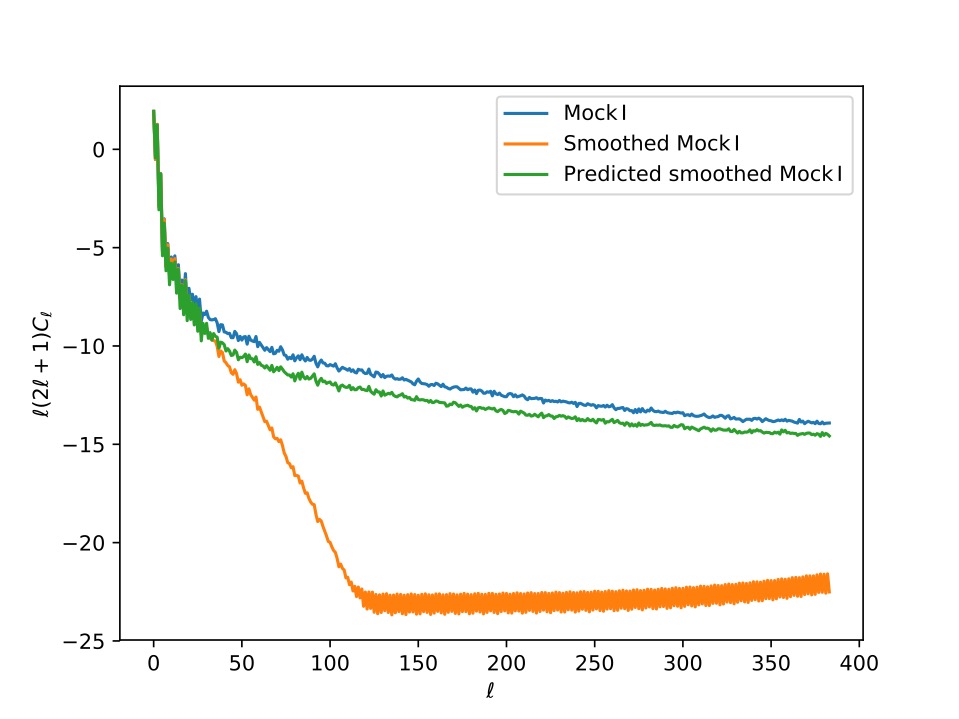}
\caption[]{Power spectra of the Mock\,I (blue), smoothed Mock\,I (orange), and the prediction of the smoothed Mock\,I map (green) in logarithmic scaling. This indicates that the prediction of the GMM for the smoothed Mock\,I map suits well in shape and structure (peaks and dips) to the original map.}
\label{fig:power}
\end{figure}
Further, we analyze the different power spectra shown in Figure~\ref{fig:power}. 
We find while the power spectra of the smoothed Mock\,I map (orange) is different from the one of the original map (blue) due to the smoothed high frequencies that the one of the prediction of the smoothed Mock\,I map (green) suits to the original one. 
In particular, the green graph has the same shape and structure (peaks and dips) as the blue one.
That the green graph is below the blue one can be traced back to the fact that the magnitudes in the resolution improved regions are slightly underestimated as discussed above.
Overall, this shows that the prediction significantly increases the resolution of the smoothed data set and that the GMM is able to predict real small-scale structures as shown by the power spectra, which is good agreement with the original map. \par
It is noteworthy that the larger the mutual information content is between a low-resolution map and higher resolved maps at other frequencies, the closer the reconstruction is to the original map. This effect of different mutual information during training process is discussed in more detail in the following section.
\subsection{Restoring unobserved areas}\label{restoring}
\begin{figure*}[h!]
\centering
   \subfloat[\label{fig:predonlydisk}]{\includegraphics[trim=0 60 0 0, clip, width=\columnwidth]{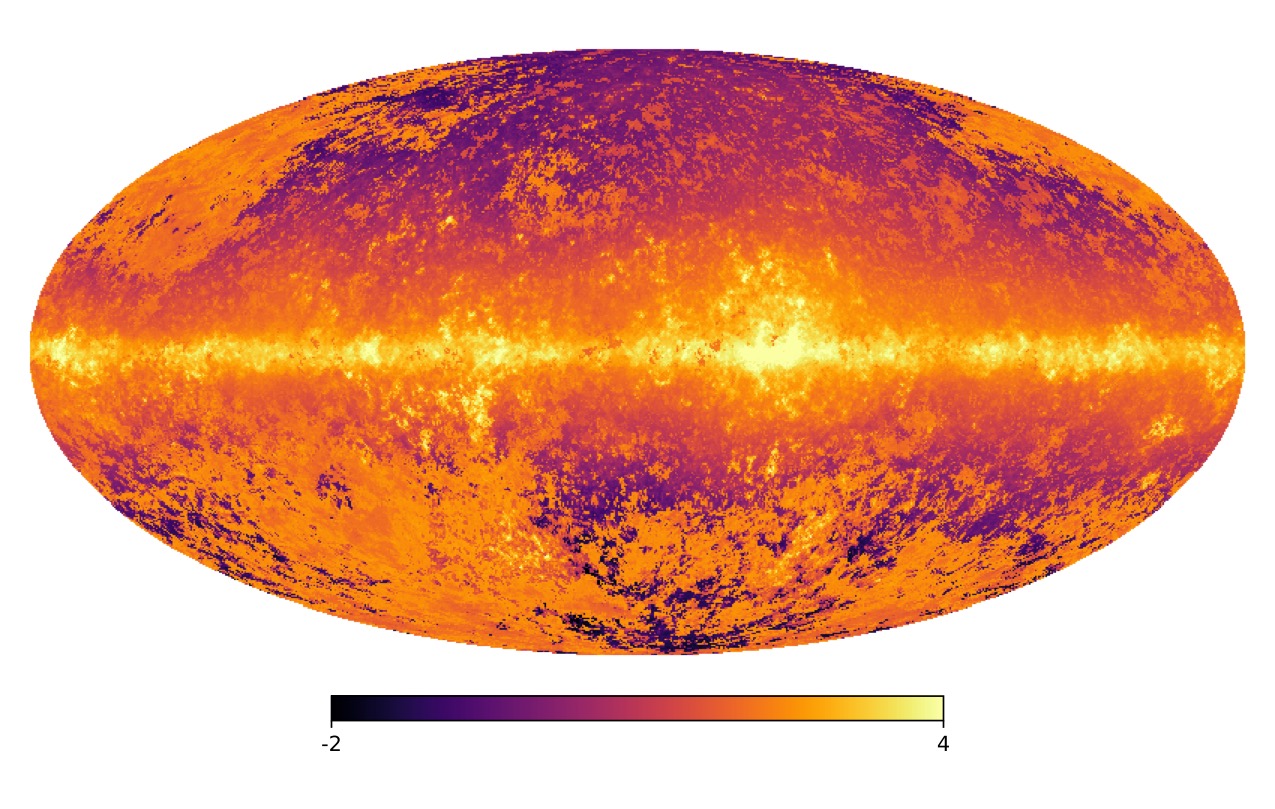}}
   \hfill
   \subfloat[\label{fig:org-predonlydisk}]{\includegraphics[trim=0 60 0 0, clip, width=\columnwidth]{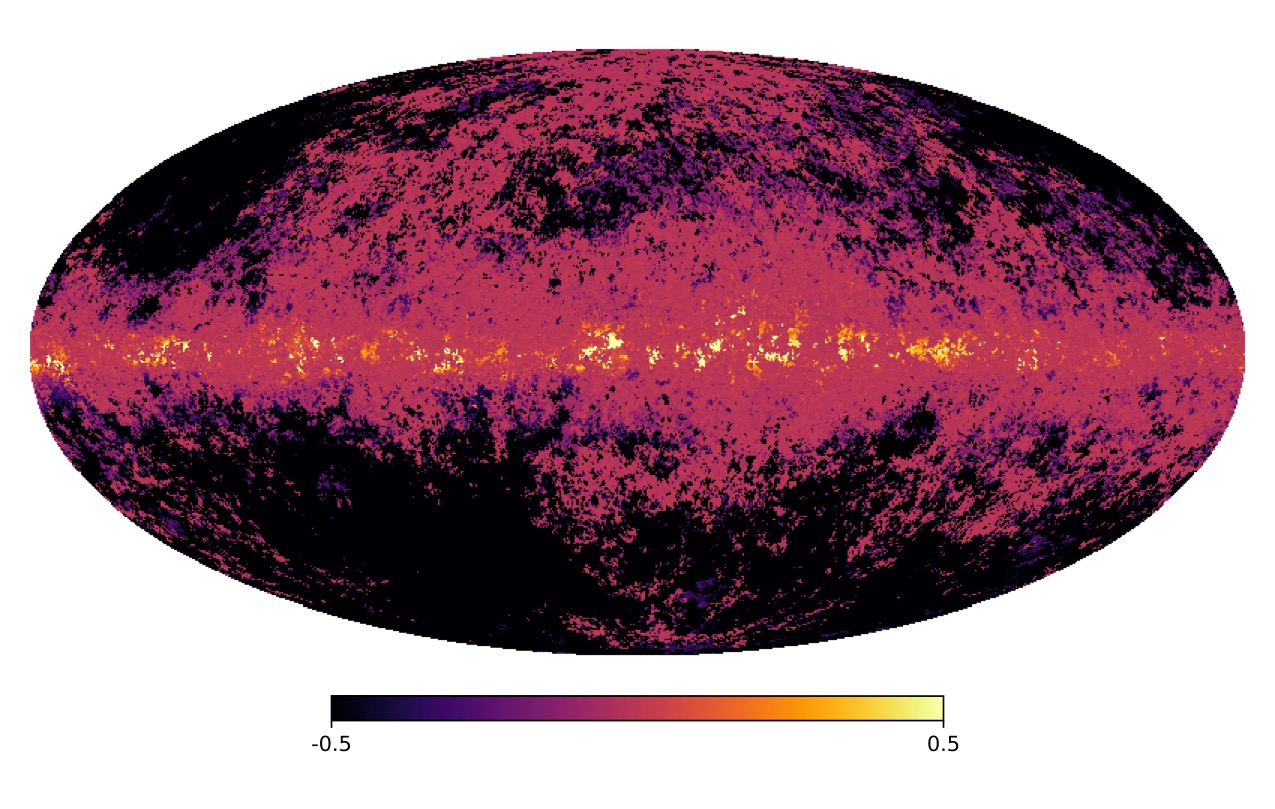}}
   \\
   \subfloat[\label{fig:preddiskback}]{\includegraphics[trim=0 60 0 0, clip, width=\columnwidth]{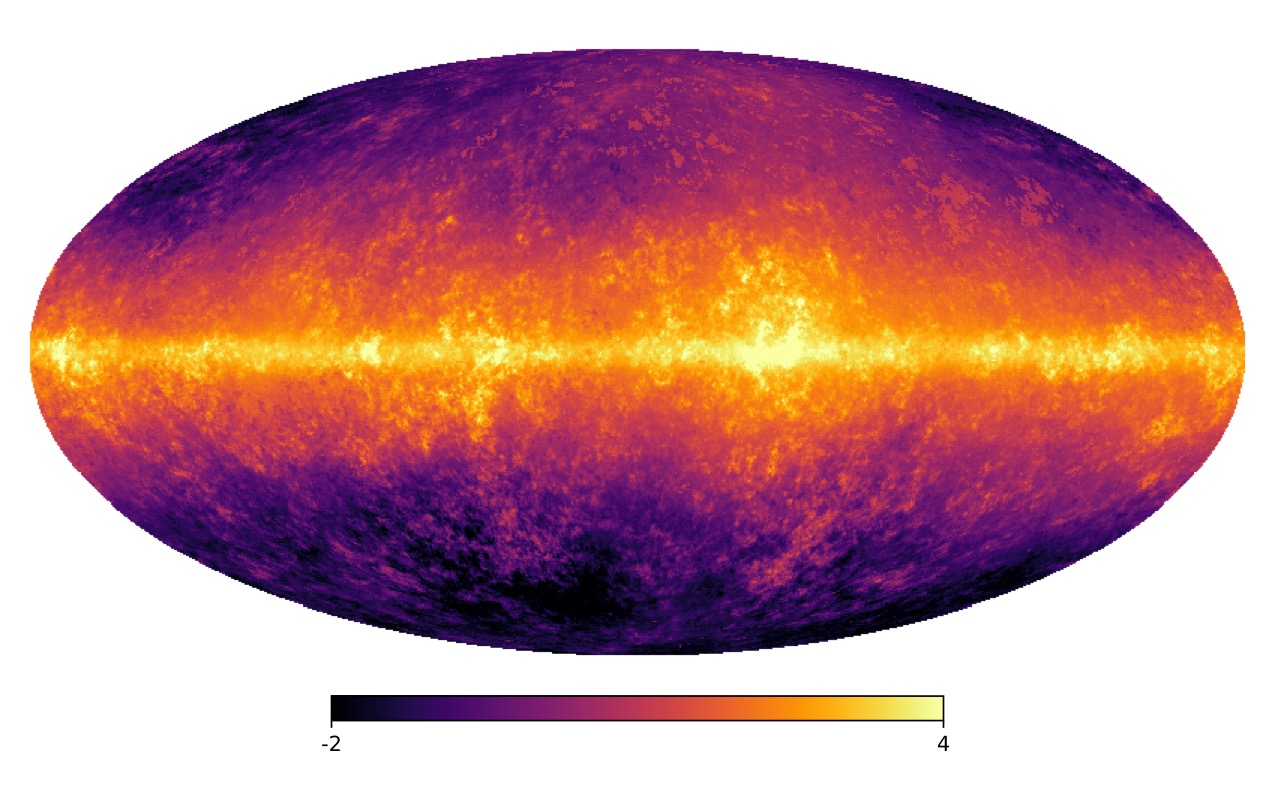}}
   \hfill
   \subfloat[\label{fig:org-preddiskback}]{\includegraphics
  [trim=0 60 0 0, clip, width=\columnwidth]{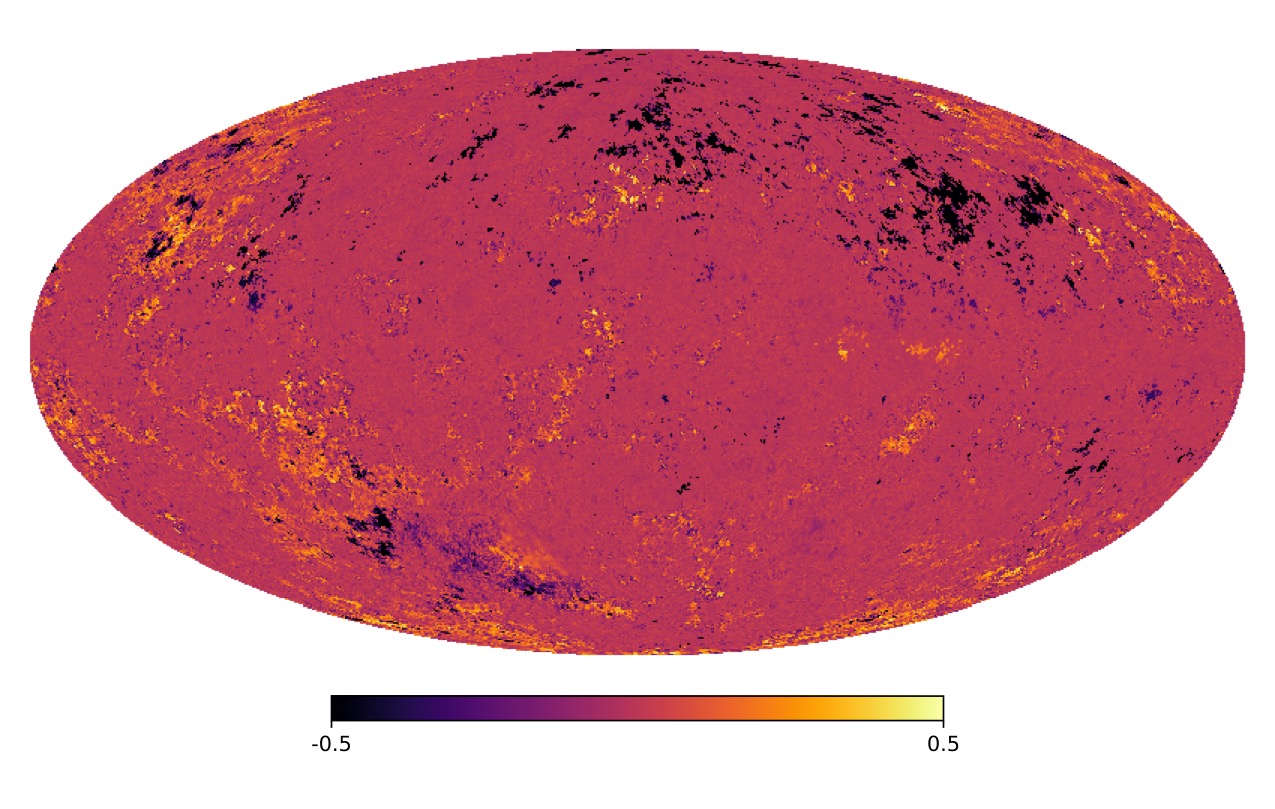}}
\\
   \subfloat[\label{fig:predhalfcut}]{\includegraphics[width=\columnwidth]{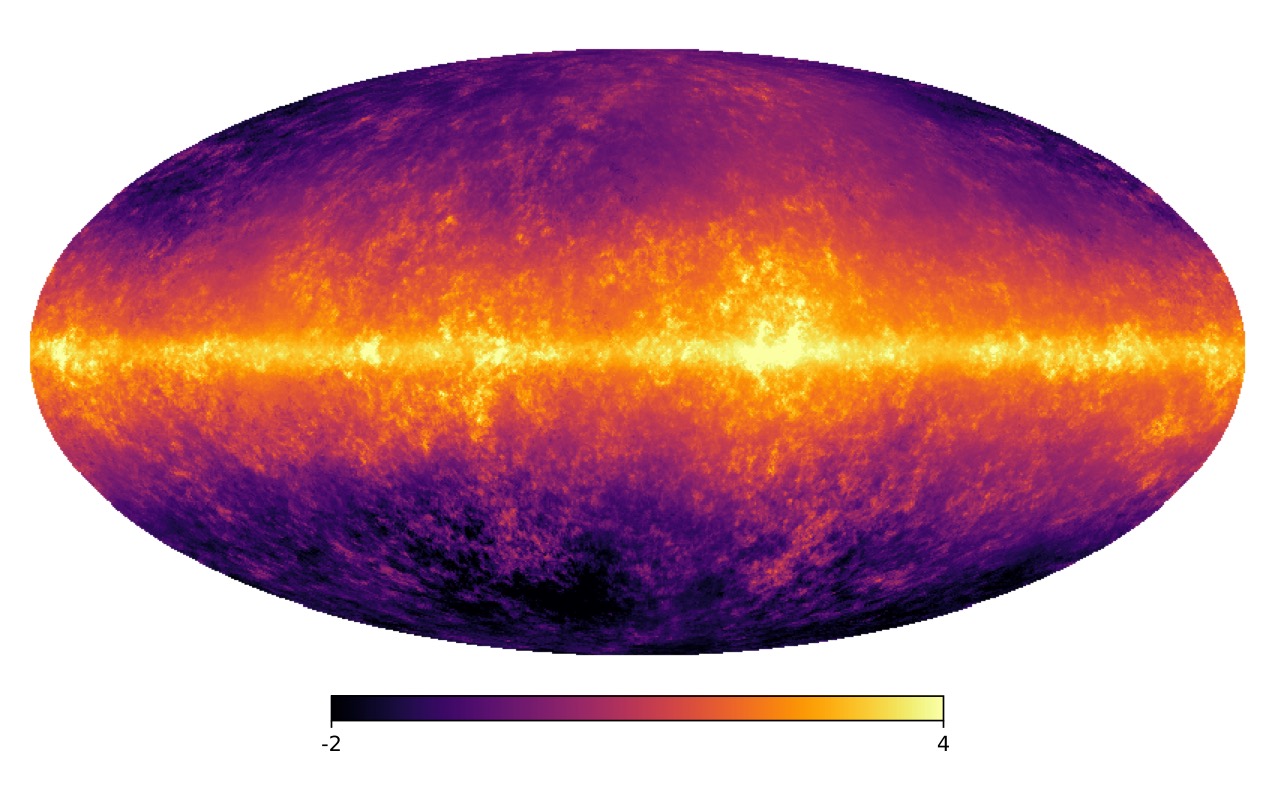}}
   \hfill
   \subfloat[Difference between the original Mock\,I map and the prediction computed from the southern hemisphere information, $\textrm{RMS}=0.07$ \label{fig:org-predhalfcut}]{\includegraphics
  [width=\columnwidth]{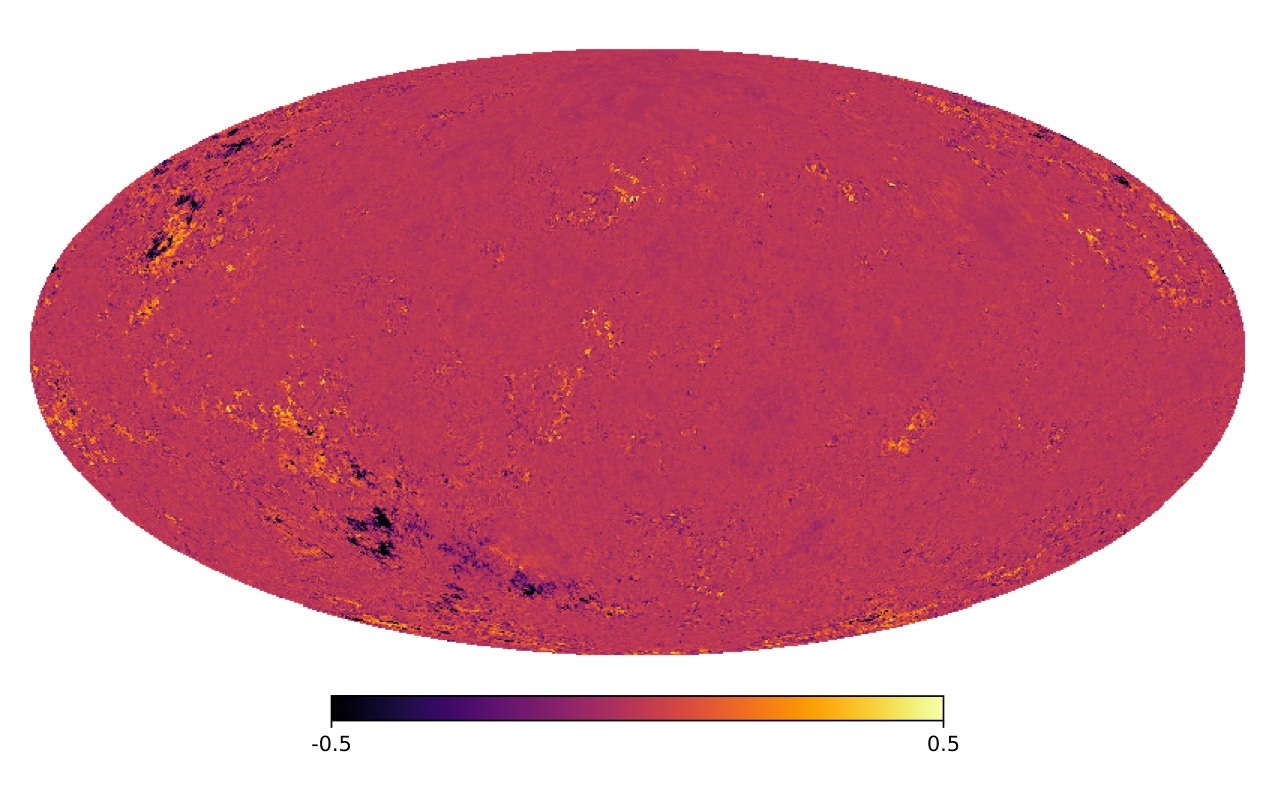}}
  \caption[]{Simulation of the restoration of unobserved areas. The GMM is trained with $K=9$ Gaussians using different observed parts of the whole sky from Mock\,I and the full maps of Mock\,II--VI. The predictions are determined from the CPD marginalized over the respective partial maps of Mock\,I. Panel (a): Prediction of Mock\,I with only the disk information as GMM input. Panel (b): Difference between the original Mock\,I map and the prediction computed from only the disk information, $\textrm{RMS}=1.64$. Panel (c): Prediction of Mock\,I with the disk and only a small part of the southern hemisphere information as GMM input. Panel (d): Difference between the original Mock\,I map and the prediction computed from the disk and a small part of the southern hemisphere background information, $\textrm{RMS}=0.14$. Panel (e): Prediction of Mock\,I with only the southern hemisphere information as GMM input. Panel (f): Difference between the original Mock\,I map and the prediction computed from the southern hemisphere information, $\textrm{RMS}=0.07$.} \label{fig:predhalfcuts}
\end{figure*}
With the mock data we are also able to simulate incomplete sky maps demonstrating the ability of the GMM to predict nonobserved parts of the sky in a given frequency range.
Therefore, only subsets of the selected Mock\,I map are used for GMM training in addition to the remaining five complete maps. Thereafter, we predict the full Mock\,I map using the GMM trained on only parts of it given the complete all-sky information of the other maps and compare the prediction with the original data set.\par
In Figure~\ref{fig:predonlydisk} Mock\,I is predicted using only the information of the disk of Mock\,I based on a 20$^\circ$ wide stripe centered on latitude zero for the GMM training.
As expected, the prediction suffers from the loss of information on the weaker emission above and below the disk, which is evident by the significant differences there to the original data (Fig.~\ref{fig:org-predonlydisk}).
Nevertheless, several distinct bulge and halo features have been reconstructed, demonstrating that the information given by the other maps is used effectively to reconstruct the missing information.
The disk profile itself is reconstructed accurately, albeit the bright structures within the disk are underestimated as can be seen in the prominent flux differences in the disk (Fig.~\ref{fig:org-predonlydisk}).
\par
Here, a limit of the reconstruction from the GMM becomes evident, since, although the local Galactic latitude profile as well as the filamentary emission profile is contained in the selected disk area for training, these profile levels are not reconstructed in the prediction.
These profile magnitudes cannot be recovered from the Galactic disk information alone.
Furthermore, there are structures in the disk area, and therefore in the training set, that are less well reconstructed than by a GMM that had been trained on the full sky (compare Fig.~\ref{fig:org-predonlydisk} to Fig.~\ref{fig:org-predfull}).
However, adding already small parts of higher Galactic latitude regions of Mock\,I to the training data enables the GMM to reproduce the original sky with a much higher fidelity as discussed in the following.\par
Figure~\ref{fig:preddiskback} displays a similar input situation for training the GMM. 
Here, we added a 10$^\circ$ wide stripe along the Galactic latitude 40$^\circ$ south of the disk from Mock\,I to the training process. This additional data already vastly improves the prediction quality of the GMM trained with it, since it contains information about the composition of the less bright areas. Also the fidelity in the disk area improved by the inclusion of some nondisk data. This is apparent especially in the corresponding difference map (Fig.~\ref{fig:org-preddiskback}) showing some inaccurate reconstruction predominantly in the northern hemisphere but overall revealing only small differences. However, the differences map also shows that the reconstruction is significantly more accurate in the southern hemisphere.
This indicates that the GMM does not perfectly learn the original relation used to construct the mock data (Eq. \ref{eq:mock}) but partly the concrete realization of the random processes used to generate a structured sky. \par
Comparing Figure~\ref{fig:org-preddiskback} with Figure~\ref{fig:org-predfull} reveals nearly the same differences in the southern sky.
We assume that they are caused by the overall missing information in the whole data sets, since most regions are mistakenly predicted for both cases.
The effect of the nonsymmetrical sky differences is minimized in the concluding mock test example.\par
For this example the complete southern hemisphere of Mock\,I has been used for training while the northern hemisphere has been omitted. Although half of the information from the selected map has therefore been ignored, the reconstruction in Figure~\ref{fig:predhalfcut} is of a comparably high fidelity as the one based on the all-sky map Mock\,I for GMM training (Fig.~\ref{fig:predfull}). This good achievement is presumably caused by the fact that all relevant latitudes are trained, and also with the correct area ratios with respect to each other. In short, the GMM could train on a fully representative sample of magnitude vectors.\par
In comparison to the prediction calculated from a full map with original resolution learned by the GMM (Fig~\ref{fig:predfull}, $\textrm{RMS}=0.05$) the quality of the prediction computed from a data set with half of the information on Mock\,I is very similar (Fig \ref{fig:predhalfcut}, $\textrm{RMS}=0.07$).
The quality of the prediction computed from a GMM trained with the information of the Galactic disk and a small part of the Galactic latitude profile (Fig~\ref{fig:preddiskback}) is half as accurate with an RMS of 0.14 as the prediction using the full southern hemisphere.
However, the GMM should not be used to predict the whole sky using only the disk information (Fig.~\ref{fig:predonlydisk}) since the RMS value with 1.64 indicates significant discrepancies.


\section{Data preparation}
\label{sec:dataprep}
In the following, the method is applied to measured all-sky data of the Milky Way. In this section we describe the steps of data selection, homogenization, cleaning, and unification that were necessary to prepare the data sets for the GMM training.
\subsection{Selection criteria}
To analyze the mutual information in the multifrequency magnitude space of the pixels ideally a total coverage of the frequency space is needed, from radio to $\gamma$-rays.
We used most of the currently available all-sky maps that are covering nearly the full measurable electromagnetic frequency space for analysis.
Our selection of all-sky maps consists of the highest-resolved maps available with at least 1$^{\circ}$ angular resolution and a spatial coverage of at least 80\,\% provided in HEALPix format.\par
If a frequency space is covered by multiple surveys, we used the highest-resolved data set. Here, we did not used, for example, the all-sky WISE 12\,$\mu$m map provided by \cite{WISE}, whose frequency regime is also covered by IRIS. While the WISE data is higher resolved, we used the IRIS map since the WISE data include further calibration artifacts that would be reproduced by the GMM. Furthermore, these small scale structures additionally measured by WISE are smeared out when reducing the resolution to the common $27^\prime.5$ (Sec. \ref{clean}). \par
For now, partly measured all-sky surveys are not use in this work to provide an as much as possible similar sky coverage of the different maps. For example, we did not used the additional information that might be embedded in the UV regime since the GALEX all-sky survey \citep{GALEX} suffers from many unobserved regions.
The choice of data sets is described and listed in Appendix \ref{App:Sds}.
\begin{figure*}[h!]
\centering
   \subfloat[\label{old3}]{\includegraphics[width=\columnwidth]{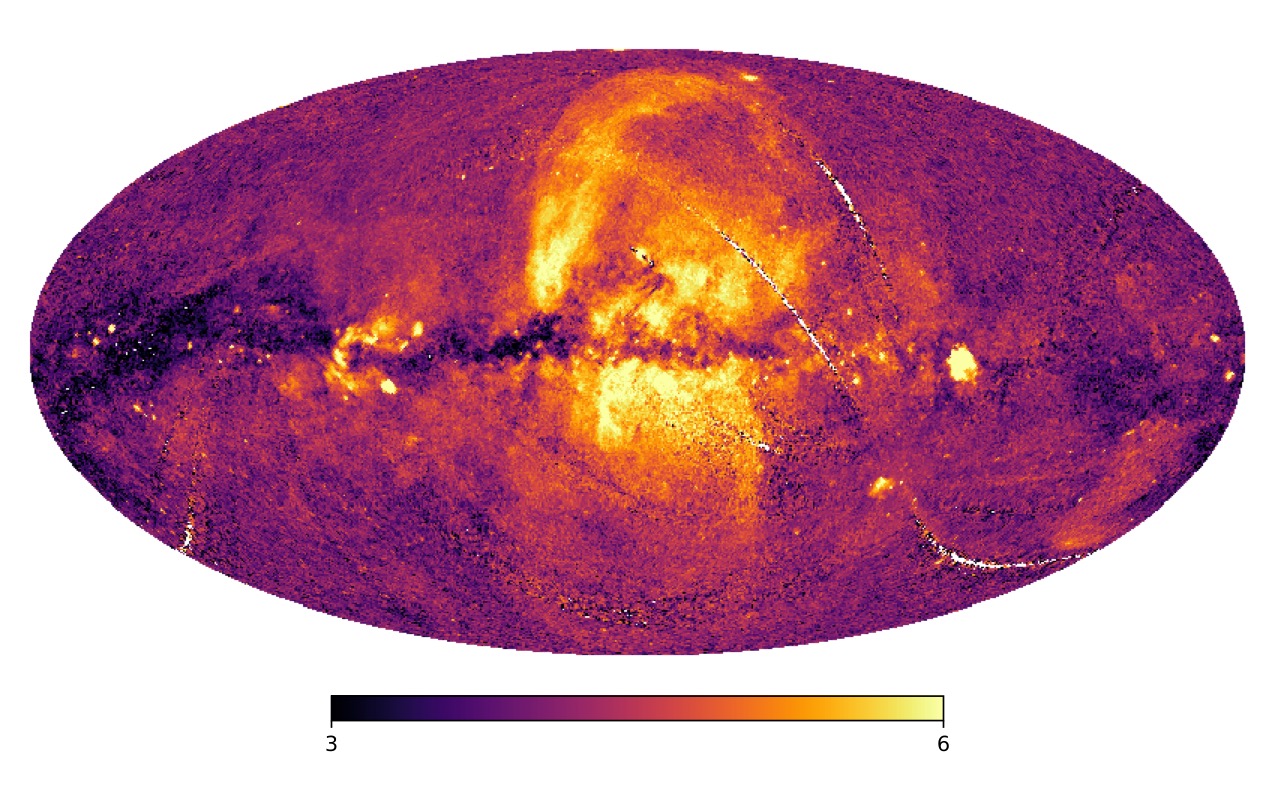}}
   \hfill
      \subfloat[\label{org3}]{\includegraphics[width=\columnwidth]{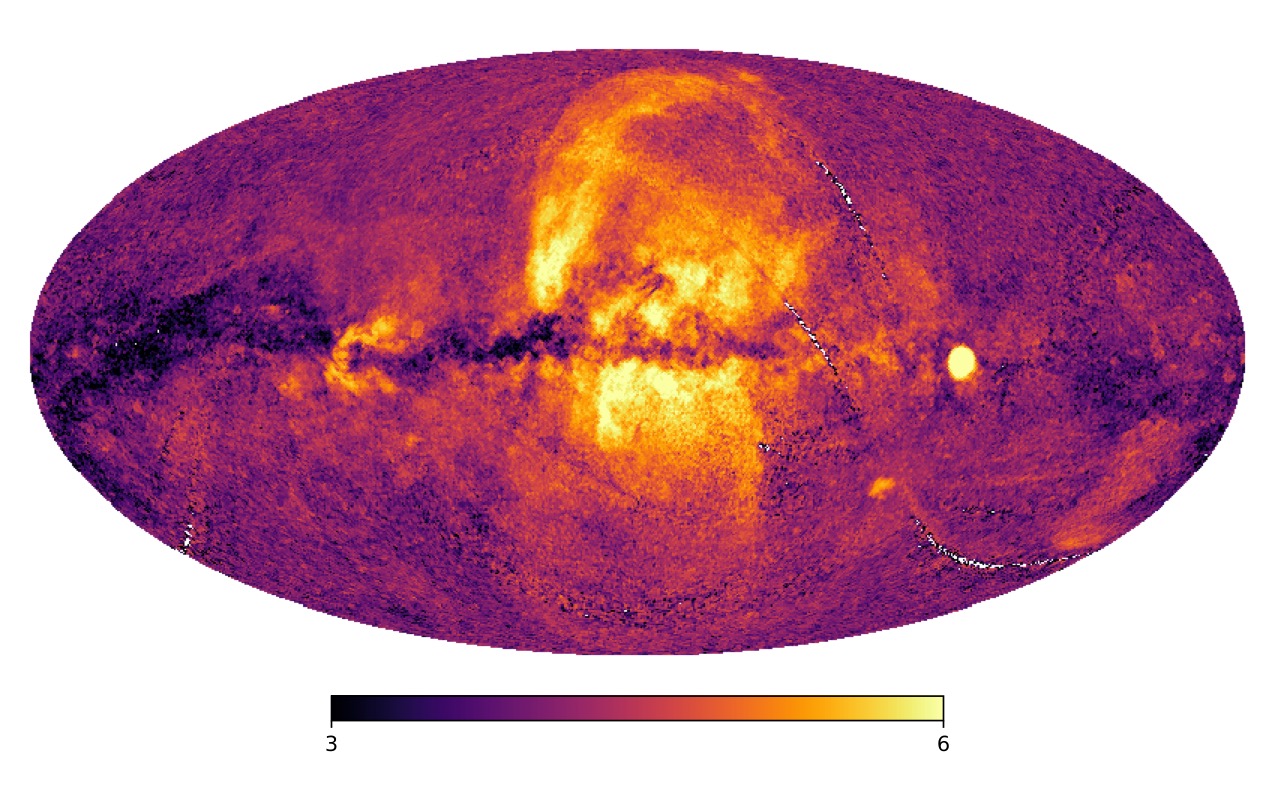}}
   \hfill
      \subfloat[\label{pred33}]{\includegraphics[width=\columnwidth]{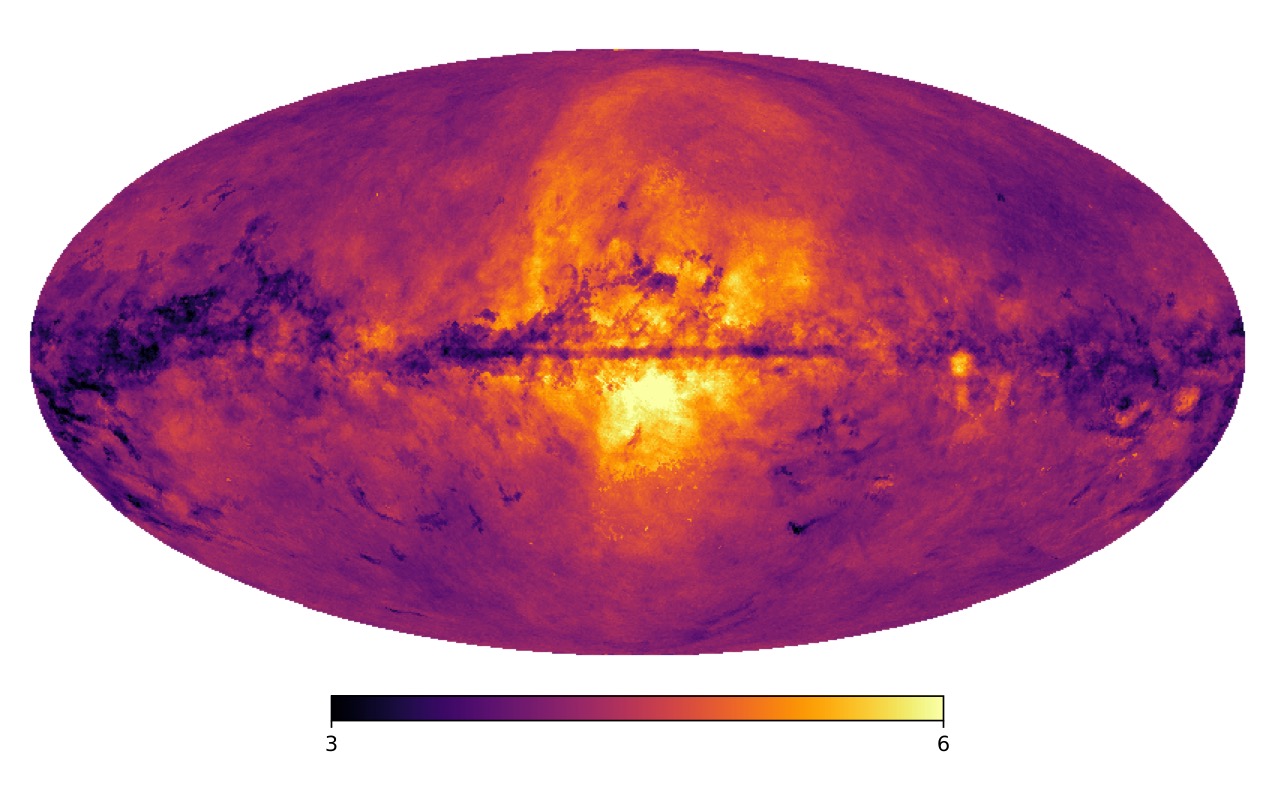}}
   \hfill
   \subfloat[\label{org3-pred33}]{\includegraphics
  [width=\columnwidth]{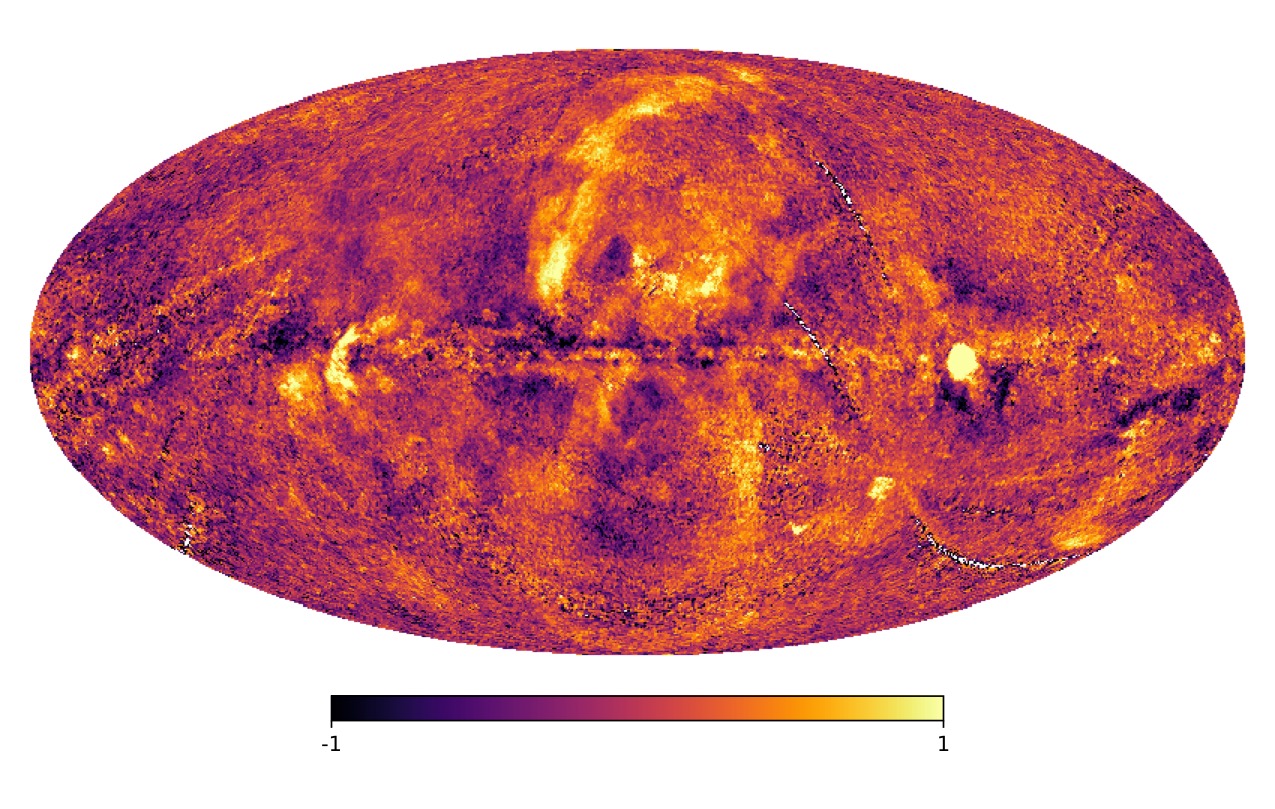}}
  \caption[org3-pred33]{Application of the point source removal routine and the GMM results for the ROSAT 0.885\,keV map; logarithmic scaling. Panel (a): Original ROSAT 0.885\,keV map. Panel (b): 0.885\,keV map after applying the point source removal routine. Panel (c): Prediction of the 0.885\,keV map computed from the GMM trained with $K=3$ Gaussians and $n=37$ data sets neglecting the $\gamma$-ray component maps. The prediction is determined from the CPD marginalized over the X-ray regime. Panel (d): Difference between the map shown in Figure \ref{org3} and the predicted 0.885\,keV map, $\textrm{RMS} = 0.35$.} \label{fig:org3}
\end{figure*}
\subsection{Homogenization}
The flux units of the original data sets were used as provided.
The sky brightness maps were converted to logarithmic magnitudes.
This is problematic for maps, where a global zero level has been subtracted by the actual measurement process.
For example, the Planck maps lack a zero level so that areas with negative flux values appear.
About 20\,\% of the values are negative in the Planck LFI data sets, hence this information would be missing in the GMM without correction.
Therefore, an offset needs to be added before a logarithm can be taken and further processing with the GMM is possible.
First, the flux minimum of a map is subtracted from all pixels resulting in the lowest pixels to be zero, which corresponds to unmeasured areas.
To circumvent this issue an adequate offset is defined as a small percentage of the flux of the lowest nonzero pixel among the zero-level corrected pixel values.
During tests the summation of an offset of 0.1\,\% of the smallest positive values seemed adequate.
By that, the structural resolution (pixel to pixel contrast) is changed only little.\par
Further, we reduced the noise level of the particularly noisy X-ray data sets of the ROSAT survey, since the signal-to-noise ratio is significantly inferior to the other data sets.
For this, we applied a smoothing with a Gaussian kernel by a width of $\sigma = 0.002$\,rad ($\sim 6.9\,^\prime$) to the data.
\subsection{Cleaning and Unification} \label{clean}
Since the provided data sets are not free of point, compact, and extragalactic sources as well as calibration artifacts, we cleaned all data sets except the $\gamma$-ray maps provided by \cite{Fermi1} (see Appendix \ref{App:Sds}). before training the GMM.
We developed and implemented a median replacement routine in Python for such disturbing pixels.
After cleaning, the resolutions of all the data sets are unified. For this, each of the $N=39$ data sets is downgraded to the data set with the smallest provided resolution given by $\textrm{nside}=128$, which corresponds to a resolution of $27.5\,^\prime$ \citep{HEALPix}.


\section{Results} \label{Chap:Res}
The results discussed in the following are computed by GMMs trained with a varying number of $n \leq N=39$ different multifrequency input data sets. Unless noted differently, the GMM uses $K=3$ Gaussian components for training. Pertinent sources are marked in an example image in Appendix~\ref{app:regions}, which we refer to when comparing the predictions to the original data.
\subsection{Reconstructing the X-ray sky} \label{Chap:xray}
The X-ray regime is represented by the all-sky observation of the ROSAT survey (see Appendix \ref{App:Sds}).
The original ROSAT map (Fig.~\ref{old3}) at a central wavelength corresponding to 0.855\,keV has been preprocessed as described in Section~\ref{sec:dataprep}; the result is shown in Figure~\ref{org3}.
By that, the point, compact, and extragalactic sources have been purged, and the scanning artifacts have been reduced.\par
The GMM is trained with $n=37$ data sets. The omitted maps are the two derived component maps of the $\gamma$-ray sky that are neglected since their information is redundant to the used Fermi maps at different photon energies (see Appendix \ref{App:Sds}).\footnote{Redundant information (data sets that are included in other data sets) is not used since it would need to be trained in the GMM additionally, which reduces its accuracy. This effect is not negligible as we only use three Gaussian to train this hypersurface of data points.}
Figure~\ref{pred33} displays the reconstructed 0.855\,keV X-ray map. It is computed by the output of the trained GMM, which is marginalized over all the X-ray maps.\par
First, we find that the reconstructed map (Fig.~\ref{pred33}) carries much fewer observational artifacts compared to the original data.
That means that the scanning effects, which can be identified by the stripe-like pattern in Figure~\ref{org3}, are eliminated entirely.
Additionally, we find that the high density gas and dust component, visible as dark filaments dominant in the outer disk regions in Fig.~\ref{pred33}, is clearly higher resolved in our reconstruction compared to the original map.
It now reveals structures that reach far into the direction of the northern Fermi bubble toward the Ophiuchus region.\par
However, there is a lack of faint X-ray structures in the Galactic high-latitude regions of the reconstruction.
This issue can be circumvented by increasing the number of Gaussian components in the GMM training, but this would also increase the influence of artifacts in the results since those are then learned as well, as our experiment with $K > 3$ revealed (see Appendix \ref{App:xray}).\par
The North Polar Spur, which is also contained in, for example, radio continuum and $\gamma$-ray observations, is reconstructed, albeit fainter than in the original data.
The spur-like outflow south of the Galactic plane, however, could not be reproduced.
Additionally, the strong outflow emission in the direction of the southern Fermi bubble has not been reconstructed correctly and shows less detail than in the original data.
The outer emission around the southern Fermi bubble is missing completely, which is visible as the significant flux discrepancy in the difference map (Fig.~\ref{org3-pred33}).
The same applies to the bright regions of Vela and Cygnus X as well as the large diffuse emission region south of the Orion region close to the right edge of the map.\par
With respect to the reproduced map computed from a GMM trained with $K=18$ Gaussians (Fig \ref{xray18} in Appendix \ref{App:xray}), these missing reconstructions show that independently of the number of Gaussians the GMM is not able to identify the correlations between the used data sets for these magnitudes.
Overall, the underestimated magnitudes of these pronounced regions in the predicted map clearly indicate that the original X-ray map contains emission from at least one independent component that is not encoded in the other frequency regimes.
This component is identified by the brightest emission in the difference map. This is also supported by comparison of the RMS values, which are 0.35 and 0.31 for $K=3$ and $K=18$ Gaussians, respectively. In case this independent component was included in the other frequency regimes we would expect the prediction with $K=18$ Gaussian components to be able to reproduce this feature, which is not the case (compare Fig.~\ref{predorg-xray18}).\par
Furthermore, the southern Fermi bubble itself is visible in the reconstructed map although, by eye, it does not seem to be present in the original data.
This can be attributed to the fact that the GMM is not learning the spatial but the magnitude correlations.
This means that the GMM is trained by each $n$-dimensional pixel vector separately excluding any spatial information.
By that, the GMM measures the distribution of the pixel vectors by their probabilities in the magnitude space.
Hence, the $\gamma$-ray magnitudes of the southern Fermi bubble classify the X-ray emission in direction of the southern Fermi bubble.
Nevertheless, the northern Fermi bubble cannot be clearly identified as a separate emission structure. However, we argue that this is due to the strong surrounding emission, which encloses the Fermi bubble.
As discussed later on, a presumably similar situation is visible in the Fermi 1.70 GeV map, where the Fermi bubble emission is present, as it is evident from the leptonic component map shown in Figure~\ref{org31}, but cannot easily be identified in the original map (Fig.~\ref{org14}).\par
Thus, we find that the multifrequency magnitude distribution includes the information that the Fermi bubbles are a feature contained also in the X-ray regime.
This finding agrees with the analysis by \citet{su2010} who claim that at least the edges of the Fermi bubbles can both be identified in the ROSAT maps by comparison to the $\gamma$-ray data of the Fermi LAT satellite.
\subsection{Sharpening up the $\gamma$-ray sky}
\begin{figure*}[h!]
\centering
      \subfloat[\label{sum_data}]{\includegraphics[width=\columnwidth]{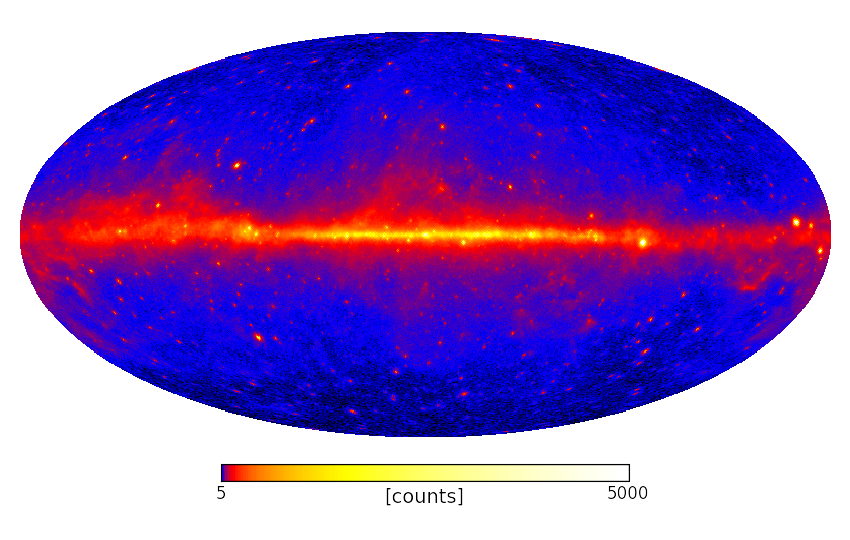}}
   \hfill
   \subfloat[\label{org14}]{\includegraphics
  [width=\columnwidth]{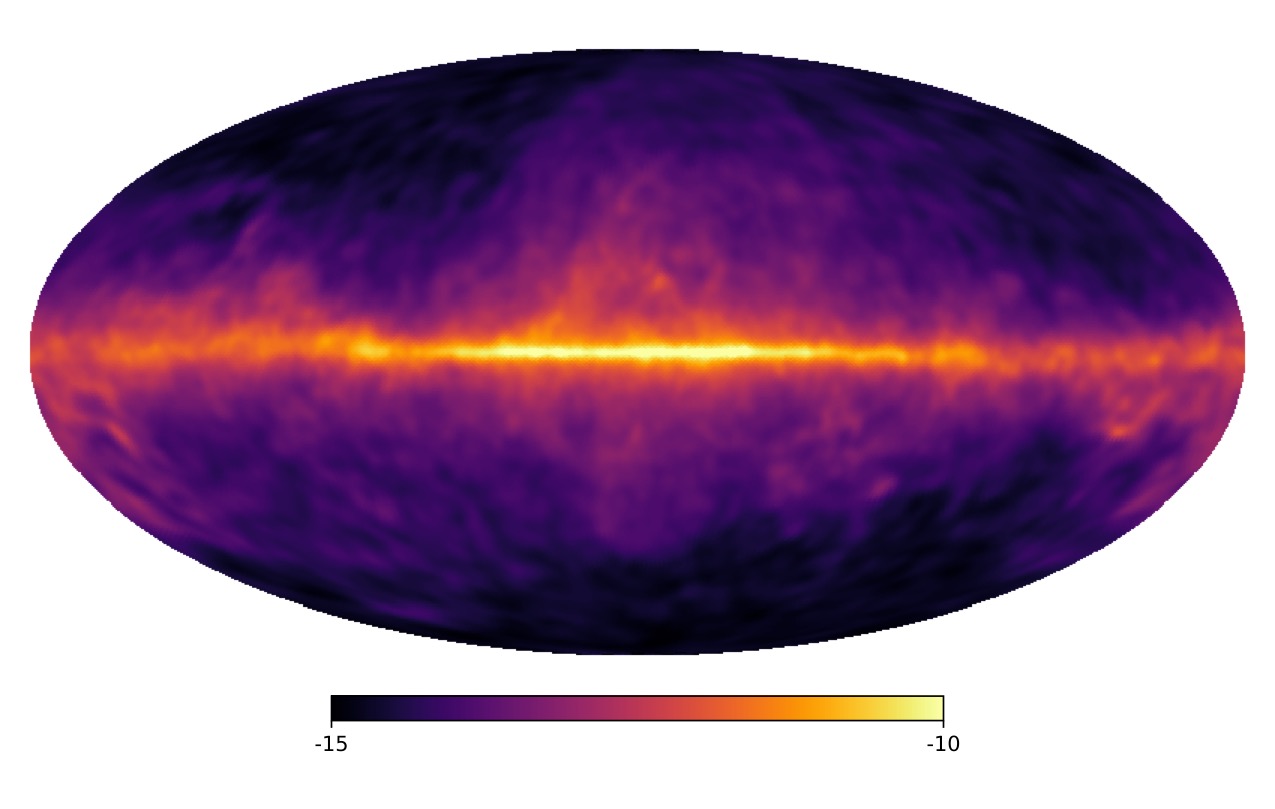}}
  \\
      \subfloat[\label{pred143whl}]{\includegraphics[trim=0 60 0 0, clip, width=\columnwidth]{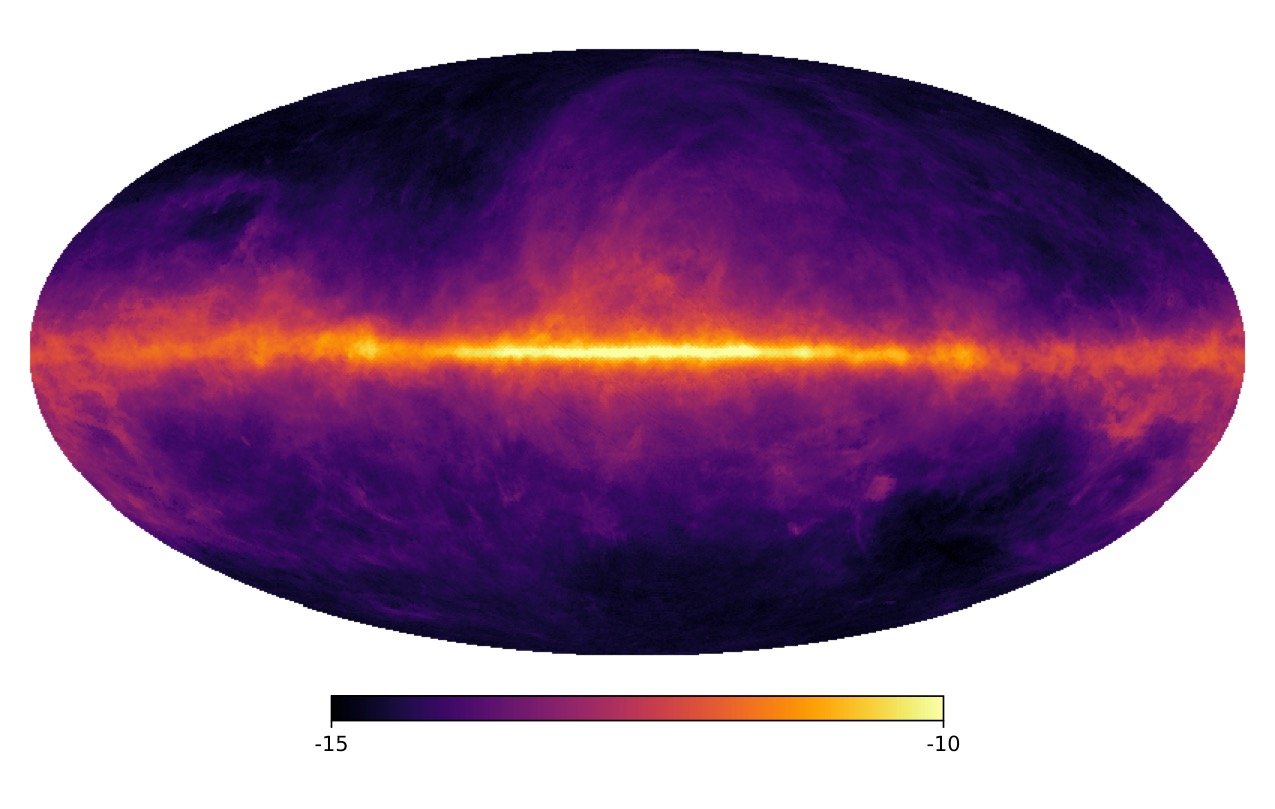}}
   \hfill
   \subfloat[\label{org-pred143whl}]{\includegraphics
  [trim=0 60 0 0, clip, width=\columnwidth]{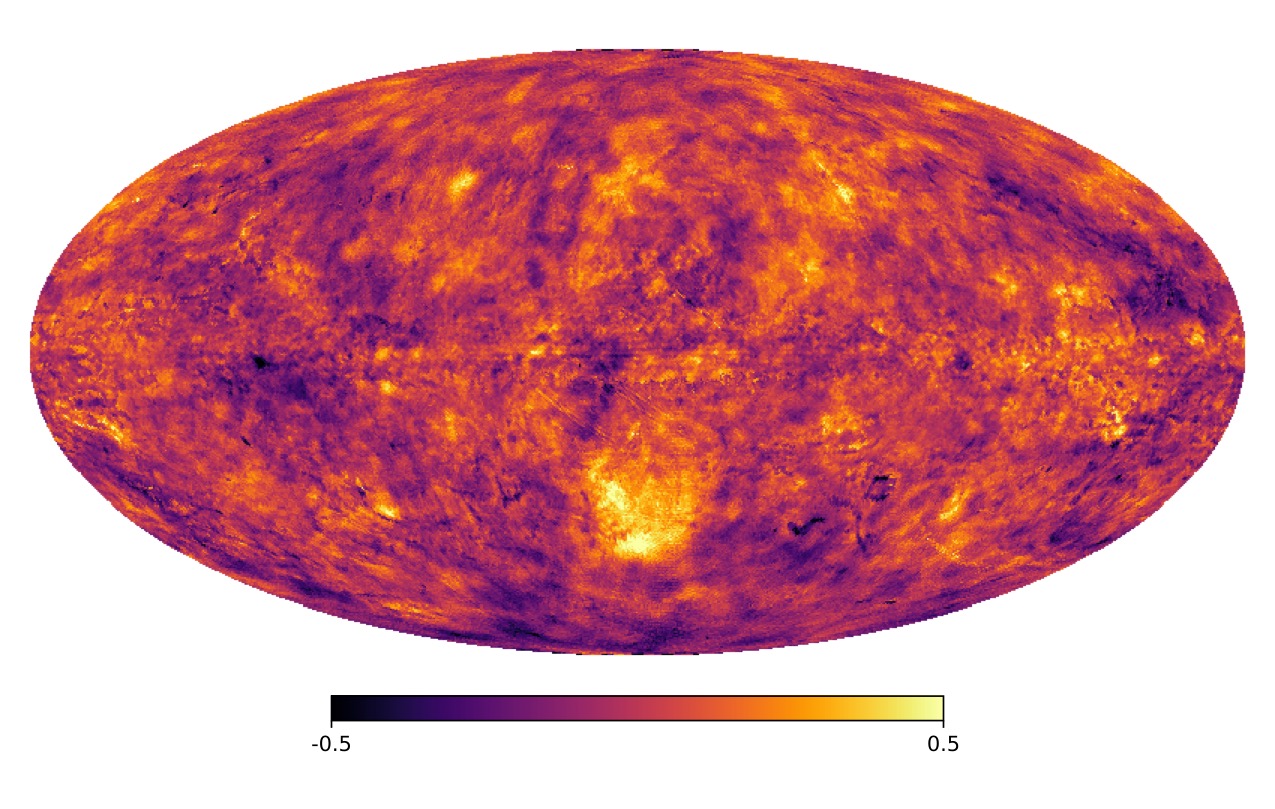}}
  \\
      \subfloat[\label{pred14_18_whl}]{\includegraphics[width=\columnwidth]{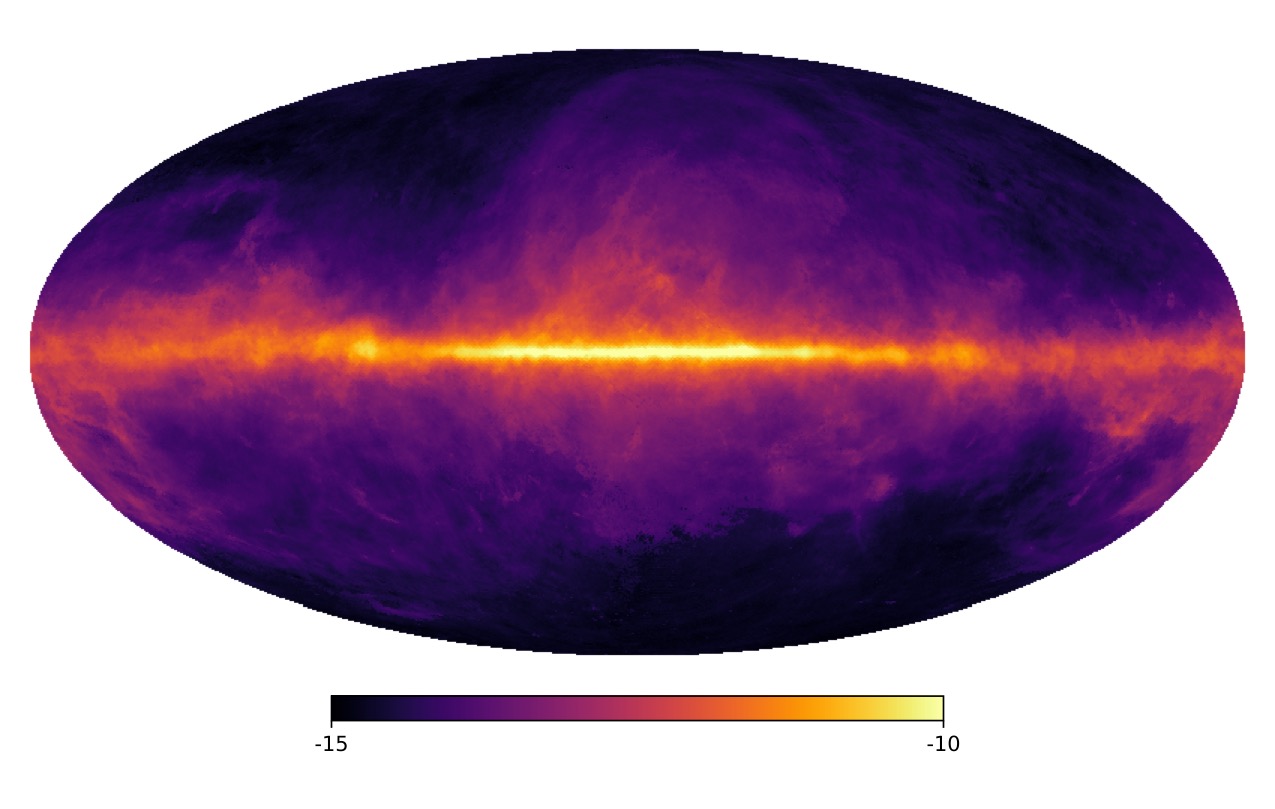}}
   \hfill
   \subfloat[\label{org-pred14_18_whl}]{\includegraphics
  [width=\columnwidth]{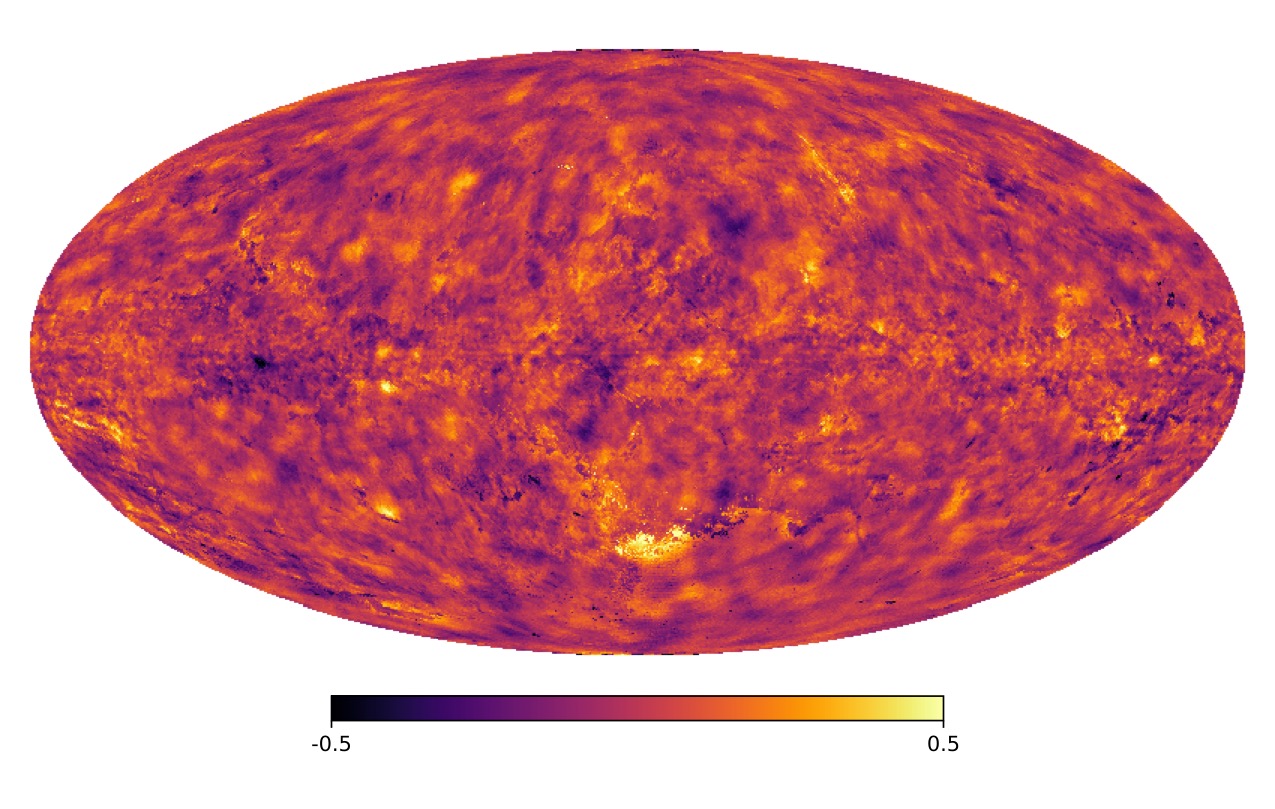}}
  \caption[]{GMM results for the Fermi 1.70 GeV map; logarithmic scaling. The predictions are computed from the GMM trained with $n=37$ data sets neglecting the $\gamma$-ray component maps. The predictions are determined from the CPD marginalized over the $\gamma$-ray regime. Panel (a): Total photon flux reconstructed from photon count data of 6.5 years mission elapsed time in the complete Fermi frequency range \citep{Fermi1}. Panel (b): Fermi 1.70 GeV map after application of the D$^{3}$PO algorithm. Panel (c): Prediction of the 1.70 GeV map with $K=3$ Gaussians. Panel (d): Difference between the original and predicted 1.70 GeV map, $\textrm{RMS}=0.13$. Panel (e): Prediction of the 1.70 GeV map with $K=18$ Gaussians. Panel (f): Difference between the original and predicted 1.70 GeV map with $K=18$ Gaussians, $\textrm{RMS}=0.10$.}
\end{figure*}
The $\gamma$-ray sky has been observed the Fermi LAT satellite.
All-sky maps computed from this observation have been published for nine different energy bands by \cite{Fermi1} (see Appendix \ref{App:Sds}). These were determined from the photon counts measured over 6.5 years (Fig. \ref{sum_data}) with D$^3$PO \citep{Fermi2}, an algorithm for denoising, deconvolving, and decomposing. \footnote{For our analysis we are dependent on the D$^3$PO-separated Fermi data sets. A more recent separation for the updated Fermi LAT observation is in preparation but not available yet.}
The map at 1.70\,GeV reconstructed from the Fermi data by these authors is displayed in Figure~\ref{org14}. It is one of the lowest-resolved maps in our data sample.
Compared to the maps at other frequencies, the small-scale variations within the diffuse emission structures are much less detailed.
However, this data set provides information about the $\gamma$-ray sky that faintly contains features such as, for example, the northern and southern Fermi bubble.\par
Here, the GMM is again trained with $n=37$ data sets, where the redundant information of the Fermi component maps is excluded from the training sample.
The GMM prediction of the 1.70\,GeV map based on all non-$\gamma$-ray maps is shown in Figure~\ref{pred143whl}.
This displays a significantly improved resolution revealing more detailed filaments and more compact appearance of slightly extended sources, such as the North Polar Spur and the Magellanic Clouds.
Here, the Large Magellanic Cloud appears more dominant and the Small Magellanic Cloud can be easily identified.
The Galactic disk is reconstructed and also higher resolved.
This is visible at the outflow structures of the outer disk regions, for example, the Perseus, Taurus, and Orion region as well as the central filaments in the direction of the Fermi bubbles.
These outflows are now clearly defined in shape and structure.\par
The emission of the southern Fermi bubble, however, is missing nearly completely from the reconstructed image.
This missing flux becomes evident in the difference map (Fig.~\ref{org-pred143whl}).
This could be attributed to the fact that the GMM has been trained with only $K=3$ Gaussians.
In case of the Fermi bubble its emission is not separately visible in the other data sets, but it could be contained in their faint diffuse background emission outshined by the other large-scale Galactic outflows.
This embedded information, however, can be extracted by training the GMM with a higher number of Gaussian components.\par
For that, the number of Gaussians has to be selected so that overfitting\footnote{Overfitting means an overly precise representation of the data by the Gaussians so that the underlying information is less accurately abstracted. For further detail see, e.g, \cite{Bishop}} is avoided.
Additionally, in case of using many Gaussians the GMM could be able to learn local information from the magnitude space input, which we also circumvent.
As an example, a reconstruction with $K=18$ Gaussians is shown in Figure~\ref{pred14_18_whl}, where the southern Fermi bubble is nearly completely recovered.
A comparison of the RMS values reveals this improvement.
While  the difference map between the original data sets and its prediction computed from a GMM trained with $K=3$ Gaussians provides an RMS of 0.13 (most of the deviation is due to the resolution improvement), the RMS of the prediction computed with $K=18$ shows less deviation with an RMS of 0.10.\par
However, the southernmost part is still missing (Fig.~\ref{org-pred14_18_whl}) indicating that the $\gamma$-ray sky contains at least one physical process that cannot be re-predicted from the other frequency regimes.
This is supported by the fact that a choice of even more Gaussians does not improve the reconstruction of the southernmost part significantly.\par
We find that the resolution of the Fermi data at $1.70$\,GeV is significantly improved using a GMM trained with only $K=3$ components.
The faintest emission structures (i.e., the Fermi bubbles) are determinable from a GMM trained with $K=18$ components.
However, we find $\gamma$-ray structures that cannot be reproduced from the other frequency regimes.
Therefore, the southernmost part of the Fermi bubble seems to be special in comparison to other areas of the sky.
\subsection{Restoring the hadronic sky}
In \cite{Fermi1} the spectral information of the diffuse $\gamma$-ray flux has been used for a sky decomposition \citep{Fermi2} into two component maps, which are claimed to represent mostly the hadronic and leptonic interactions of cosmic rays with the ISM, respectively.
The spatial resolution of these maps is low compared to our maps in other frequency regimes (Fig.~\ref{org28}, \ref{org31}).\par
For restoring the hadronic component map the GMM is trained with $n=30$ data sets and $K=3$ Gaussians.
Here, the information of the nine Fermi frequency maps is excluded from training, since they are redundant to the derived component maps.
The magnitudes in Figure~\ref{pred28_3} are computed from the determined CPD marginalized over the leptonic component.
It shows a significant increase in resolution similar to the improvement we found in the reconstructed 1.70 GeV map.
However, here we find an even stronger quality increase in structural detail all over the map.
Small compact regions such as the Taurus, Perseus, and Orion regions become visible in their filament structures.
The Magellanic Clouds, the Ophiuchus region, and the central Galactic outflows into the southern hemisphere emerge from the local diffuse background revealing detailed small-scale variations.
The filament structures of the North Polar Spur are now highly resolved.
Additionally, the predicted map is free of artifacts.\par
The original map contained an empty region, displayed in Figure~\ref{org28}.
This region was assumed to be of solely leptonic origin in the original decomposing process and no flux in that region was attributed to the hadronic component.
We are able to predict this missing part of the sky without visible artifacts yielding a completely covered high-resolution hadronic component map.
Most of the structures seen in the difference map (Fig.~\ref{org28-pred28_3}) are attributed to the fact that these are more extended in the lower-resolved original map in comparison to the higher-resolved reconstruction.
This disparity in resolution leaves residues in the difference map.
\subsection{Predicting the leptonic sky}
\begin{figure*}[h!]
\centering
   \subfloat[\label{org28}]{\includegraphics[trim=0 60 0 0, clip, width=\columnwidth]{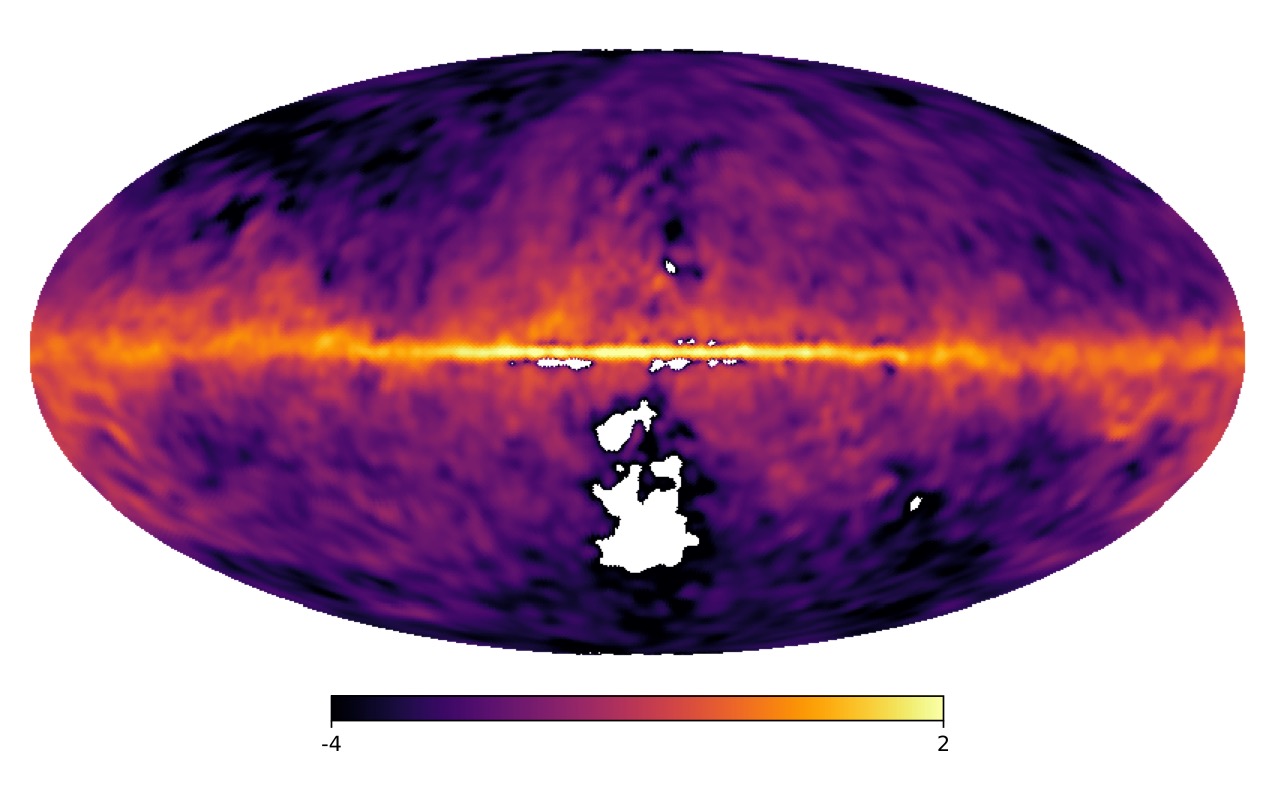}}
   \hfill
   \subfloat[\label{org31}]{\includegraphics[trim=0 60 0 0, clip, width=\columnwidth]{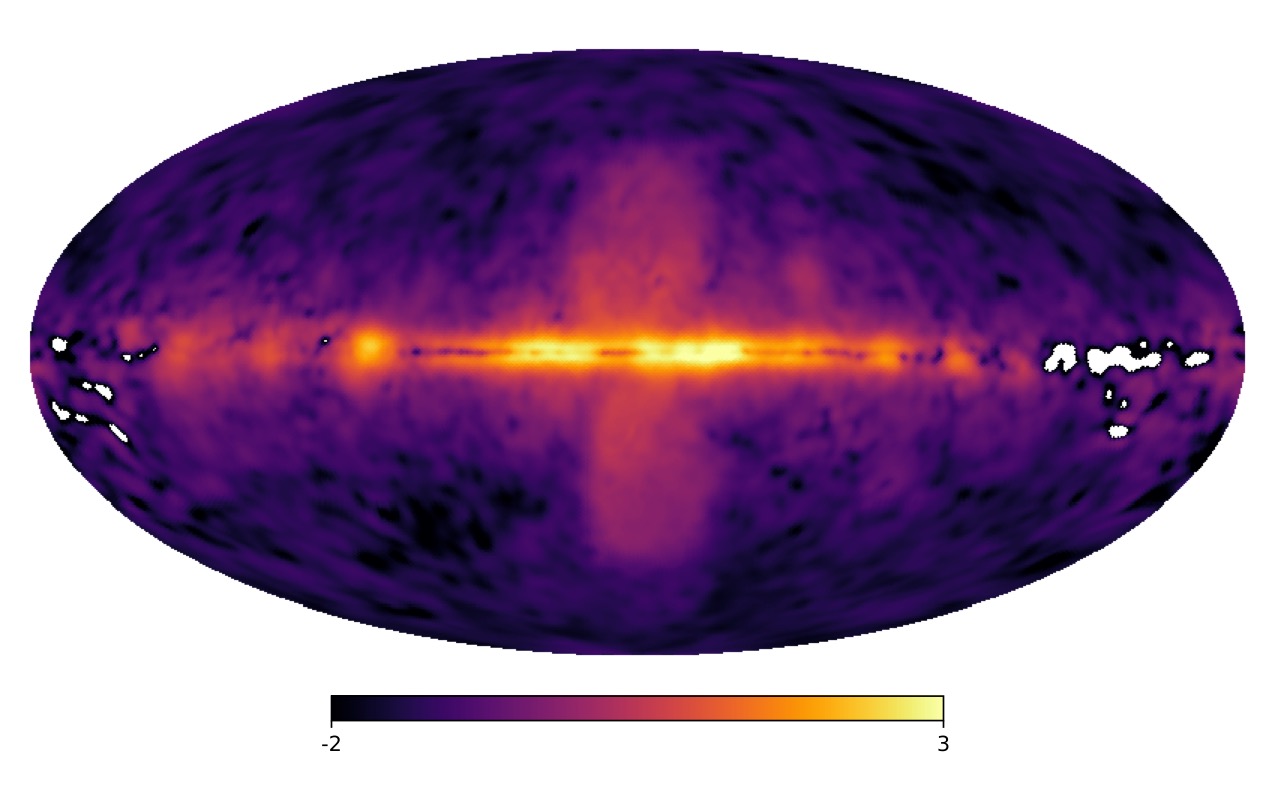}}
   \\
   \subfloat[\label{pred28_3}]{\includegraphics[width=\columnwidth]{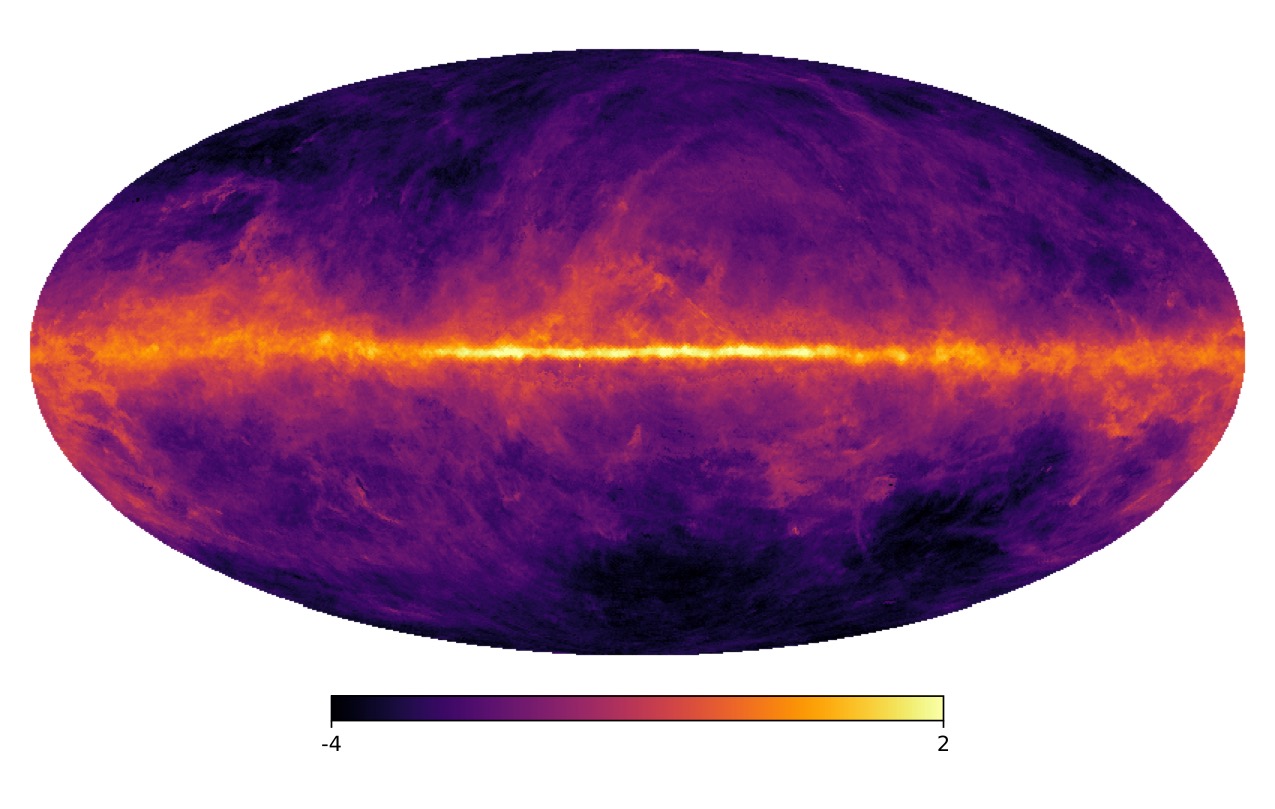}}
   \hfill
   \subfloat[\label{pred31_3}]{\includegraphics
  [width=\columnwidth]{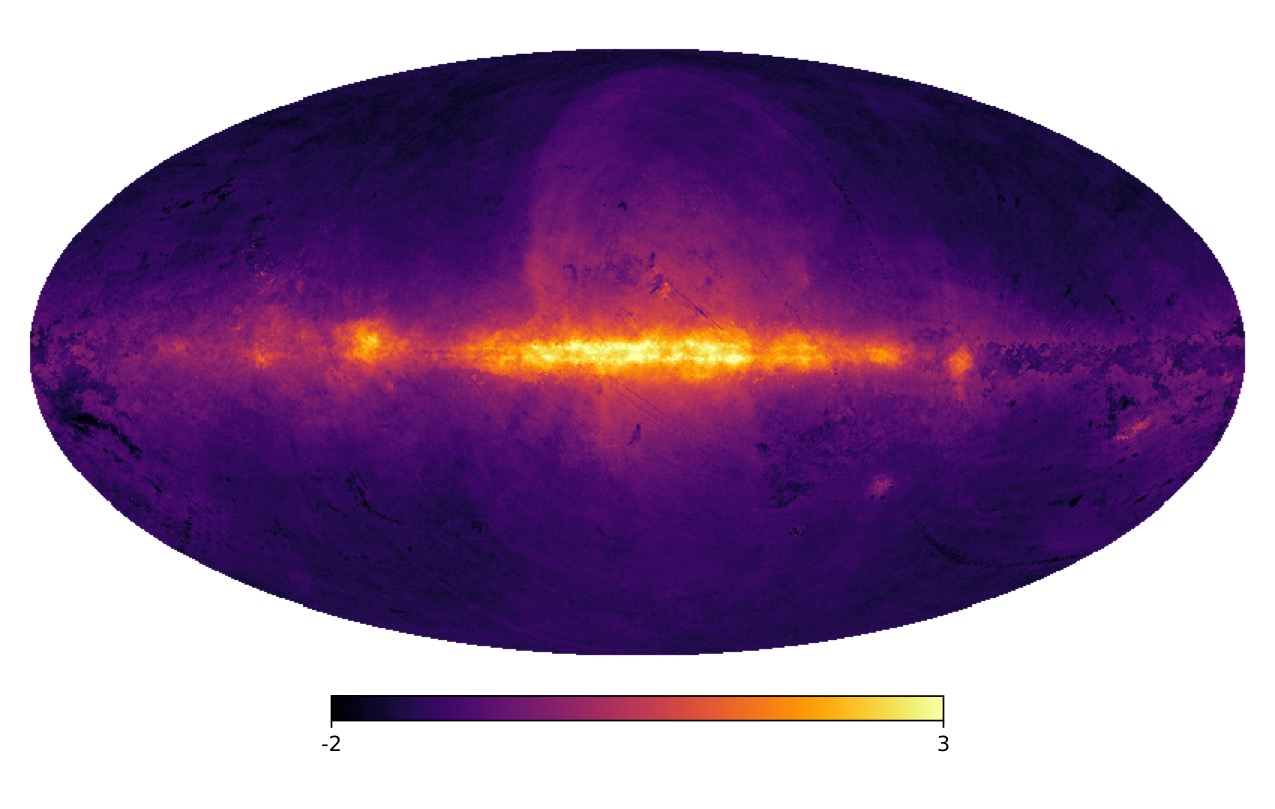}}
\\
   \subfloat[\label{org28-pred28_3}]{\includegraphics[width=\columnwidth]{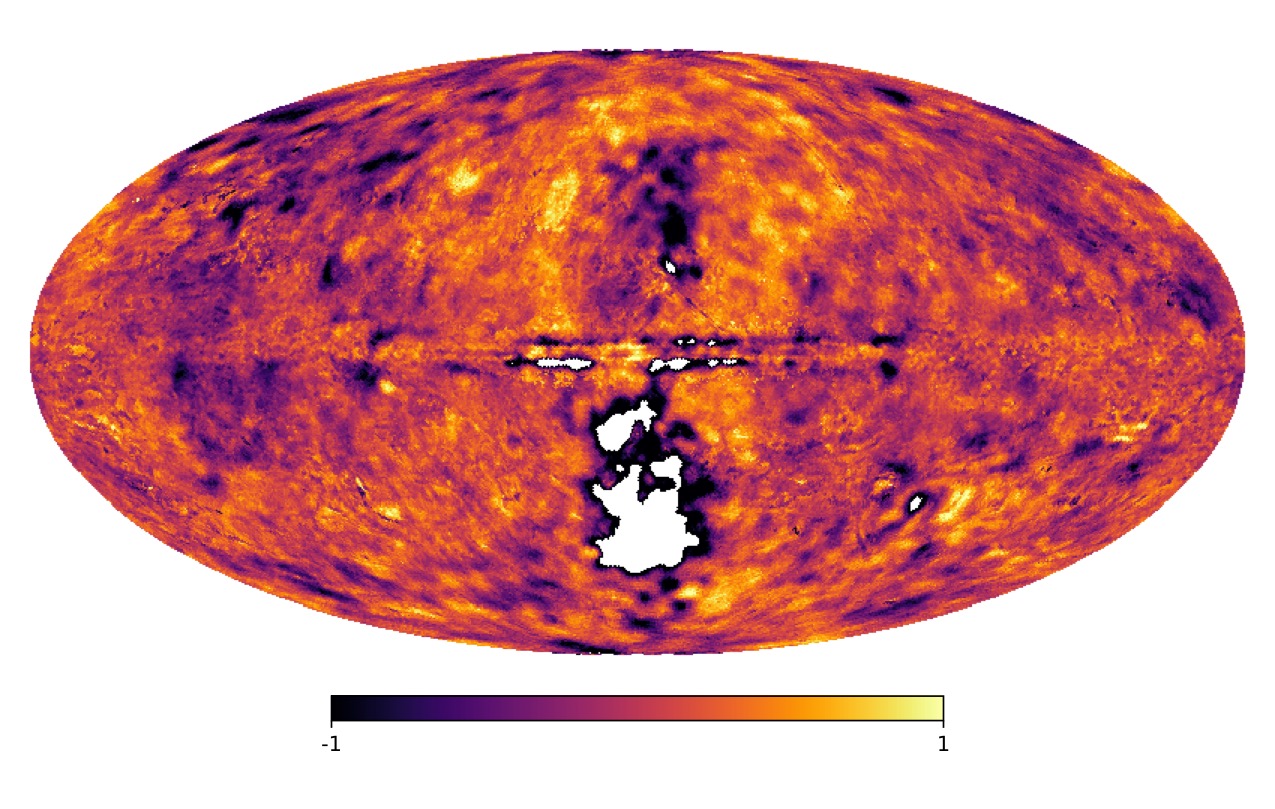}}
   \hfill
   \subfloat[\label{org31-pred31_3}]{\includegraphics
  [width=\columnwidth]{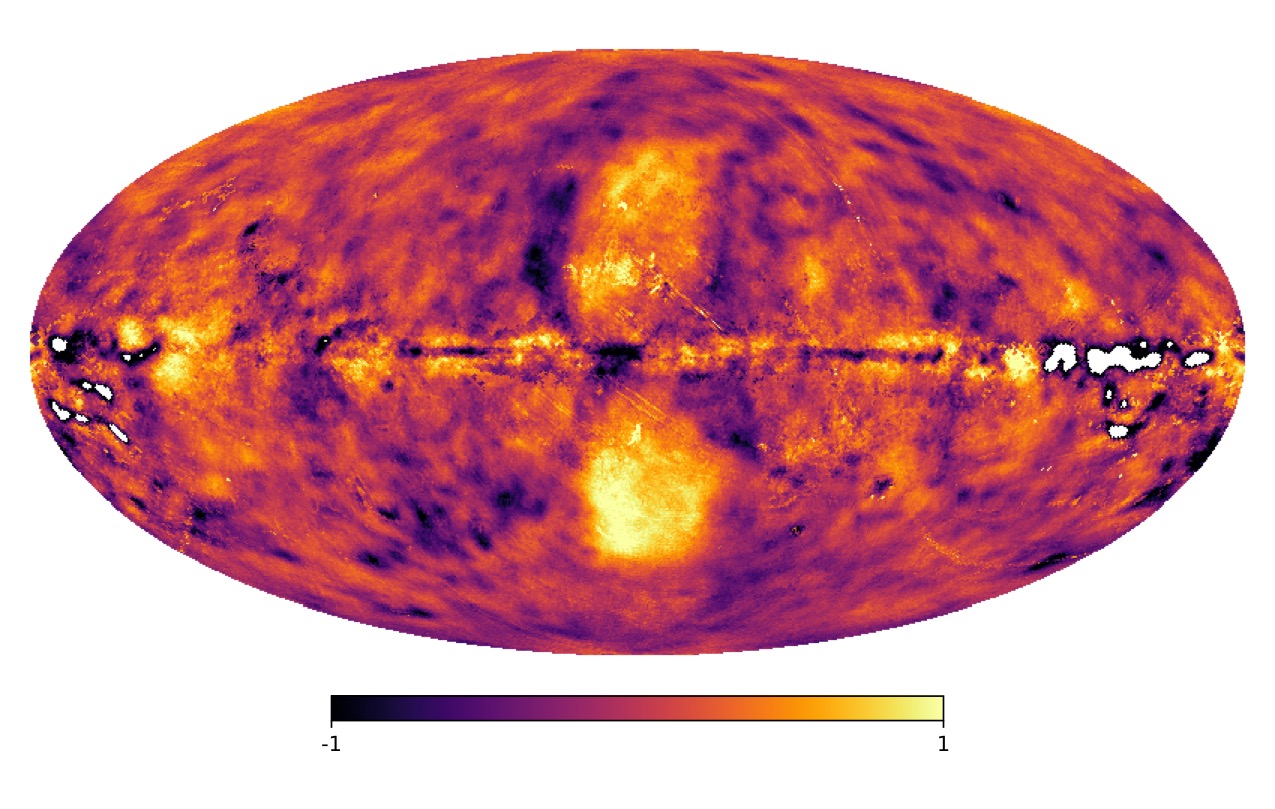}}
  \caption[]{GMM results for the claimed hadronic and leptonic component; logarithmic scaling. The predictions are computed from the GMM trained with $K=3$ Gaussians and $n=30$ data sets neglecting the $\gamma$-ray signal maps. The predictions are determined from the CPD marginalized over the $\gamma$-ray component maps. Panel (a): Original hadronic component. Panel (b): Original leptonic component. Panel (c): Predicted hadronic component. Panel (d): Predicted leptonic component. Panel (e): Difference between the original and predicted hadronic component. Panel (f): Difference between the original and predicted leptonic component.}
\end{figure*}
Figure~\ref{org31} shows the $\gamma$-ray sky presumably from leptonic origin, as provided by \cite{Fermi1}.
The northern and southern Fermi bubbles are clearly visible.
This map carries imaging artifacts, most prominently the splitting of the Galactic emission into two layers with a gap in the middle.
We will study how the GMM deals with such unusual structures.
To this end, we again use the GMM trained with $n=30$ data sets as in the previous section.
The prediction of the leptonic component is determined from a CPD marginalized over the hadronic component and conditioned on all non-$\gamma$-ray maps.
The resulting prediction shown in Figure~\ref{pred31_3} has a higher spatial resolution as the original leptonic $\gamma$-ray map.
The artificial splitting of the original map is largely reduced.
The Cygnus X and Vela regions and their surroundings are higher resolved and show considerable detail.
These regions are pronounced in the leptonic prediction compared to the hadronic prediction.
This would attribute these features to mainly leptonic origin as has been assumed from the decomposition by \cite{Fermi1}.
The gap in the Galactic disk of the original leptonic map (Fig.~\ref{org31}) is now filled as well.
However, in this case we find clear residuals of the gap in contrast to the smooth prediction of the hadronic emission at locations where the original map clearly shows artifacts (see Fig.~\ref{pred31_3}).\par
The Fermi bubbles, which are pronounced in the original map, however, have not been recovered by the GMM in their distinct shape. This can be seen clearly in the difference map (Fig.~\ref{org31-pred31_3}).
The $\gamma$-ray emission of the North Polar Spur, on the other hand, is predicted as part of the leptonic emission, although it is almost absent in the original leptonic component map and is rather seen as a clear feature of the original hadronic sky.
While the component separation yielded a clear separation attributing the North Polar Spur to solely hadronic interactions, the GMM identified a correlation of the magnitudes that partly ascribes the structure forming the North Polar Spur also to leptonic interactions.\par
Thus, the GMM prediction partly contradicts here the original decomposition by \cite{Fermi1}.
That original decomposition was based on spectral similarity in the $\gamma$-ray regime with the southern Fermi bubble for the leptonic component, and with a western Galactic disk cloud complex for the hadronic component, respectively.
The GMM, on the other hand, searched for correlations of these $\gamma$-ray components with structures in other frequency regimes.
It therefore seems that the material forming the North Polar Spur lets the GMM expect predominantly leptonic emission from there, although the $\gamma$-ray spectrum points toward hadronic emission.
Since the used spectral templates for the original decomposition had different spectral slopes, with the hadronic template being steeper, a cooled electron population in the North Polar Spur region might just have mimicked a hadronic component in the original decomposition.
However, from the GMM predictions we find that these structures are not congruent but rather a tripartition of a common larger structure building the North Polar Spur.
The hadronic component constitutes the inner and outer arch shape, and the leptonic component the space in between.\par
Furthermore, the Large Magellanic Cloud emerges in the leptonic prediction as a small, dense region.
In the hadronic prediction, however, it is seen as a more extended structure.
We assume that this difference in apparent morphology coincides with our impression from the Galactic decomposition.
That means that it might be reasonable to assume that the leptonic emission is more concentrated to the central region propagating along the Galactic latitude, while the hadronic emission is more diffuse and spread along the Galactic plane with a smaller perpendicular extent.
This separation might hold for the Milky Way as well as for the Large Magellanic Cloud.
Further, the GMM is dominated by the magnitude correlation of the Milky Way such that this interpretation might be biased by the distribution of the Galactic hadronic and leptonic component.\par
We find additional artifacts all over the predicted map, seen most noticeably in the central region north and south of the Galactic plane.
These can be attributed to the other maps, such as the unobserved stripes of the AKARI survey or the scanning lines of the ROSAT survey (see Appendix \ref{App:Sds}).\par
Concluding, we find that, while we produced a higher-resolved leptonic component map, this prediction reveals some limitations of the GMM applied to this data set.
Since this data set is affected by several artifacts, the GMM is reproducing their distribution instead of weak embedded emission of, for example, the Fermi bubbles.
The usage of more Gaussians might circumvent the issue of weak structures, however the usage of less Gaussians is reducing the contamination of the artifacts.
The GMM prediction of the proposed leptonic $\gamma$-ray sky of \cite{Fermi1} might have revealed a misclassification of parts of the North Polar Spur $\gamma$-ray flux to the hadronic component.
Although the GMM was trained on a more hadronic North Polar Spur model, it clearly prefers a more leptonic emission model.
\begin{figure*}[h!]
\centering
   \subfloat[\label{pred20_3_pico}]{\includegraphics[width=\columnwidth]{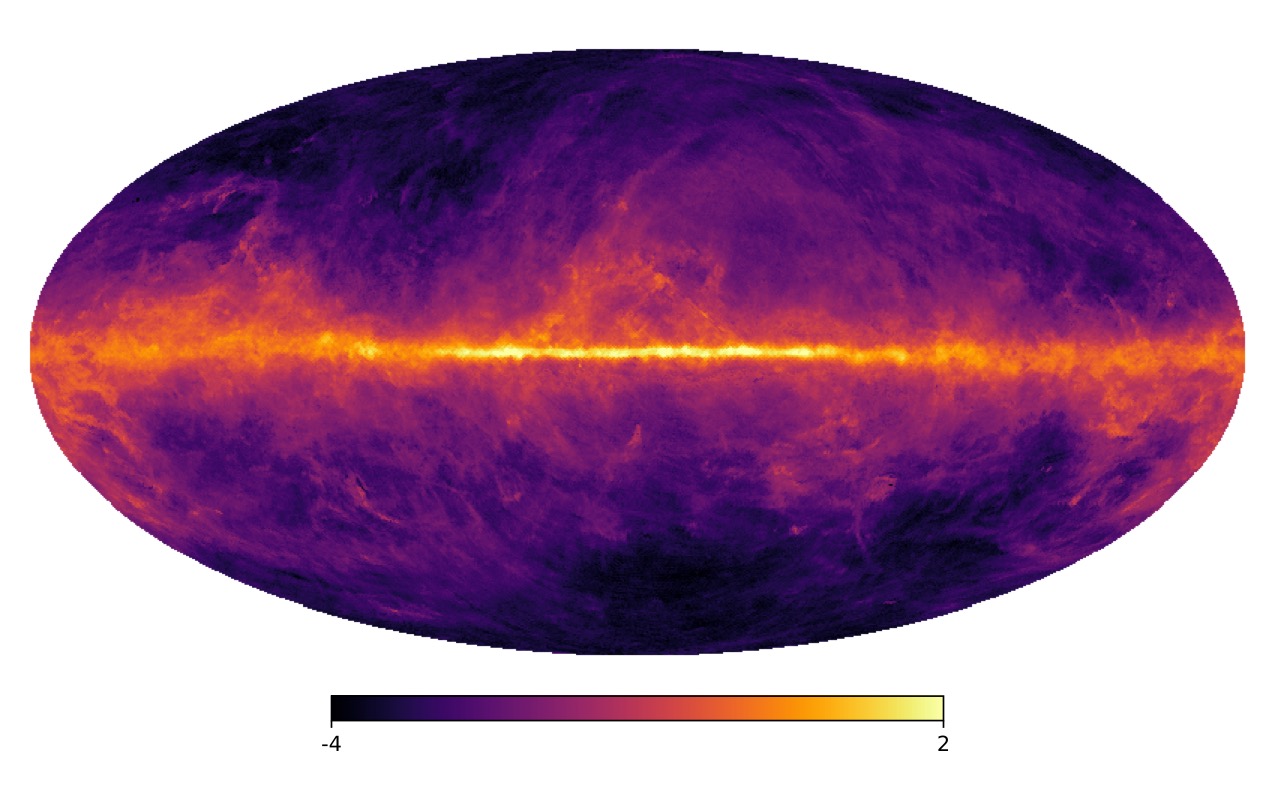}}
   \hfill
   \subfloat[\label{predhad-pred20_3_pico}]{\includegraphics[width=\columnwidth]{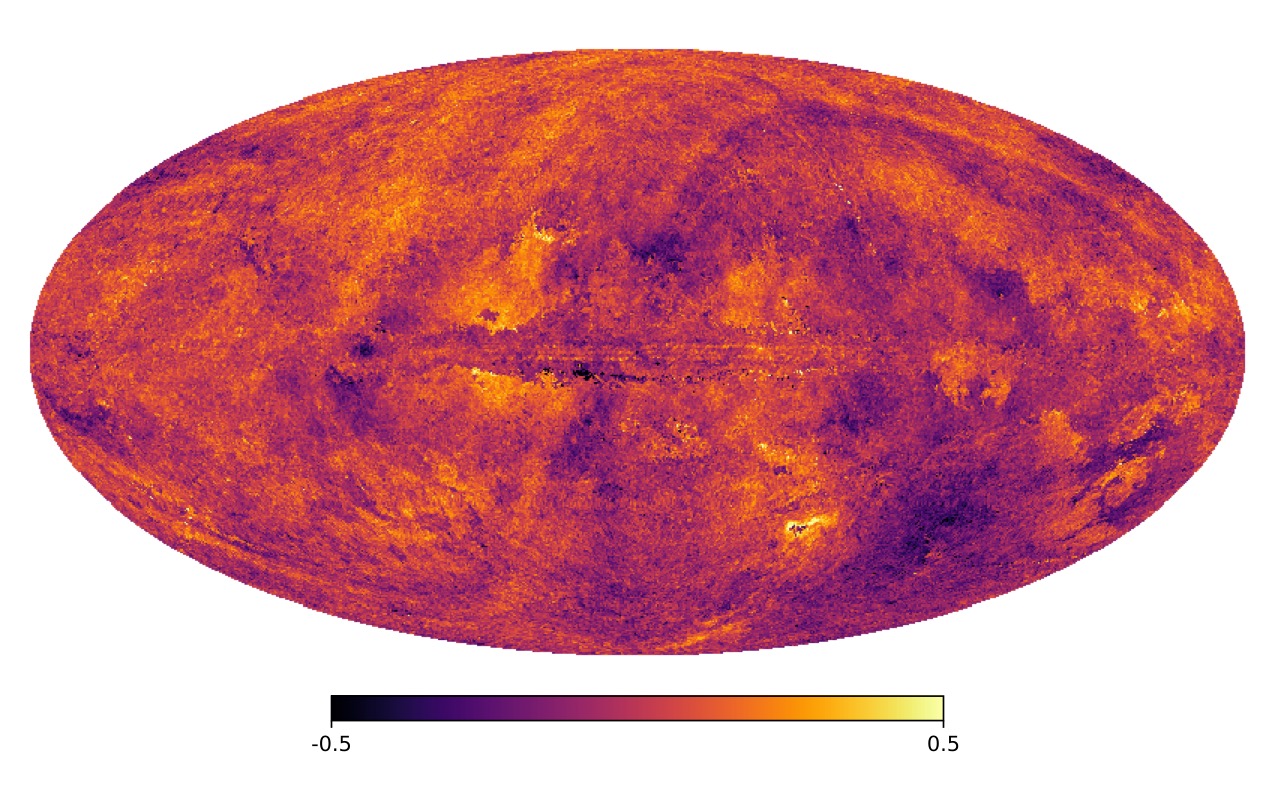}}
   \\
      \subfloat[\label{pred20}]{\includegraphics[trim=0 60 0 0, clip, width=\columnwidth]{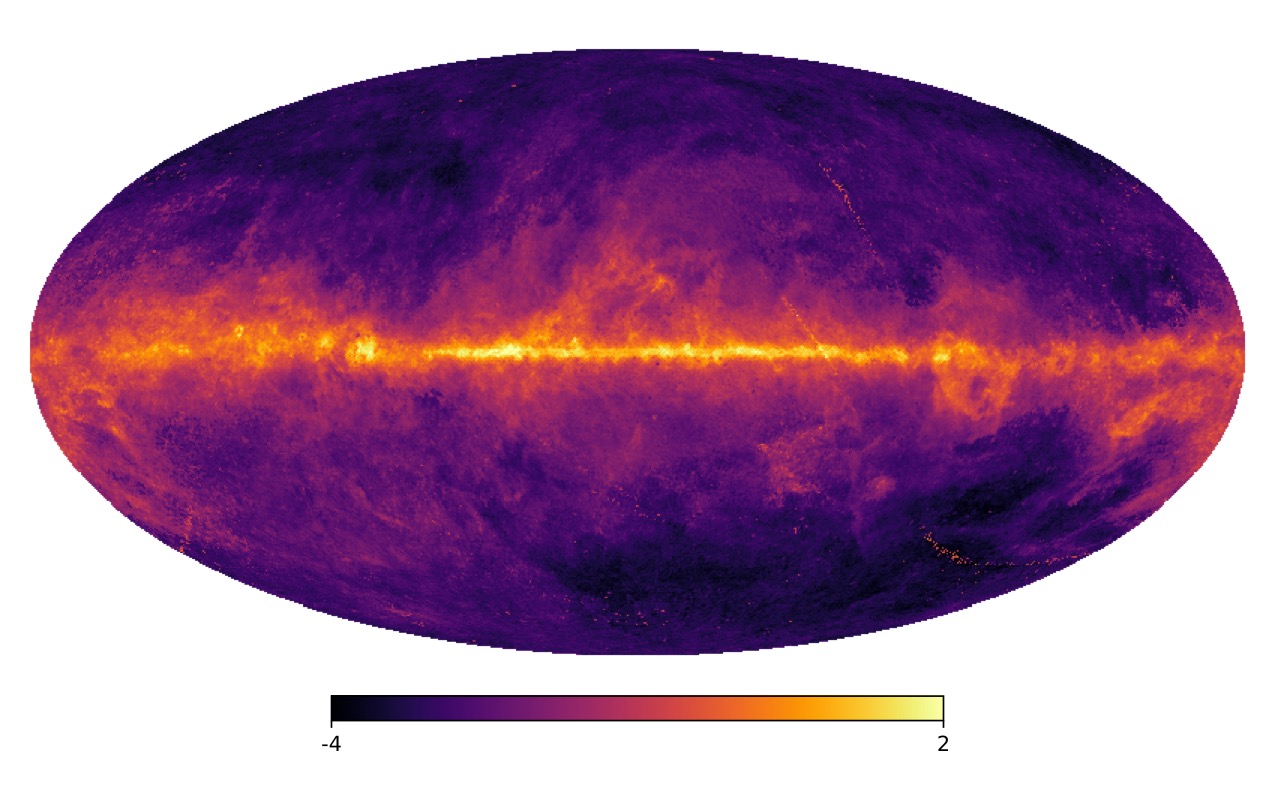}}
   \hfill
   \subfloat[\label{predall20-pred20}]{\includegraphics[trim=0 60 0 0, clip, width=\columnwidth]{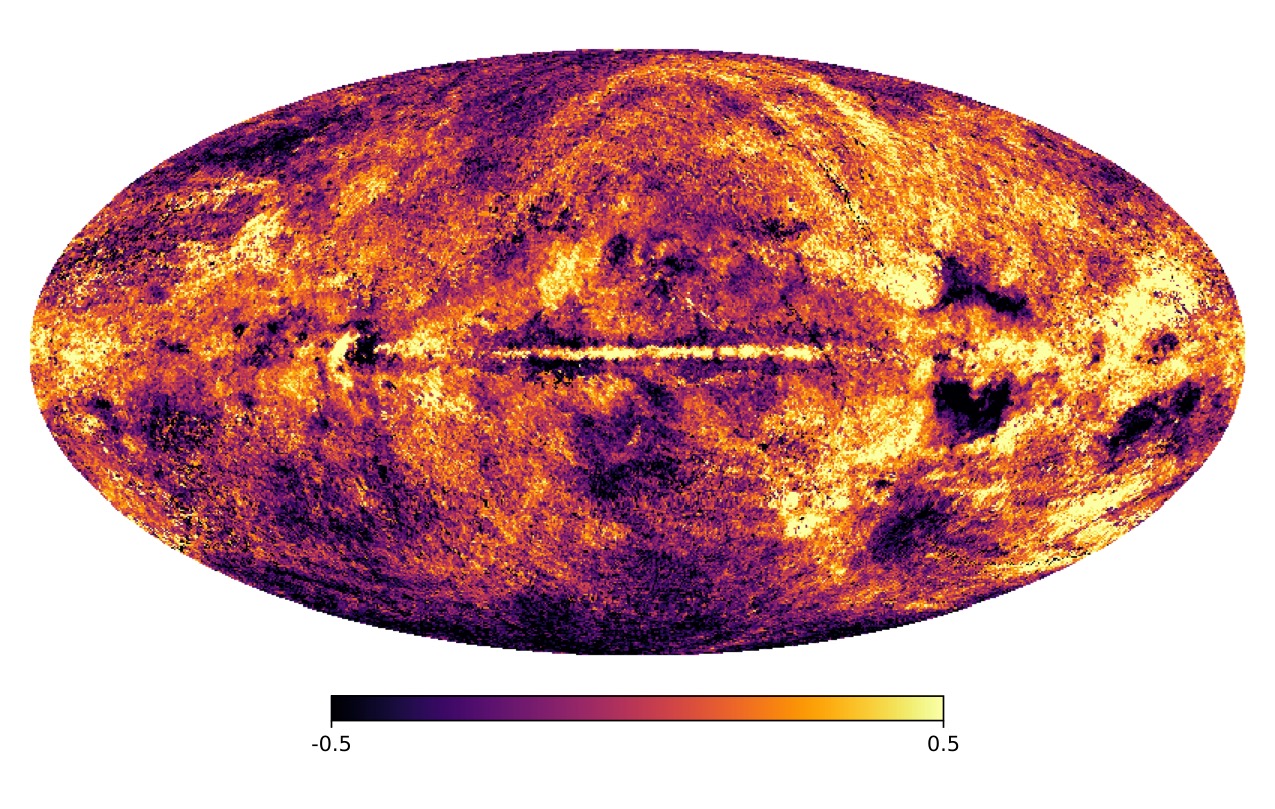}}
   \caption[]{GMM results for the composition of the hadronic component; logarithmic scaling. The predictions are computed from the GMM trained with $K=3$ Gaussians and $n=30$ data sets neglecting the $\gamma$-ray signal maps. The predictions are determined from the CPD marginalized over the $\gamma$-ray component maps. Panel (a): Preditiction of the hadronic component given the CO-line emission, the Planck HFI maps, and the infrared regimes. Panel (b): Difference between the prediction of the hadronic component given all frequency regimes except the $\gamma$-ray regime (Fig \ref{pred28_3}) and the selection of frequency regimes of Figure \ref{pred20_3_pico}, $\textrm{RMS}=0.12$. Panel (c): Preditiction of the hadronic component given the radio continuum emission, the Planck LFI maps, and the X-ray regime. Panel (d): Difference between the prediction of the hadronic component given all frequency regimes except the $\gamma$-ray regime (Fig \ref{pred28_3}) and the selection of frequency regimes of Figure \ref{pred20}, $\textrm{RMS}=0.31$.}
\end{figure*}
\subsection{Inspecting physical connections}
\label{physicalconnections}
\begin{figure*}[h!]
\centering
   \subfloat[\label{pred22_3_rx}]{\includegraphics[width=\columnwidth]{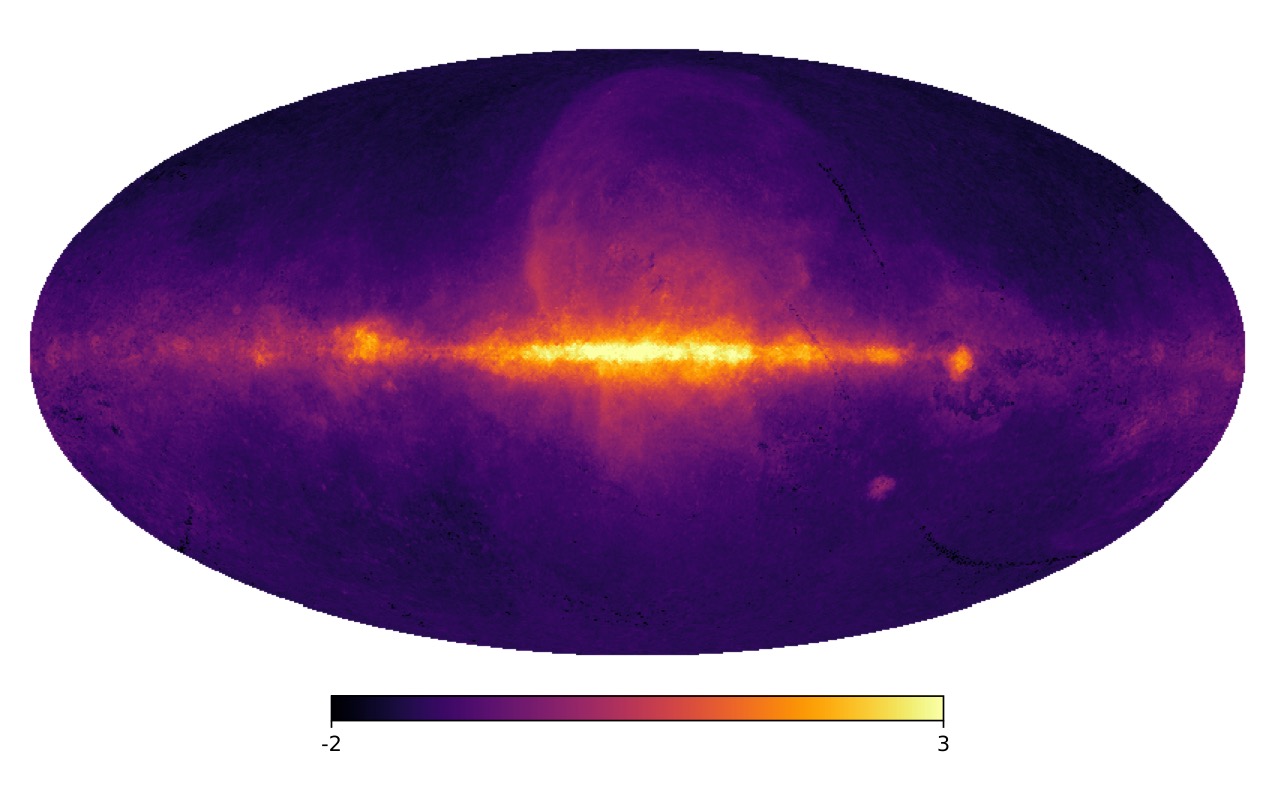}}
   \hfill
   \subfloat[\label{predlep-pred22_3_rx}]{\includegraphics
  [width=\columnwidth]{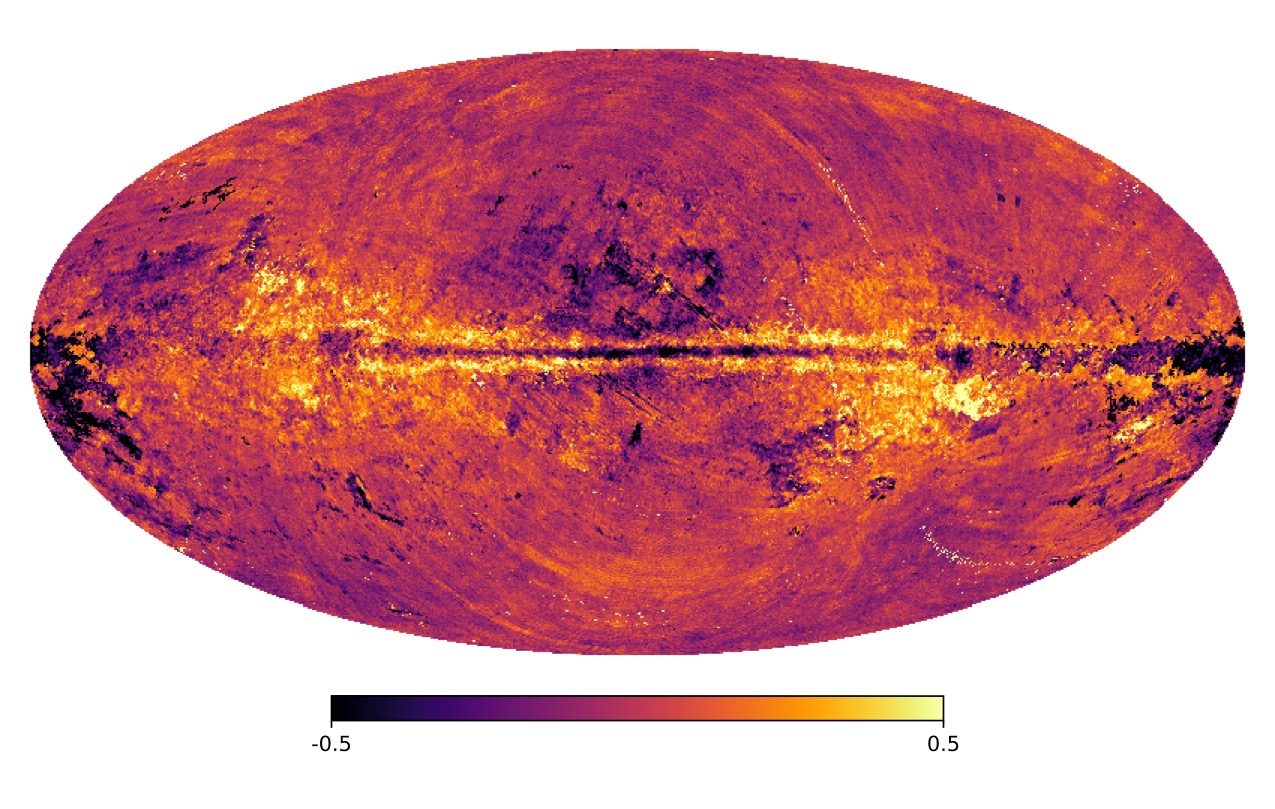}}
    \\
   \subfloat[\label{pred22}]{\includegraphics[trim=0 60 0 0, clip, width=\columnwidth]{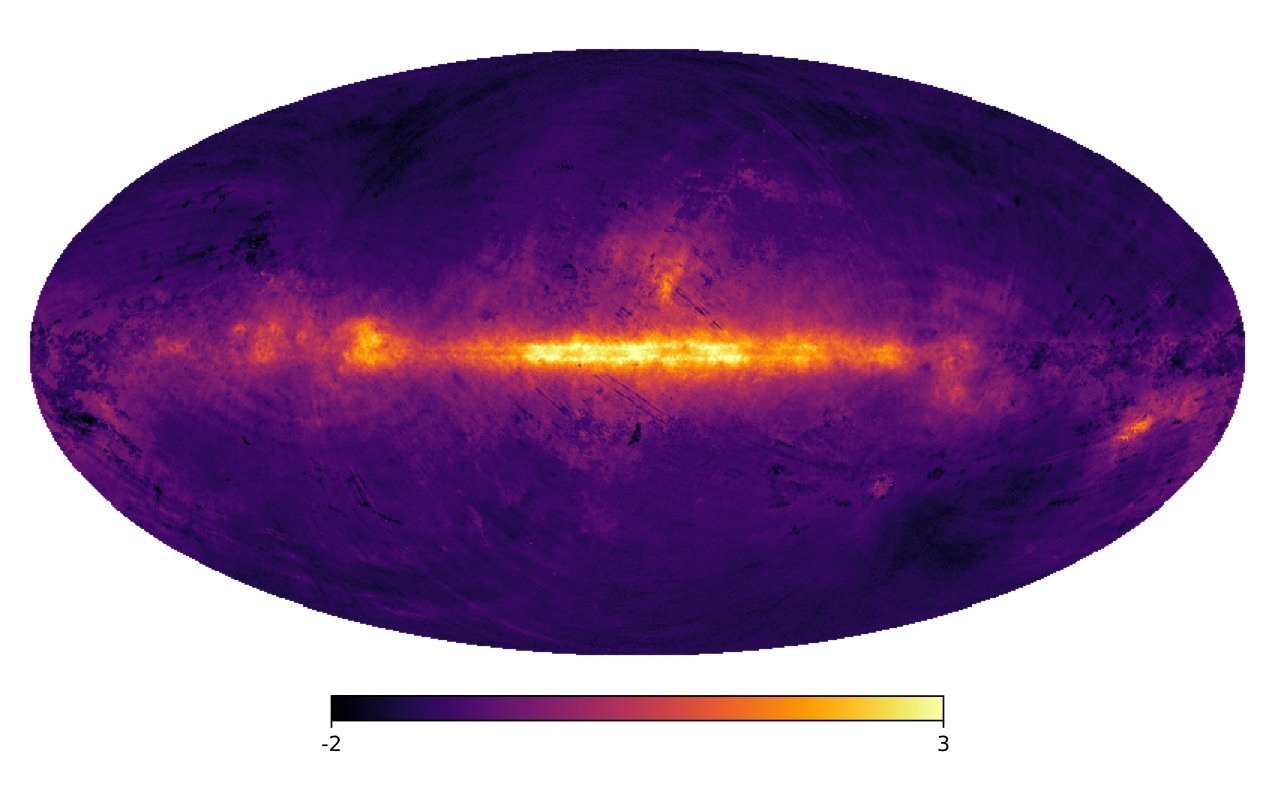}}
   \hfill
   \subfloat[\label{predall22-pred22}]{\includegraphics
  [trim=0 60 0 0, clip, width=\columnwidth]{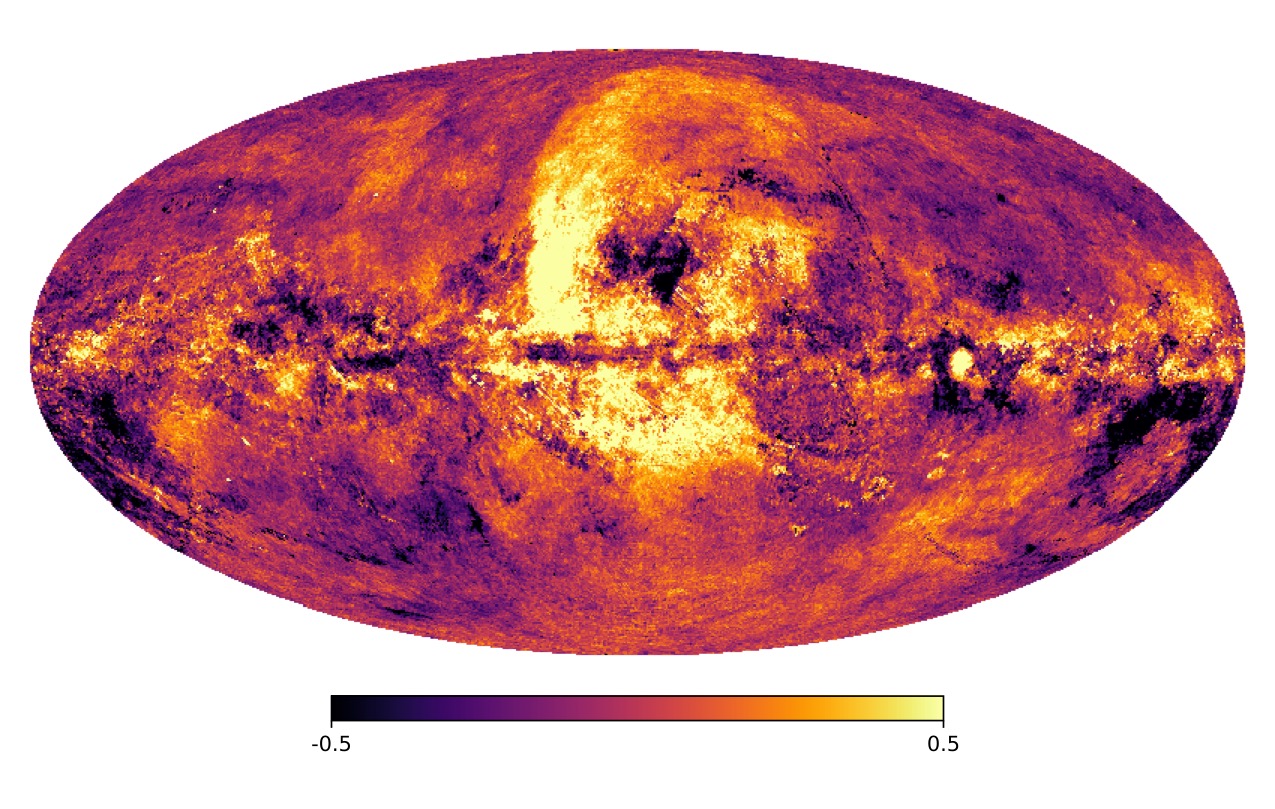}}
  \caption[]{GMM results for the composition of the leptonic component; logarithmic scaling. The predictions are computed from the GMM trained with $K=3$ Gaussians and $n=30$ data sets neglecting the $\gamma$-ray signal maps. The predictions are determined from the CPD marginalized over the $\gamma$-ray component maps. Panel (a): Preditiction of the leptonic component given the radio continuum emission, and the Planck LFI and the X-ray regime. Panel (b): Difference between the prediction of the leptonic component given all frequency regimes except the $\gamma$-ray regime (Fig \ref{pred31_3}) and the selection of frequency regimes of Figure \ref{pred22_3_rx}, $\textrm{RMS}=0.16$. Panel (c): Preditiction of the leptonic component given the CO-line emission, and the Planck HFI and the infrared regimes. Panel (d): Difference between the prediction of the leptonic component given all frequency regimes except the $\gamma$-ray regime (Fig \ref{pred31_3}) and the selection of frequency regimes of Figure \ref{pred22}, $\textrm{RMS}=0.23$.}
\end{figure*}
The GMM predictions of the individual sky maps and frequency regimes work relatively well since the information about the presence and magnitude of a component is also stored in other frequency maps.
We ask which other frequency maps carry the relevant information of the hadronic and leptonic component maps.
To this end, we assess how much the two $\gamma$-ray component maps are determined by the separate other frequency regimes.
Therefore, we again use the GMM trained with $n=30$ data sets to predict each $\gamma$-ray component.\par
We find that combining the Planck HFI and infrared data together with the CO dust component (see Appendix \ref{App:Sds}) permits the GMM output marginalized over the remaining frequency regimes to predict the hadronic component map (Fig.~\ref{pred20_3_pico}) with almost identical accuracy as when using all available frequencies except for the $\gamma$ regime (Fig.~\ref{pred28_3}).
Including the CO line emission yielded a more accurate reconstruction of the Galactic disk emission.
This is in perfect agreement with the interpretation that the origin of the hadronic radiation coincides with the location of high dust concentration as dust reveals dense environments in which hadronic interactions of cosmic rays are abundant (\citealt{hadronic2} and \citealt{hadronic1}, and references therein).
This also agrees well with the findings of \cite{Fermi1}.\par
The difference map (Fig.~\ref{predhad-pred20_3_pico}) shows only small variation with some flux missing especially around the Small Magellanic Cloud.
Hence, the radiation of the ISM observed in the other used frequency regimes seems to play a minor role.
To substantiate this claim, we predicted the hadronic component map given only the radio continuum, the Planck LFI, and the X-ray data sets, which we find to be mainly connected to the leptonic radiation, as demonstrated later on.
This prediction (Fig. \ref{pred20}) shows an inferior reconstruction of only a part of the original features.
Especially, this prediction suffers from artifacts contained in the original data sets.
The influence of the artifacts as well as the low accuracy of the prediction is significantly reflected in the strong remaining magnitudes in the difference map shown in Figure \ref{predall20-pred20} as well as its RMS value of 0.31, which is much larger than the RMS of 0.12 of Figure~\ref{pred22_3_rx}.\par
The spatial information for the re-predicted leptonic emission, however, can be traced back to the contributions of mainly the X-ray, the Planck LFI, and the radio continuum (i.e., synchrotron) emission, as can be seen in Figure~\ref{pred22_3_rx} using only these parts of the multifrequency data sets for prediction.
Compared to the overall similar prediction using all frequency ranges except the $\gamma$-ray regime (Fig.~\ref{pred31_3}) the disk emission is better reconstructed disposing of the splitting along the Galactic plane. The influence of artifacts, especially in the outer disk regions, is highly reduced yielding an overall smooth prediction. Both effects are visible in the difference map shown in Figure~\ref{predlep-pred22_3_rx}.
Including also the Planck LFI synchrotron emission yielded a slight increase in resolution and contrast such that the southern Fermi bubble is marginally better separated from the local background.
The regions around Cygnus X and Vela are better reproduced.\par
The leptonic $\gamma$-ray emission is generated by inverse Compton scattering of the Galactic photon field by relativistic electrons \citep{leptonic, hadronic1}.
The presence of the latter might coincide with hot gas visible in the X-ray regime and with radio synchrotron emission if these relativistic electrons reside in magnetized areas.\par
To test this assumption that the leptonic component is predominantly describable by the radio continuum emission, the Planck LFI, and the X-ray regime we compute the prediction of the leptonic component map given only the data containing the thermal dust emission.
This prediction of the leptonic component from a CPD that includes the CO-line emission, the infrared, and the Planck HFI regime marginalized over the remaining frequency regimes is shown in Figure~\ref{pred22}.
Here, the unwanted reconstruction of the disk splitting becomes visible again, and the emission from the outflow regions above and below the disk, especially the North Polar Spur and the Fermi bubbles, is missing almost entirely, which becomes evident in the difference map (Fig.~\ref{predall22-pred22}).
Additionally, the influence of the artifacts contained in the original data sets becomes extensive.
This influence is reflected in the significantly differing RMS values of 0.16 for the prediction composed of the synchrotron emission and X-ray maps and 0.23 for the dust emission maps.
We conclude that the leptonic $\gamma$-ray emission is not well encoded with tracers of the dense ISM.\par
%


\section{Conclusions}

\subsection{Machine learning lessons}

We investigated the mutual information of Galactic all sky maps in
various frequency ranges.
For this, we trained a GMM on the magnitude vectors of the pixels of a common sky pixelization into which we transformed a set of 39 maps.
The GMM representation of the magnitude vector distribution could then be asked to reconstruct individual maps pixel by pixel from subsets of other maps.
Thereby, we could test how well the magnitudes of the other, complementary maps
encode the original one and therefore how well the physical phenomena
visible in the maps are connected.\footnote{All the presented maps can be downloaded from \href{https://wwwmpa.mpa-garching.mpg.de/ift/data/sharpening_up_galactic_sky_maps/}{https://wwwmpa.mpa-garching.mpg.de/ift/data/sharpening\_up\_galac-tic\_sky\_maps/}}
This method managed to
\begin{itemize}
\item partially heal imaging artifacts,
\item complete incomplete maps to a certain degree,
\item improve the resolution of low-resolution maps (by transferring the information of the image structure from complementary maps to the target map),
\item reveal sky structures genuine to some frequency regime (albeit failing
to reconstruct them fully), and
\item point out potential inconsistencies in maps (by proposing that certain
structures are absent or present in disagreement with the training
data).
\end{itemize}
All these achievements of the method are based to a varying degree
on its failure to reconstruct the original map exactly.
In the training phase the GMM abstracted relations between the different magnitudes and
then used these relations to complete the incomplete magnitude vectors
from subsets of the original training data. The abstraction was coarse,
as we used only three Gaussian components to represent the magnitude
distribution in 30 to 37 dimensional magnitude spaces. Peculiarities,
as generated for example by imaging artifacts in individual bands,
which do not have counterparts in other bands, where thereby erased
from the relation in the coarsely grained representation of magnitude
vector distributions by the GMM. This allowed the GMM to correct image
artifacts, to fill in empty regions, to transfer resolution between
maps, and to deviate in the reconstruction from the original maps
in areas where a map seems to be inconsistent with the conditions
indicated by the complementary set of other maps. \par
The results of the GMM should be taken with a grain of salt. As the
GMM does not have a real understanding of the astrophysical relations,
it can not really distinguish between image artifacts and real emission. It just happens that many of such artifacts are difficult
to be represented by the GMM and therefore erased in the procedure
of predicting a trained map from other maps. With increasing sophistication
of the modeling of the magnitude distribution, which can be achieved
by increasing the number of Gaussian components in the internal representation
of the GMM, the GMM starts to reproduce more and more of the artifacts
of the original map. As this is not wanted, two strategies to regularize
the solutions can be used:\par
The one applied here is \textbf{regularization by stupidity}. We just
did not provide the GMM with sufficient degrees of freedom, in other words the number of Gaussian components, to accurately represent the training distribution,
so that it ignored peculiarities. The right level of intelligence
of our GMM had to be found by trial and error and on the basis of
our physical insight into the correctness of the reconstructed maps.
Our choice of three Gaussians was based upon that this number gave sensible
looking and physically plausible results. It would be better to include
our physical insight directly and explicitely into the method.\par
Thus, a better approach would be regularization by physical insight.
That would require that training of the GMM would be guided by priors
that encode physical knowledge on the phenomena under study. Here,
only an inverse Wishard prior was present that gave a moderate preference
for compact Gaussians while preventing individual Eigenvalues of the
covariances of the Gaussians to collapse to zero. This is only a minimalist's
prior, mostly aimed at stabilizing the numerical machinery of GMM
training. The inclusion of physical priors, however, would require
much more conceptual work and argumentation. We therefore leave it
for future studies.

\subsection{Astrophysical findings}

Although the warnings spelled out above, that the GMM not necessarily
abstracts the physical relations but only what it is able to identify in the
data and to represent by its internal knowledge representation,
our experiments seem to provide a number of astrophysical insights.
These should be regarded as suggestions of a simple mind, our GMM with
only three Gaussian components, which, however, explored a high dimensional
magnitude space that is not that easily accessible for human brains.
These suggestions should be investigated by other means. However,
we think that many of these suggestions are plausible from a physical
point of view. The fact that the GMM has literally no understanding
of physical relation but can only reveal relations in data sets
in combination with the physical plausibility of these findings can be counted as
some support for them being real.
The astrophysical findings our GMM experiments suggest are:\par
\textbf{Firstly}, there exist sky structures that are genuinely represented in individual
frequency bands. The brightness of the North Galactic Spur or the
Vela region in X-rays and in particular the southern Fermi bubble
in $\gamma$-rays are not well encoded into the magnitudes at other
frequencies. They seem to be genuine phenomena of the frequency bands
in which they appear.\par
\textbf{Secondly}, the spectral decomposition of the $\gamma$-ray sky into a predominantly
hadronic and predominantly leptonic component by \cite{Fermi1}
is largely confirmed. The proposed hadronic sky map is mostly encoded
into tracers of the dense interstellar medium, such as dust and CO lines,
as it is expected for $\gamma$-rays produced by cosmic ray protons
hadronically interacting with atoms \citep{hadronic1, hadronic2}. The proposed leptonic sky map
is mostly encoded in the synchrotron emission of the radio, the microwave (Planck LFI), and the X-ray maps. This is also plausible,
as the leptonic emission is due to inverse Compton scattering of
Galactic photons by cosmic ray electrons. As the Galactic photon field
is only smoothly structured, the leptonic map largely follows the
cosmic ray distribution in the Galaxy that follows the hot ISM as
seen in X-ray, and reveals itself via radio synchrotron radiation.\par
\textbf{Thirdly}, the spectral decomposition of the $\gamma$-ray emission of the North
Galactic Spur, however, seems to be incorrect according to the magnitude
relations our GMM found. This expects for the North Galactic Spur
a predominantly leptonic component in contrast to the original decomposition, which
found a predominately hadronic component there. This might indeed
point to a real classification error that could have a simple physical
explanation. The $\gamma$-ray-producing electrons in the North Galactic
Spur could have a steepened spectrum (due to radiative cooling processes)
with respect to other typical locations in the Milky Way, and in particular
with respect to the southern Fermi bubble, from which the spectral
template of the leptonic component was taken. For that reason, the
simplistic spectral fitting of the decomposition performed by \cite{Fermi1}
could have assigned its $\gamma$-ray flux to the spectrally steeper hadronic
component.\par
\textbf{Finally}, the hadronic $\gamma$-ray sky is expected to exhibit the filigree
structures of the cold, dense ISM visible in the Planck and infrared dust maps
if observed at higher resolution.

\section{Outlook}

Our study should be regarded as an exploration of what machine learning
can do with astrophysical data sets. It showed promising results
but also shortcomings. In order to overcome the latter, further studies
and research is necessary. We anticipate the following research directions:\par
\begin{itemize}
\item[1.] In our work, the regularization used was technically motivated (Wishard prior) or ad-hoc (only three
GMM components used). The \textbf{inclusion of physical knowledge} (prior knowledge) to guide
training should lead to more reliable results that can also be extrapolated
beyond the training data.\par
\item[2.] We had hoped that the association of the
individual pixels with Gaussian components, which the GMM internally
does, would allow to decompose the sky into line of sights dominated
by individual ISM phases, such as the hot, warm, and cold ISM. Visual
inspection of such main component maps (see Appendix~\ref{App:color}) does not support this. Nevertheless, an \textbf{automatic classification} of different astrophysical
environments might be possible, maybe by using other machine learning tools such as self organized maps \citep{SOM1982, SOMBI}. \par
\item[3.] The relation found in the data and exploited so far for the (re-)prediction of maps were purely statistical.
An alternative route to putting in physical priors could be the \textbf{detection
of physical relations} (abstraction). Deep neuronal networks \citep{deeplearning} and in particular auto-encoder \citep{autoencoder} are a promising route for this.\par
\item[4.] The (re-)predicted maps lacked certain features.
In principle, the information on them is available, as the original
maps having those features exist as well. Thus, the \textbf{data fusion} of
original and predicted maps could in principle combine the advantages
of both. This would require a stringent information-theoretical treatment,
as the original maps are used twice in this process, once to train
the machine and once to guide the reconstruction. Information field theory
\citep{ensslin09} is a suitable language for the necessary bookkeeping of uncertainties and information forces. \par
\end{itemize}

\begin{acknowledgements}
We acknowledge valuable contributions, useful discussions, and comments on the manuscript by Max-Niklas Newrzella, Martin Reinecke, Reimar Leike, Jakob Knollm\"uller, Björn Adebahr, Stefan Blex, and Alexander Becker.
We also thank the anonymous referee for the insightful comments and suggestions.
We thank Patricia Reich for providing the HEALPix map of the 1420\,MHz radio continuum observations.
\end{acknowledgements}
\bibliographystyle{aa} 
\bibliography{Literatur} 
\begin{appendix} 
\section{Additional information for the simulation}\label{App:Mat}
Here, additional information is given on the simulated Galactic all-sky maps.
In particular, we present the mixture matrix $B$ and the setup of the maps schematically.
As described in Section~\ref{sec:simulation} we use three different angular power spectra to simulate Galactic all-sky maps mimicking diversity of real observations.
Figure~\ref{flowchart} summarizes the different steps that are applied to the spectra.\\
The matrix $B$, which mixes the Gaussian random fields into the six mock maps, is given by
\begin{equation*}
B=
\renewcommand*{\arraystretch}{1.5}
\setlength\arraycolsep{10pt}
\begin{pmatrix}
0.18 & 0 & 0.09  \\
0 & 2.74\cdot 10^{-7} & 9.14\cdot 10^{-8}  \\
2 & 3 & 0  \\
14.72 & 30.32 & 2.60\\
104 & 49 & 0.5 \\
0 & 53.97 & 0.71
\end{pmatrix}
.
\end{equation*}
\textbf{We choose this parameters via trial-and-error to mimic Galactic all-sky maps with as diverse structures as possible.}

\tikzstyle{line} = [draw, -latex']
\tikzstyle{cloud} = [ellipse, draw, fill=red!20, text width=7em, node distance=3cm,  text centered, minimum height=3em]
\tikzstyle{ball} = [circle, draw, fill=blue!20,  align=center, anchor=north, inner sep=3]
\begin{figure}
\centering
\begin{tikzpicture}[%
    >=triangle 60,              
    start chain=going below,    
    node distance=2cm, 
    auto,   
    ]
    \node [ball] (input1) {$C_1(\ell)$};
	\node [ball, right of=input1] (input2) {$C_2(\ell)$};
	\node [ball, right of=input2] (input3) {$C_3(\ell)$};
     \node [cloud, below of=input2, node distance=2cm] (Gauss) {Gaussian random fields};
     \node [ball, below of=Gauss, node distance=2cm] (g2) {$g_2(x)$};
	\node [ball, left of=g2] (g1) {$g_1(x)$};
	\node [ball, right of=g2] (g3) {$g_3(x)$};
     \node [cloud, below of=g2, node distance=2cm] (M) {Galactic disk profile $h(x)$};
    \node [ball, below of=M, node distance=2cm] (rho2) {$\rho_2(x)$};
	\node [ball, left of=rho2] (rho1) {$\rho_1(x)$};
	\node [ball, right of=rho2] (rho3) {$\rho_3(x)$};
	\node [cloud, below of=rho2, node distance=2cm] (mixture) {Mixture matrix $B$};
     \node [ball, left of=mixture, node distance=3cm, fill=green!20] (f1) {$d_1(x)$};
     \node [ball, right of=mixture, node distance=3cm, fill=green!20] (f6) {$d_6(x)$};
     \node [ball, below of=mixture, node distance=2cm, xshift=-1cm, fill=green!20] (f3) {$d_3(x)$};
     \node [ball, left of=f3, fill=green!20] (f2) {$d_2(x)$};
	 \node [ball, right of=f3, fill=green!20] (f4) {$d_4(x)$};
     \node [ball, right of=f4, fill=green!20] (f5) {$d_5(x)$};
    \path [line] (input1) -- (Gauss);
    \path [line] (input2) -- (Gauss);
    \path [line] (input3) -- (Gauss);
    \path [line] (Gauss) -- (g1);
    \path [line] (Gauss) -- (g2);
    \path [line] (Gauss) -- (g3);
    \path [line] (g1) -- node[draw,dotted, left, pos=0.7, xshift=-22pt] {$+$}(M);
    \path [line] (g2) -- node{}(M);
    \path [line] (g3) -- node[draw,dotted, pos=0.4, xshift=15pt] {$+$}(M);
    \path [line] (M) -- node [draw,dotted, left, pos=0.4, xshift=-15pt]{exp}(rho1);
    \path [line] (M) -- node {}(rho2);
    \path [line] (M) -- node [draw,dotted, pos=0.7, xshift=8pt]{exp}(rho3);
    \path [line] (rho1) -- node [draw,dotted, left, pos=0.7, xshift=-22pt]{$\times$}(mixture);
    \path [line] (rho2) -- node {}(mixture);
    \path [line] (rho3) -- node [draw, dotted, pos=0.4, xshift=15pt]{$\times$}(mixture);
    \path [line] (mixture) -- (f1);
    \path [line] (mixture) -- (f2);
    \path [line] (mixture) -- (f3);
    \path [line] (mixture) -- (f4);
    \path [line] (mixture) -- (f5);
    \path [line] (mixture) -- (f6);
\end{tikzpicture}
\caption[]{Schematical representation of the setup of the simulated Galactic all-sky maps, the green circles highlight the output that is used for further analysis in Section \ref{Chap:Ver}}
\label{flowchart}
\end{figure}
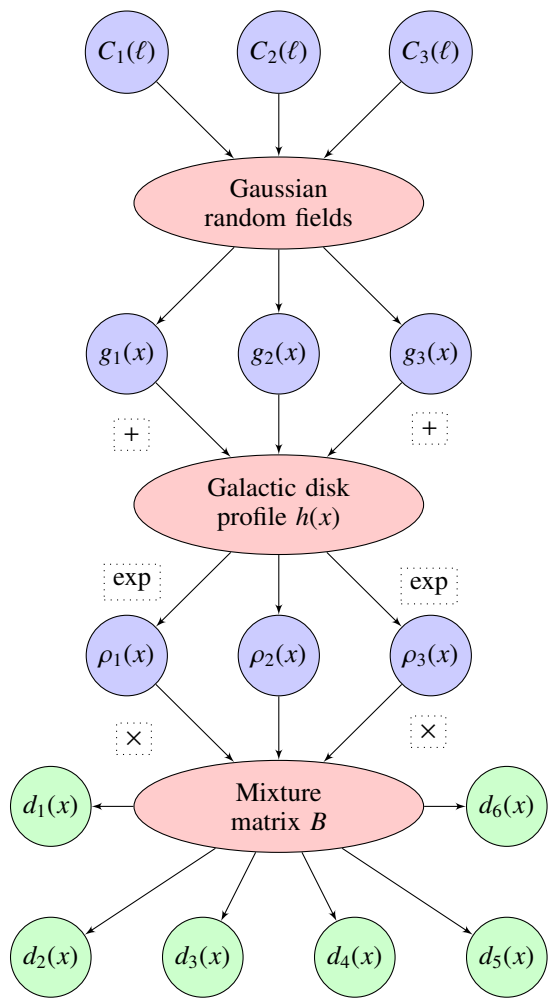
\section{Uncertainty estimate of the GMM}
\label{App:std}
\begin{figure}[h!]
\centering
 \subfloat[\label{difference}]{\includegraphics[trim=0 60 0 0, clip, width=\columnwidth]{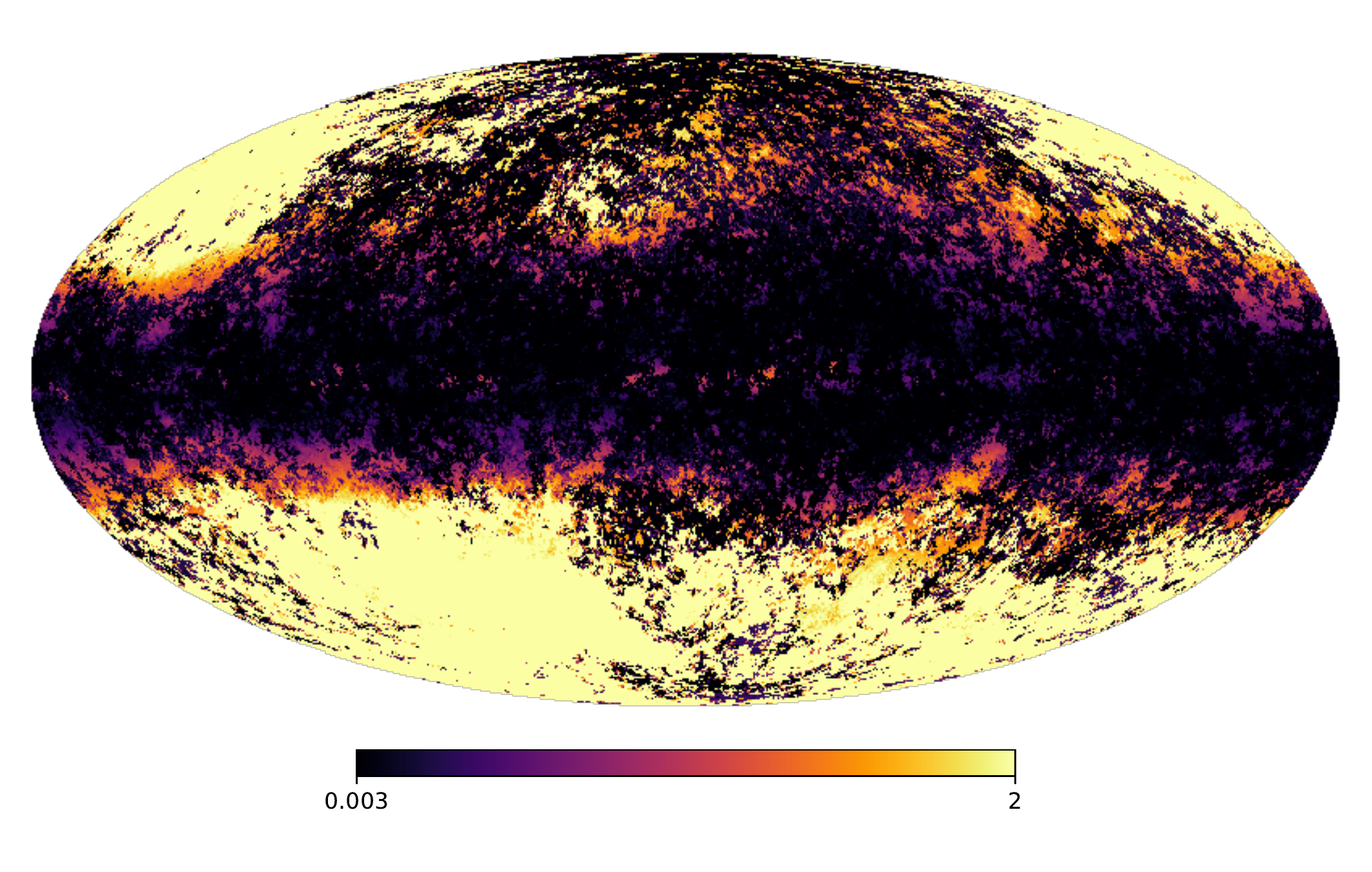}}
 \\
   \subfloat[\label{error}]{\includegraphics[width=\columnwidth]{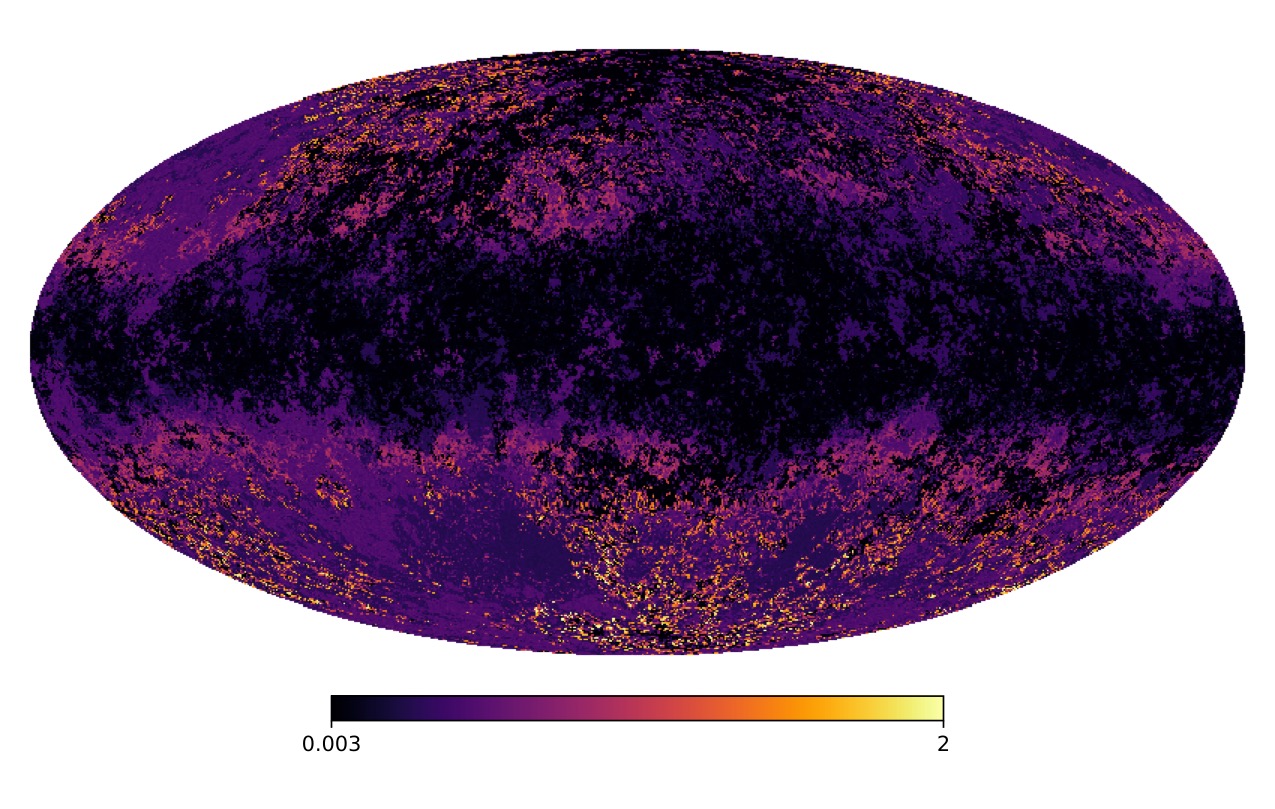}}
  \caption[]{Comparison of the difference and error map directly computed from the GMM; logarithmic scaling. Panel (a): Absolute values of the difference between the original Mock\,I map and the prediction computed from only the disk information. Panel (b): Standard deviation pixel-wise computed for the prediction of Mock\,I computed from only the disk information.}
  \label{estimate}
\end{figure}

The GMM not only provides the posterior mean of a pixel magnitude given other magnitudes but also posterior uncertainty information in form of a standard deviation.
In Figure \ref{estimate} we compare the absolute reconstruction errors (absolute values of the original magnitudes minus reconstructed magnitudes) in panel (a) to the GMM estimate of the posterior uncertainty in panel (b).
It is apparent that the GMM in this case is drastically underestimating the uncertainties. 
This overconfidence of the GMM in its results is probably caused by the fact that Gaussians are not well suited to represent fat-tail distributions.
\section{Selected data sets} \label{App:Sds}
We use a large set of currently available all-sky maps at frequencies distributed over the whole electromagnetic spectrum to be analyzed with the GMM. These maps are briefly described in the following, and their relevant parameters are summarized in Table~\ref{tab:data}.\\[1ex]
\textbf{Haslam Continuum Survey} ($\sim 10^8$\,Hz)\\
In \cite{Haslam1} and \cite{Haslam2} observations of the Jodrell Bank MkI (1965--1966) and MkIA (1973--1975), Bonn 100\,m (1971--1972), and Parkes 64\,m (1978) telescopes were combined to a 408\,MHz all-sky map.
It represents the radio continuum synchrotron radiation at 408\,MHz characterizing the magnetic field perpendicular to the line of sight.
The provided data set is denoised, destriped, and only partly point source removed.\\[1ex]
\textbf{HI4PI Line Emission Survey} ($\sim 10^9$\,Hz)\\
The HI4PI column density map has been observed by two surveys, which share similar angular resolution and match well in sensitivity \citep{HI4PI}.
Measurements from these two surveys, the Effelsberg-Bonn HI survey (EBHIS, 2008--2013) and the Galactic All Sky Survey (GASS, 2005--2006), are combined to a column density all-sky map observing the 21\,cm atomic neutral hydrogen line.
The $N_{\textrm{HI}}$ map contains Milky Way disk material, features residing in the halo, clouds, and extragalactic objects.\\[1ex]
\textbf{1420\,MHz Continuum Survey} ($\sim 10^9$\,Hz)\\
The measurements of the Stockert observatory (northern sky) are combined with those of the Villa Elisa observatory (southern sky) to an all-sky map \citep{Bonn1, Bonn2, Bonn4}.
These surveys also include the Galactic synchrotron emission but at 1420\,MHz measuring the radio continuum around the 21\,cm atomic neutral hydrogen line.
Therefore, the 21\,cm line emission is blocked using a bandstop filter \citep{Bonn3}.
The provided map is contaminated by point, compact, and extragalactic sources over the whole sky.\\[1ex]
\textbf{Planck Continuum and Line Emission Survey} ($\sim 10^{10}- 10^{11}$\,Hz)\\
These and the two following data samples (AKARI and IRIS infrared surveys) are affected by zodiacal light and are background corrected for its influence. However, considerable residuals are still present in the data.\\
The Planck data was taken by two different instruments, the LFI and HFI.
They observed the microwave range in total intensity and polarization \citep{Planck2}.
Here, we only use the microwave emission in total intensity.
With respect to the analysis done in \cite{2015PhRvL.114j1301B} the LFI survey is predominantly observing synchrotron emission while the HFI survey is dominated by dust emission.
The provided all-sky maps are based on measurements from 2009 to 2013.
All maps used here were corrected for the (CMB) temperature fluctuations.
Since the zero-level calibration does not measure the absolute sky background, negative values appear by CMB subtraction \citep{Planck1, Planck3, Planck5, Planck6}.
Hence, an offset needs to be added before further processing with the GMM.
The maps are not provided point source corrected.
There is a routine to extract the sources used to generate foreground maps by \cite{CO}, but its application to the single frequency data is not published and therefore needs to be done again.\\[1ex]
\textbf{Planck CO Determined Line Emission} ($\sim 10^{10}- 10^{11}$\,Hz)\\
Several foreground maps have been determined by \cite{CO} measuring physical phenomena separately.
Since the CO line emission is implicitly measured by the single frequency measurements it has been determined by the algorithm described in \cite{CO}.\\[1ex]
\textbf{AKARI Continuum Survey} ($\sim 10^{12}$\,Hz)\\
The AKARI satellite \citep{Akari1, Akari2} is a far infrared (FIR) survey in a regime, which is not observed with IRAS.
It contains the thermal radiation of heated interstellar dust radiating in the FIR.
Although the data are provided with point sources removed, there are still distinct sources visible. Additionally, there are observational artifacts visible in form of thin stripes.\\[1ex]
\textbf{IRIS Continuum Survey} ($\sim 10^{12}-10^{13}$\,Hz)\\
The InfraRed Astronomical Satellite (IRAS) provides all-sky observations in the mid infrared \citep{Iris2}.
In the actual IRIS maps the latest data sets based on the measurement from 1983 are published with several calibration advances \citep{Iris1}. It includes a correction for zodiacal light, moving objects, and residual glitches, however leaving considerable zodiacal background residuals in the data (except for the 100\,$\mu$m measurement).\\[1ex]
\textbf{SHASSA, VTSS and WHAM Line Emission Survey} \linebreak ($\sim 10^{14}$\,Hz)\\
The heterogeneous data of the Southern H-Alpha Sky Survey Atlas (SHASSA, 1997--2000), the Virginia Tech Spectral-line Survey (VTSS, before 2001), and Wisconsin H-Alpha Mapper (WHAM, 1996--1998) were combined to an all-sky H$\alpha$ map by \cite{finkbeiner}.
Hence, the provided all-sky map shows a diversity in resolution and zero point calibration.
This data set is representing an all-sky description of the diffuse ISM distribution, measuring the hydrogen line emission at 656.3\,nm.\\[1ex]
\begin{figure*}[h!]
\centering
\includegraphics[width=\linewidth]{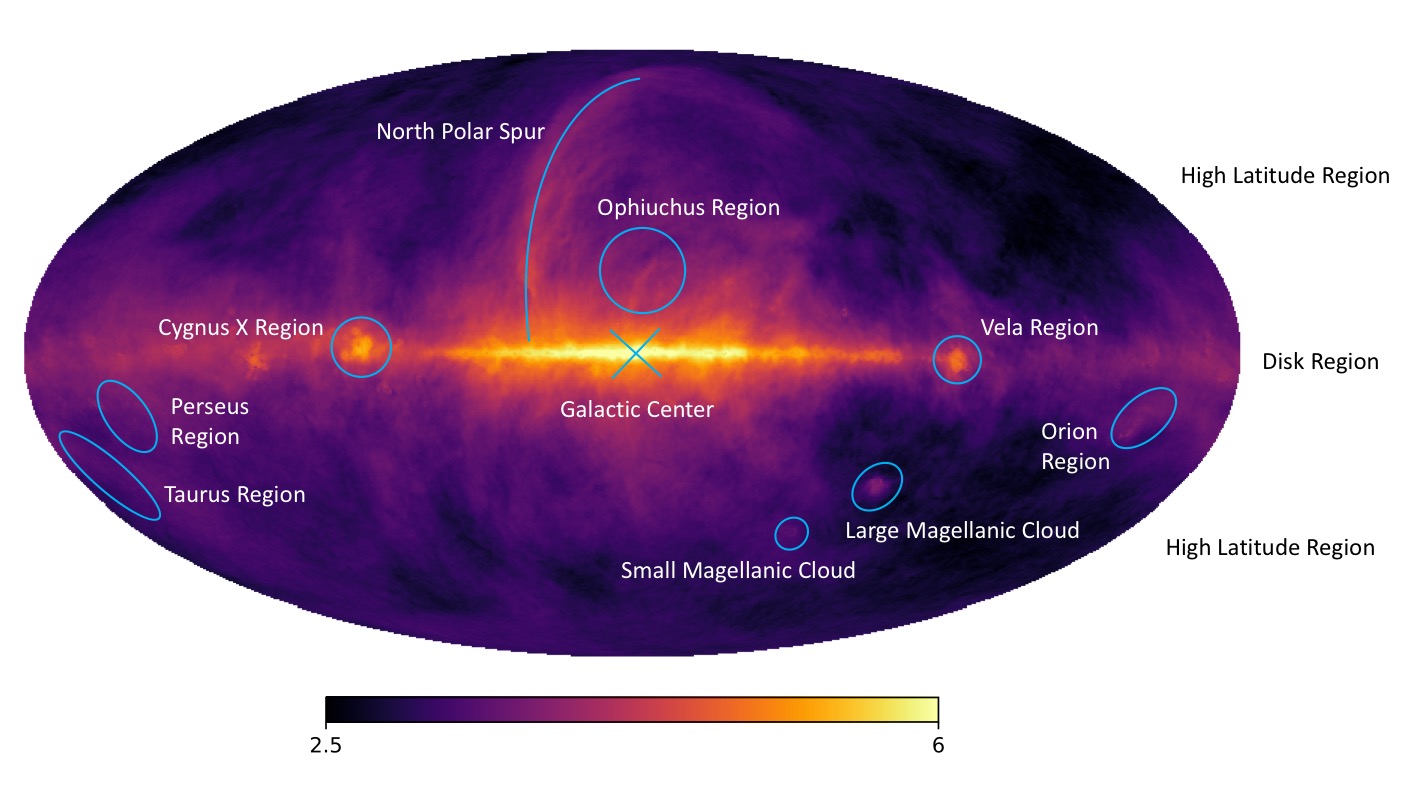}
\caption[]{Pertinent regions of the Milky Way marked for further reference on top of the 408\,MHz Haslam map; logarithmic scaling}
\label{fig:regions}
\end{figure*}
\textbf{ROSAT Soft-X-ray Continuum Survey} ($\sim 10^{19}- 10^{20}$\,Hz)\\
The ROentgen SATellite (ROSAT) is an X-ray survey measuring the diffuse components and point sources as well as scattered solar photons and the particle background \citep{Rosat5,Rosat4,Rosat3,Rosat2} over the whole sky.
These maps are derived from the data that were collected between 1990 and 1991.
Due to the observation method stripes appeared parallel to the scan direction that are unfortunately not corrected in the maps provided in HEALPix format by \cite{Rosat1}, \textbf{which are based on the original data \citep{Rosat5}}.
These stripes have been smoothed using a Gaussian kernel with a width of $\sigma = 0.002$\,rad ($\sim 6.9\,^\prime$).\\[1ex]
\textbf{FGST} ($\sim 10^{23}-10^{25}$\,Hz)\\
The Fermi $\gamma$-ray Space Telescope (FGST) measures the $\gamma$-ray sky from 0.55\,GeV to 294\,GeV \textbf{\citep{Fermi3}}.
Most photons in the GeV range originate from cosmic rays, hadronic interaction of cosmic ray nuclei with the ISM, and inverse Compton scattering of electrons with the background photons.
With the D$^{3}$PO algorithm described by \cite{Fermi2} the data has been denoised as well as deconvolved and decomposed providing nine different diffuse emission maps (point-source free) from the so-called hadronic and leptonic component for the $\gamma$-ray sky \citep{Fermi1} that were extracted by spectral template fitting.\\[1ex]
 \textbf{FGST--Derived Component Maps} ($\sim 10^{23}-10^{25}$\,Hz)\\
As mentioned above the provided FGST measurements are used to decompose the $\gamma$-ray sky.
This results in a two-component description of which the authors claim that one component is predominantly composed of leptonic and the other of hadronic emission \citep{Fermi1}.
\longtab[1]{
\begin{landscape}
\def\arraystretch{2}
{\footnotesize
\begin{longtable}[h]{p{2cm}p{1.5cm}>{\raggedleft}p{2cm}>{\raggedleft}p{1.6cm}>{\raggedleft}p{1.2cm}p{1cm}p{1cm}>{\centering}p{1cm}>{\centering}p{1cm}>{\centering}p{1cm}>{\centering}p{1.9cm}r}
    \hline
\hline
    \textbf{survey} & \textbf{domain} & \textbf{$\lambda$} & \textbf{$\Delta \lambda$ } & \textbf{res} & \textbf{nside}  & \textbf{pub} & \textbf{noise} & \textbf{sources} & \textbf{flux} & \textbf{background} & \textbf{coverage}
    \\
\hline
        Haslam & Radio & 408\,MHz& --- & 56$^\prime$.0 & 512 & 2014  & \ding{51} & \ding{51} & \ding{51} & \ding{55} & 100\,\% \\
     HI4PI & Radio & 21\,cm $\rightarrow$ $N_{\textrm{HI}}$ & 100\,MHz 8\,MHz & 16$^\prime$.2 & 1024 & 2016  & \ding{51} & \ding{55}  & \ding{51} & \ding{55} & 100\,\% \\
       Stockert, Villa Elisa & Radio & 1420\,MHz 1435\,MHz 1420\,MHz& 18\,MHz 14\,MHz 13\,MHz&  35$^\prime$.4  &   512     &  ---     &  \ding{51}     &   \ding{55}    &  \ding{51}   &    \ding{55}  &  100\,\% \\
        Planck LFI & Microwave & 30\,GHz 44\,GHz 70\,GHz & $\Delta \nu/ \nu= 0.2$ & 24$^\prime$.0 & 1024  & 2015  & \ding{51} & \ding{55} & \ding{51} & \ding{51} & 100\,\% \\
          Planck HFI & Microwave & 100\,GHz 143\,GHz 217\,GHz 353\,GHz 545\,GHz 857\,GHz & $\Delta \nu / \nu = 0.3$ & 5$^\prime$.5  & 2048 & 2015  & \ding{51}& \ding{55}  & \ding{51} & \ding{51} & 100\,\% \\
          Planck CO & Microwave & 115.3\,GHz  & --- & 15$^\prime$  & 2048 & 2015  & \ding{51}& \ding{55}  & \ding{51} & \ding{51} & 100\,\% \\
         AKARI & FIR   & 65\,$\mu$m \\90\,$\mu$m \\140\,$\mu$m \\160\,$\mu$m & 30\,$\mu$m 50\,$\mu$m 70\,$\mu$m 40$\mu$m & 1$^\prime$.3 & 4096 & 2015  & \ding{51} & \ding{51}  & \ding{51} & \ding{51} & $>99$\,\% \\
         IRIS  & IR    & 12\,$\mu$m\\ 25\,$\mu$m \\60\,$\mu$m \\100\,$\mu$m & 6.5\,$\mu$m 11.0$\mu$m 40.0\,$\mu$m 37.0\,$\mu$m & 4$^\prime$.0  & 1024   & 2006  & \ding{55} & \ding{55}  & \ding{51} & \ding{51} & 98\,\% \\
             Finkbeiner & H$\alpha$ & 656.3\,nm & 1.7\,nm 3.2\,nm 4.0\,nm & 6$^\prime$.0 & 512  & 2003  & \ding{55} & \ding{55}  & \ding{51} & \ding{55} & 100\,\% \\
    ROSAT & X-ray & 0.197\,keV 0.212\,keV 0.725\,keV 0.885\,keV 1.145\,keV 1.545\,keV & 0.174\,keV 0.144\,keV 0.570\,keV 0.650\,keV 0.830\,keV 0.990\,keV & 120$^\prime$.0   & 512   & 2015  & \ding{55} & \ding{55}  & \ding{51} & \ding{51} & 98\,\% \\
    Fermi & $\gamma$-ray & 0.85\,GeV 1.70\,GeV 3.40\,GeV 6.79\,GeV 13.58\,GeV 27.15\,GeV 54.31\,GeV 108.61\,GeV 217.22\,GeV & 0.6\,GeV 1.2\,GeV 2.4\,GeV 4.8\,GeV 9.6\,GeV 19.2\,GeV 38.4\,GeV 76.8\,GeV 153.6\,GeV & --- & 128  & 2015  & \ding{51} & \ding{51}  & \ding{51} & \ding{51}& 100\,\% \\
     Fermi & hadronic, leptonic & --- &  --- & --- &  128   &  2015    & \ding{51} & \ding{51}  & \ding{51} & \ding{51}& < 90\,\% \\
    \hline
  \caption[An Overview of the used data sets]{An overview of the used data sets representing parts of the whole electromagnetic spectrum; unfilled gaps mean that the effect is not relevant for the survey. \textbf{Table header}: $\lambda$ stands for central wavelength or frequency, $\Delta \lambda$ stands for bandwidth, res is short for resolution of the instrument (not the resolution defined by the nside), nside is defining the actual resolution, pub is short for the year of publication; further it is stated whether the data is noise corrected and therefore destriped, partly point and extragalactic source removed, flux calibrated, and background corrected or not; coverage stands for the total all-sky coverage.}
   \label{tab:data}%
\end{longtable}}%
\end{landscape}
}
\section{Milky Way Regions} \label{app:regions}
We present the 408\,MHz Haslam map in Figure~\ref{fig:regions} to highlight separate regions.
These emission structures are marked in blue and labeled by their literature name\footnote{The designations follow the Galactic and extragalactic sources from \cite{2008MNRAS.388..247D}, \url{http://sci.esa.int/jump.cfm?oid=47340}, and \url{https://www.esa.int/spaceinimages/Images/2007/07/All-sky_map_in_infrared_light_with_constellations_and_star_forming_regions}}.
For the by-eye comparison in Section \ref{Chap:Res} we focused on the Galactic center, the Perseus, Taurus, Ophiuchus, Orion, Cygnus X, and Vela region, and the Large and Small Magellanic Cloud as well as the North Polar Spur.
Furthermore, the areas we refer to as disk and high-latitude regions are marked.

\section{X-ray data reproduced with more Gaussians} \label{App:xray}
\begin{figure}[h!]
\centering
 \subfloat[\label{predxray18}]{\includegraphics[width=\columnwidth]{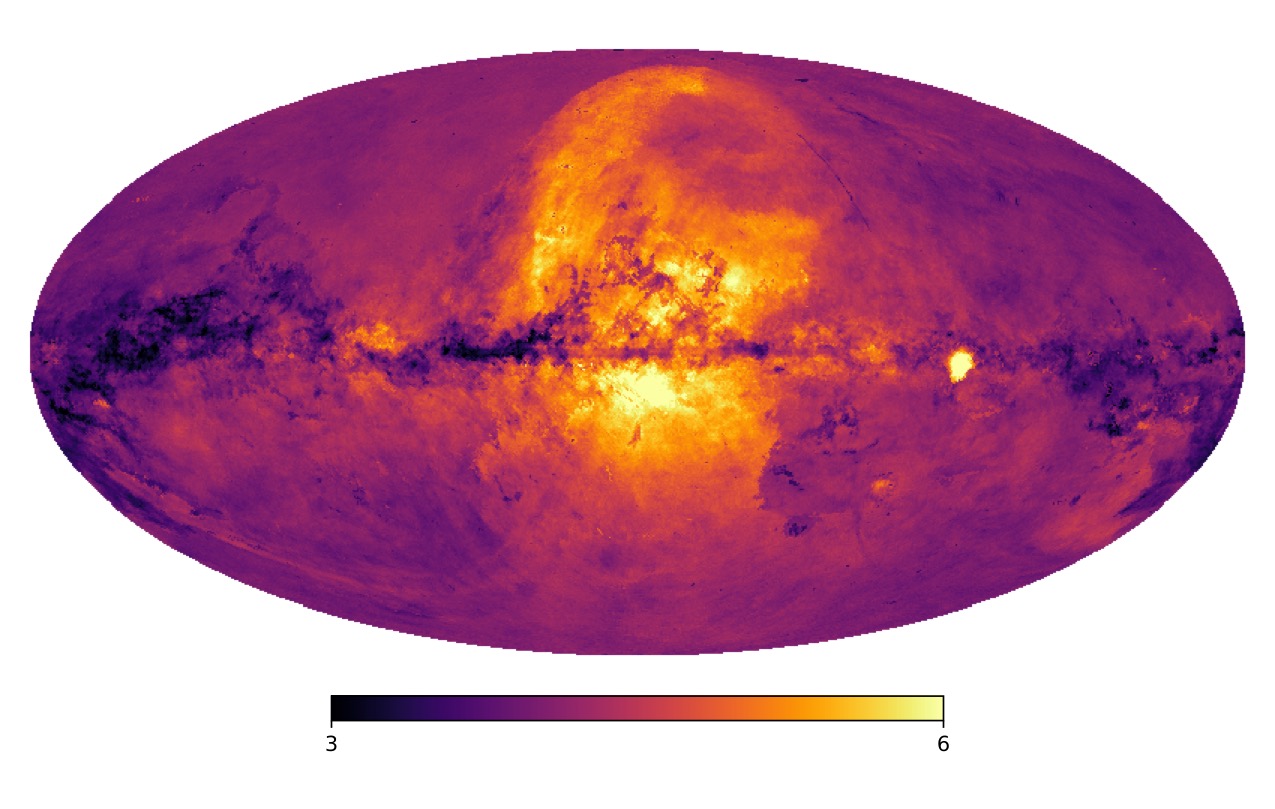}}
 \\
   \subfloat[\label{predorg-xray18}]{\includegraphics[width=\columnwidth]{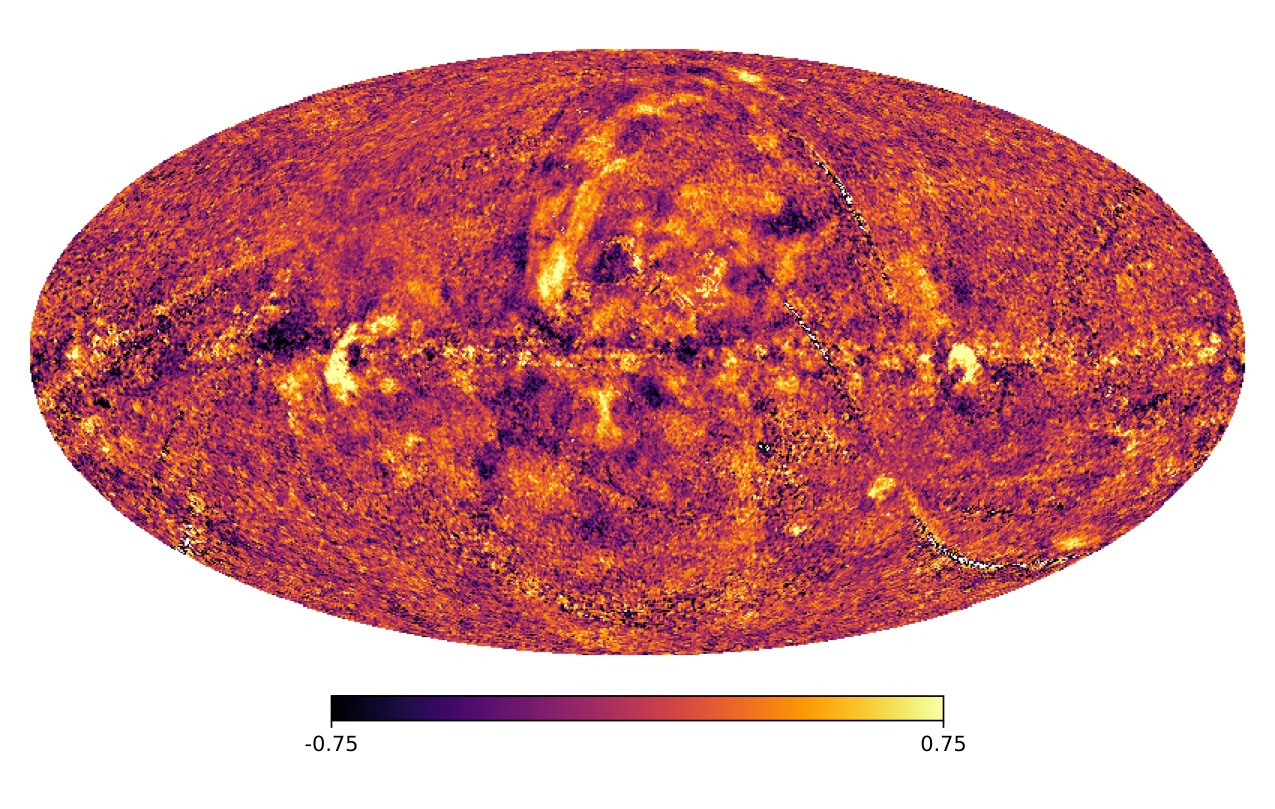}}
  \caption[]{GMM results for the ROSAT 0.885\,keV map computed with $K=18$ Gaussians and $n=37$ data sets neglecting the $\gamma$-ray component maps. The prediction is determined from the CPD marginalized over the X-ray regime; logarithmic scaling. Panel (a): Prediction of the 0.885\,keV map. Panel (b): Difference between the original and predicted 0.885\,keV map, $\textrm{RMS} = 0.31$.}
  \label{xray18}
\end{figure}

Here, we show the reproduction of the X-ray data set at 0.855\,keV computed from a GMM trained with $K=18$ components in comparison to the reproduction computed with $K=3$ components discussed in Section \ref{Chap:xray}.
We use the $n=37$ input data sets specified in Section \ref{Chap:xray}.
The reproduced map in Figure~\ref{predxray18} is showing minor improvements in resolution of the magnitudes close to the noise level.
Further, the magnitudes of the missing component are slightly better reproduced, however as discussed before a higher number of Gaussians cannot measure this emission.\par
The most noticeable difference to the reproduction shown in Figure \ref{pred33} is the influence of different artifacts such as the zodiacal light residuals and the unobserved stripes of the AKARI survey, as well as reproduction residuals (see, e.g, the region north-east of the Large Magellanic Cloud).\\
Overall, the difference map in Figure~\ref{predorg-xray18} does not show a significantly improved reproduction of the X-ray data set in comparison to the difference map computed from a GMM trained with $K=3$ Gaussians (Fig.~\ref{org3-pred33}). The reproduction computed with $K=18$ Gaussians is in comparison affected by artifacts.
\section{Gaussian component analysis}
\label{App:color}
\begin{figure}[h!]
\centering
\includegraphics[width=\columnwidth]{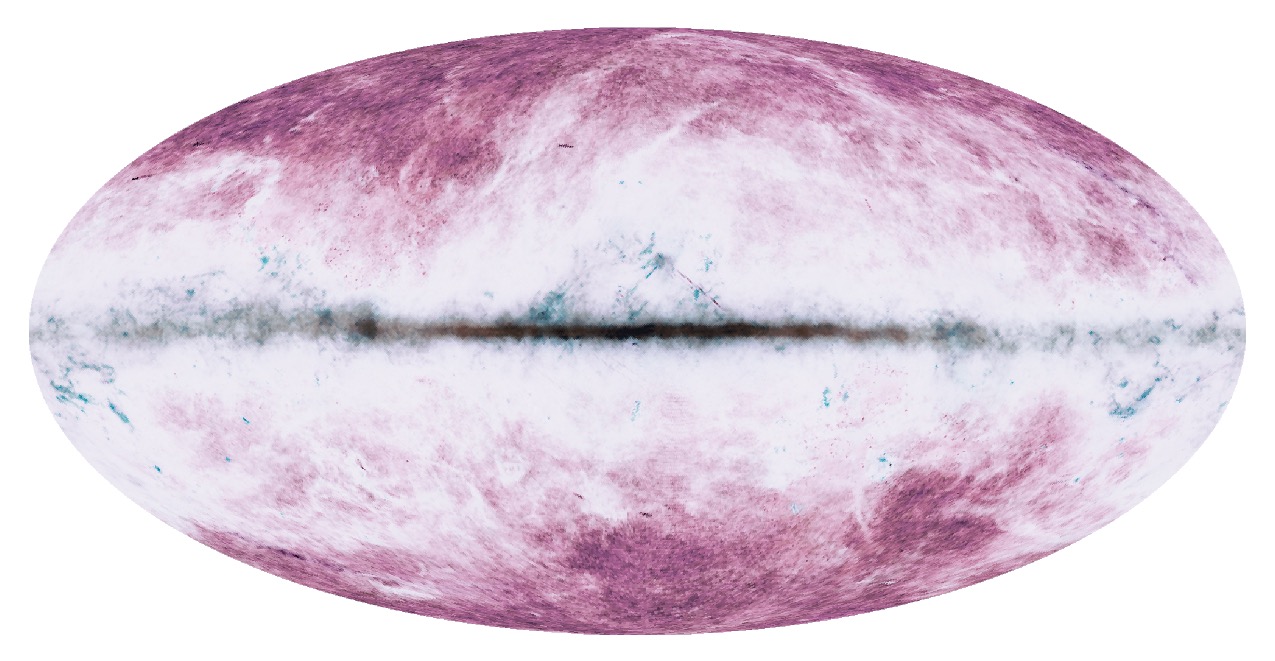}
\caption[]{Color coding of the probability of the individual GMM components for the magnitude vector of a pixel. Each GMM component is associated with the color (red, green, blue). The intensity of the color is chosen according to Equation \ref{eq:color}.
A high log-probability of a pixel magnitude vector resulting from a specific component is here represented by a high numerical value in the commonly used RGB color scheme, which encodes brightness. White pixels are likely for all GMM components, black pixels for none. The galactic disk appears black due to the extreme values of its magnitudes that are unlikely for all components. The red, green, and blue regions are likely for the corresponding components.}
\label{fig:rgb}
\end{figure}

Here, we analyze in which way the three GMM components represent the data.
The hope is that the different components can be associated with different phases of the ISM, such as cold, warm and hot phases.
To visualize the relation of pixels with GMM components, we attribute each component $k$ to a special $\textrm{color}\in \textrm\{R, G, B\}$ (with R, G, B representing red, green, and blue) by computing the affiliation of each pixel $i$ to each component $k$ via
\begin{equation}
\textrm{color}_k=\textrm{ln}\left(\pi _k \, \frac{1}{\sqrt{|2 \pi \Sigma_k|}}\, \textrm{exp} \left\{- \frac{1}{2}(d_i - \mu _k) \Sigma_k^{-1} (d_i - \mu _k)^\textrm{T} \right\}\right)
\label{eq:color}
\end{equation}
with $\pi_k$ being the weights of the $K$ Gaussian components, $\mu _k$ their means, and $\Sigma _k$ their covariances as well as $d_i$ as multifrequency vector for each pixel.
We do not find a simple association of the components with the physical phases of the ISM.
Here, we show the $n=37$ dimensional pixel-wise color representation of the $K=3$ Gaussian components in Figure \ref{fig:rgb}.
\end{appendix}
\end{document}